\documentclass[12pt]{article}
\usepackage{array}
\usepackage{tabularx}
\usepackage{amsmath,amssymb,amsfonts,epsfig,cite,setspace,bigstrut,framed}
\usepackage[all]{xy}
\usepackage{color}
\usepackage{pifont}
\usepackage{xcolor}
\usepackage{tikz}
\usepackage{tikz-cd}
\usetikzlibrary{shapes.geometric}
\usepackage[colorlinks=true,urlcolor=blue,anchorcolor=blue,citecolor=blue,filecolor=blue,linkcolor=blue,menucolor=blue,linktocpage=true,pdfproducer=medialab,pdfa=true]{hyperref}	 
\usetikzlibrary{positioning}
\usepackage[perpage]{footmisc}
\usepackage[compat=1.1.0]{tikz-feynman}
\usepackage{colortbl}
\definecolor{dgreen}{rgb}{0, 0.55, 0}
\definecolor{llightyellow}{rgb}{1.0, 0.95, 0.7}
\definecolor{llightblue}{rgb}{0.7, 0.9, 1.0}
\definecolor{llightpink}{rgb}{1.0, 0.85, 0.95}
\definecolor{llightgreen}{rgb}{0.7, 1.0, 0.4}
\colorlet{lightyellow}{llightyellow!50!white}
\colorlet{lightblue}{llightblue!50!white}
\colorlet{lightgreen}{llightgreen!50!white}
\colorlet{lightpink}{llightpink!50!white}

\usepackage{mathtools}

\newcommand{\bra}[1]{\langle{#1}|}
\newcommand{\ket}[1]{|{#1}\rangle}

\newcommand{\dsl}{\pa \kern-0.5em /}

\newcommand{\pa}{\partial}

\newcommand{\bit}{\begin{itemize}}
\newcommand{\eit}{\end{itemize}}

\newcommand{\bea}{\begin{eqnarray}}
\newcommand{\eea}{\end{eqnarray}}
\newcommand{\be}{\begin{equation}}
\newcommand{\ee}{\end{equation}}

\newcommand{\ba}{\begin{array}}
\newcommand{\ea}{\end{array}}

\makeatletter \@addtoreset{equation}{section} \makeatother
\renewcommand{\theequation}{\thesection.\arabic{equation}}

\begin{document}

\vskip 0.25in

\newcommand{\todo}[1]{{\bf\color{blue} !! #1 !!}\marginpar{\color{blue}$\Longleftarrow$}}
\newcommand{\comment}[1]{}
\newcommand\T{\rule{0pt}{2.6ex}}
\newcommand\B{\rule[-1.2ex]{0pt}{0pt}}

\newcommand{\CO}{{\cal O}}
\newcommand{\cI}{{\cal I}}
\newcommand{\cM}{{\cal M}}
\newcommand{\cW}{{\cal W}}
\newcommand{\cN}{{\cal N}}
\newcommand{\cR}{{\cal R}}
\newcommand{\cH}{{\cal H}}
\newcommand{\cK}{{\cal K}}
\newcommand{\cT}{{\cal T}}
\newcommand{\cZ}{{\cal Z}}
\newcommand{\cO}{{\cal O}}
\newcommand{\cQ}{{\cal Q}}
\newcommand{\cB}{{\cal B}}
\newcommand{\cC}{{\cal C}}
\newcommand{\cD}{{\cal D}}
\newcommand{\cE}{{\cal E}}
\newcommand{\cF}{{\cal F}}
\newcommand{\cA}{{\cal A}}
\newcommand{\cX}{{\cal X}}
\newcommand{\IA}{\mathbb{A}}
\newcommand{\IP}{\mathbb{P}}
\newcommand{\IQ}{\mathbb{Q}}
\newcommand{\IH}{\mathbb{H}}
\newcommand{\IR}{\mathbb{R}}
\newcommand{\IC}{\mathbb{C}}
\newcommand{\IF}{\mathbb{F}}
\newcommand{\IS}{\mathbb{S}}
\newcommand{\IV}{\mathbb{V}}
\newcommand{\II}{\mathbb{I}}
\newcommand{\IZ}{\mathbb{Z}}
\newcommand{\re}{{\rm Re}}
\newcommand{\im}{{\rm Im}}
\newcommand{\tr}{\mathop{\rm Tr}}
\newcommand{\ch}{{\rm ch}}
\newcommand{\rk}{{\rm rk}}
\newcommand{\ext}{{\rm Ext}}
\newcommand{\bi}{\begin{itemize}}
\newcommand{\ei}{\end{itemize}}
\newcommand{\beq}{\begin{equation}}
\newcommand{\eeq}{\end{equation}}

\newcommand{\CN}{{\cal N}}
\newcommand{\y}{{\mathbf y}}
\newcommand{\z}{{\mathbf z}}
\newcommand{\C}{\mathbb C}\newcommand{\R}{\mathbb R}
\newcommand{\CA}{\mathbb A}
\newcommand{\CP}{\mathbb P}
\newcommand{\cP}{\mathcal P}
\newcommand{\tmat}[1]{{\tiny \left(\begin{matrix} #1 \end{matrix}\right)}}
\newcommand{\mat}[1]{\left(\begin{matrix} #1 \end{matrix}\right)}
\newcommand{\diff}[2]{\frac{\partial #1}{\partial #2}}
\newcommand{\gen}[1]{\langle #1 \rangle}

\pdfstringdefDisableCommands{\def\eqref#1{(\ref{#1})}}

\newtheorem{theorem}{\bf THEOREM}
\newtheorem{proposition}{\bf PROPOSITION}
\newtheorem{observation}{\bf OBSERVATION}
\newtheorem{statement}{\bf STATEMENT}

\def\theequation{\thesection.\arabic{equation}}
\newcommand{\setall}{
	\setcounter{equation}{0}
}
\renewcommand{\thefootnote}{\fnsymbol{footnote}}

\begin{titlepage}
\vfill
\begin{flushright}
{\tt\normalsize KIAS-P23043}\\

\end{flushright}
\vfill
\begin{center}
{\Large\bf Symmetry TFT for Subsystem Symmetry}

\vskip 1.5cm

Weiguang Cao$^{1,2}$ and Qiang Jia$^3$
\vskip 5mm

\it{
\begin{tabular}{ll}
		$^1$&Kavli Institute for the Physics and Mathematics of the Universe,\\
		& University of Tokyo,  Kashiwa, Chiba 277-8583, Japan\\
		$^2$&Department of Physics,  University of Tokyo, Tokyo 113-0033, Japan\\
        $^3$&School of Physics,
Korea Institute for Advanced Study, Seoul 02455, Korea
	\end{tabular}
}

\end{center}
\vfill

\begin{abstract}
We generalize the idea of symmetry topological field theory (SymTFT) for subsystem symmetry. We propose the 2-foliated BF theory with level $N$ in $(3+1)$d as subsystem SymTFT for subsystem $\mathbb Z_N$ symmetry in $(2+1)$d. 
Focusing on $N=2$, we investigate various topological boundaries. The subsystem Kramers-Wannier and Jordan-Wigner dualities can be viewed as boundary transformations of the subsystem SymTFT and are included in a larger duality web from the subsystem $SL(2,\mathbb Z_2)$ symmetry of the bulk foliated BF theory. Finally, we construct the condensation defects and twist defects of $S$-transformation in the subsystem $SL(2,\mathbb Z_2)$, from which the fusion rule of subsystem non-invertible operators can be recovered.

\end{abstract}

\vfill
\end{titlepage}

\tableofcontents

\section{Introduction}
Fracton excitation, a new kind of quasiparticle with restricted mobility, has appeared in new types of exotic phases of matter and received attention from both condensed matter physics~\cite{Haah:2011drr, Haah:2013eia, Vijay:2016phm,Ma:2017aog,Shirley:2017suz,Shirley:2018nhn,Shirley:2018hkm,Slagle:2020ugk, Tian:2018opt, Shen:2021rct} and high energy physics~\cite{Seiberg:2020wsg, Seiberg:2020bhn, Seiberg:2020cxy,SanMiguel:2020xxh,You:2020ykc}. Fracton phases of matter, originally constructed as a candidate for quantum memory~\cite{Haah:2011drr, Haah:2013eia}, are famous for their extensive ground state degeneracy~\cite{Vijay:2016phm, Haah:2011drr, Haah:2013eia}, restricted mobility of excitation~\cite{Vijay:2016phm} and large subleading corrections to the entanglement entropy~\cite{2018PhRvB..97l5102H, 2018PhRvB..97l5101M}. More detail can be found in the reviews~\cite{Nandkishore:2018sel,Pretko:2020cko}. Models with fractons also attract the interest of field theorists because their low energy effective description allows discontinuous field configurations and exhibits exotic UV-IR mixing behavior~\cite{Seiberg:2020wsg, Seiberg:2020bhn, Seiberg:2020cxy,SanMiguel:2020xxh,You:2020ykc}, which challenges our conventional understanding of field theory.

One valid construction of fracton models arises from generalizing the ordinary gauge principal\cite{Pretko:2018jbi} by introducing the tensor gauge theories\cite{Pretko:2016kxt,Pretko:2016lgv,Ma:2018nhd,Bulmash:2018lid}, where gauge fields are tensor representations of the symmetry group. There is another foliation construction~\cite{Slagle:2018swq,Shirley:2018hkm,Shirley:2018nhn,Shirley:2018vtc,Shirley:2019uou,Slagle:2020ugk} where the spacetime manifold is a foliation of lower dimensional submanifold. The gauge invariant operators have restricted mobility in the foliated directions but are topological in the other directions without foliation.\footnote{In \cite{Pace:2022wgl,Oh:2023bnk}, the restricted mobility and UV-IR mixing are also found in rank 2 gauge theory, resulting from the subsystem higher form symmetry.} The two constructions are equivalent through the exotic-foliated duality~\cite{Ohmori:2022rzz,Spieler:2023wkz}. 

From a symmetry point of view, fracton models are often realized by gauging the subsystem symmetry~\cite{ Shirley:2018vtc,Ibieta-Jimenez:2019zsf} or dipole symmetry~\cite{Gorantla:2022eem,Bidussi:2021nmp} 
which generalizes the notion of symmetry by relaxing the topologicalness of the symmetry operators. Therefore, studying these generalized symmetries is of equal importance and will shed light on the underlying structure of fracton models.
In this paper, we will focus on the subsystem symmetry. Subsystem symmetry allows symmetry transformations acting on rigid spatial submanifolds and it is sometimes referred to as ``gauge-like" symmetry~\cite{Batista:2004sc,Nussinov:2006iva,Nussinov:2009zz}. However,  it should be viewed as a global symmetry rather than gauge symmetry because the subsystem symmetry operator acts nontrivially on the Hilbert space. 
It is natural to study subsystem symmetry by generalizing corresponding ideas in ordinary global symmetry, like selection rules~\cite{Gorantla:2021bda}, spontaneously breaking~\cite{Qi:2020jrf,Distler:2021qzc,Rayhaun:2021ocs}, anomaly inflow~\cite{Burnell:2021reh} and constraints on IR dynamics~\cite{Seiberg:2020wsg, Seiberg:2020bhn, Seiberg:2020cxy,SanMiguel:2020xxh,You:2020ykc}.  In particular, we will study the duality web and the generalization of symmetry topological field theory (SymTFT) for subsystem symmetry. 

Duality is a powerful tool in theoretical physics, where the two apparently different Lagrangians describe the same theory. Here we focus on $(1+1)$d quantum field theories (QFTs) where the duality web has been revisited recently from the perspective of gauging a discrete symmetry~\cite{Hsieh:2020uwb, Fukusumi:2021zme,Ebisu:2021acm, Karch:2019lnn,Ji:2019ugf,Gaiotto:2014kfa, Duan:2023ykn}. We are interested in the duality transformation generated by symmetry manipulations such as gauging and stacking invertible phases~\cite{Kapustin:2014dxa,Kapustin:2017jrc, Senthil:2018cru}. For example, gauging the non-anomalous $\mathbb Z_2$ symmetry of $(1+1)$d Ising conformal field theory (CFT) is a self-duality and the corresponding duality defect gives the simplest example of non-invertible symmetry~\cite{Frohlich:2004ef,Aasen:2016dop}. Another famous example is the boson-fermion duality~\cite{Coleman:1974bu,Haldane_1981, Witten:1983ar}, where the Ising CFT is dual to a free Majorana fermion by first stacking a topological phase given by the Arf-invariant (Kitaev Majorana chain) and then gauging the diagonal $\mathbb{Z}_2$ symmetry.  Recently, generalizations of Kramers-Wannier (KW) and Jordan-Wigner (JW) duality has been studied in the context of subsystem symmetry~\cite{Tantivasadakarn:2020lhq,2020arXiv200212026S,Cao:2023doz,Cao:2022lig}, where a new subsystem non-invertible symmetry has been found.

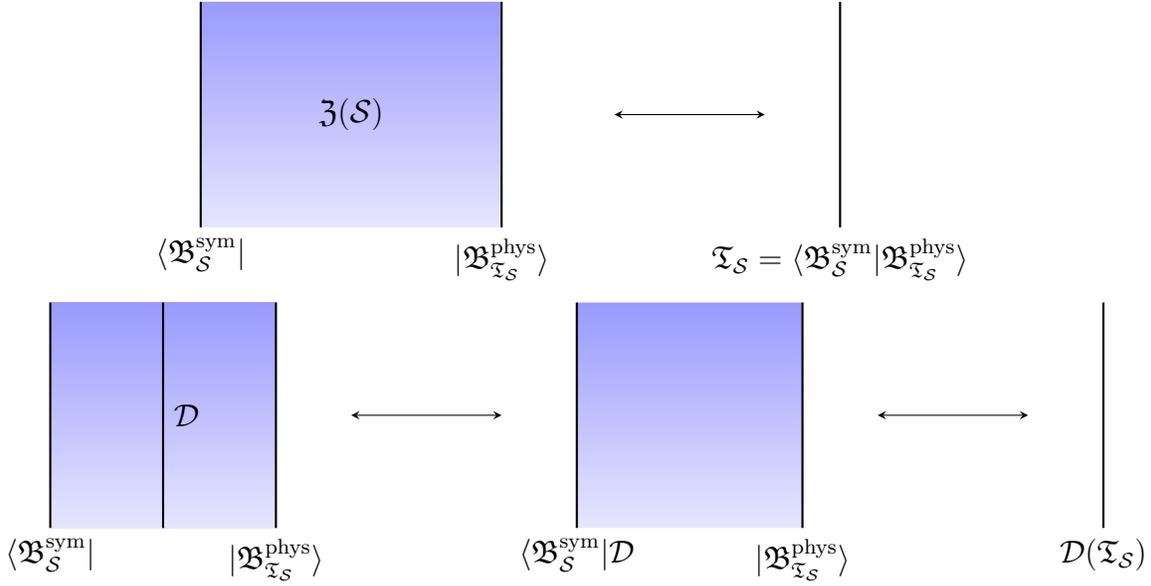
\begin{figure}[htbp]
\centering
\begin{tikzpicture}
        \shade[top color=blue!40, bottom color=blue!10]  (0+2,4) -- (0+2,7) -- (4+2,7) -- (4+2,4)-- (0+2,4);
        \draw[thick] (0+2,4) -- (0+2, 7);
        \draw[thick] (4+2,4) -- (4+2, 7);
        \node[below] at (0+2,4) {$\bra{\mathfrak{B}^{\textrm{sym}}_{\mathcal{S}}}$};
        \node[below] at (4+2,4) {$\ket{\mathfrak{B}^{\textrm{phys}}_{\mathfrak{T}_{\mathcal{S}}}}$};
        \node[] at (2+2,5.5) {$\mathfrak{Z}(\mathcal{S})$};
        \draw [stealth-stealth](5.5+2,5.5) -- (7.5+2,5.5);
        \draw[thick] (8.5+2,4) -- (8.5+2, 7);
        \node[below] at (8.5+2,4) {$\mathfrak{T}_{\mathcal{S}}=\langle \mathfrak{B}^{\textrm{sym}}_{\mathcal{S}}|\mathfrak{B}^{\textrm{phys}}_{\mathfrak{T}_{\mathcal{S}}}\rangle$};
        \shade[top color=blue!40, bottom color=blue!10]  (0,0) -- (0,3) -- (3,3) -- (3,0)-- (0,0);
        \draw[thick] (0,0) -- (0, 3);
        \draw[thick] (1.5,0) -- (1.5, 3);
        \node[right] at (1.5,1.5){$\mathcal{D}$};
        \draw[thick] (3,0) -- (3, 3);
        \node[below] at (0,0) {$\bra{\mathfrak{B}^{\textrm{sym}}_{\mathcal{S}}}$};
        \node[below] at (3,0) {$\ket{\mathfrak{B}^{\textrm{phys}}_{\mathfrak{T}_{\mathcal{S}}}}$};
        \draw [stealth-stealth](4,1.5) -- (6,1.5);
        \shade[top color=blue!40, bottom color=blue!10]  (7,0) -- (7,3) -- (10,3) -- (10,0)-- (8,0);
        \draw[thick] (7,0) -- (7, 3);
        \draw[thick] (10,0) -- (10, 3);
        \node[below] at (7,0) {$\bra{\mathfrak{B}^{\textrm{sym}}_{\mathcal{S}}}\mathcal{D}$};
        \node[below] at (10,0) {$\ket{\mathfrak{B}^{\textrm{phys}}_{\mathfrak{T}_{\mathcal{S}}}}$};
        \draw [stealth-stealth](11,1.5) -- (13,1.5);
        \draw[thick] (14,0) -- (14, 3);
        \node[below] at (14,0) {$\mathcal{D}(\mathfrak{T}_{\mathcal{S}})$};
    \end{tikzpicture}
    \caption{Illustration of the SymTFT. We will get boundary theory after shrinking the slab. When fusing on the boundary, a co-dimension one symmetry defect $\mathcal{D}$ in the SymTFT will change the boundary condition, which corresponds to a symmetry manipulation/duality transformation of the boundary theory. }
    \label{fig:SymTFT}
\end{figure}

SymTFT is another powerful tool that provides a unified picture to study duality transformations and symmetry manipulations\cite{Gaiotto:2020iye,Apruzzi:2021nmk,Lin:2022dhv,Kaidi:2022cpf,vanBeest:2022fss,Kaidi:2023maf,Bhardwaj:2023ayw,Chen:2023qnv,Freed:2022qnc,Duan:2023ykn,Antinucci:2023ezl,Cordova:2023bja,Apruzzi:2023uma,Antinucci:2022vyk}. The idea of SymTFT is illustrated in Fig.~\ref{fig:SymTFT}. Given a $d$-dimensional theory $\mathfrak{T}_{\mathcal{S}}$ with a finite symmetry  $\mathcal{S}$, the SymTFT is a $(d+1)$-dimensional topological quantum field theory $\mathfrak{Z}(\mathcal{S})$ that allows a topological boundary $\mathfrak{B}^{\textrm{sym}}_{\mathcal{S}}$ encoding the symmetry $\mathcal{S}$ of the original theory $\mathfrak{T}_{\mathcal{S}}$.
The original theory $\mathfrak{T}_S$ can be expressed as an interval compactification of $\mathfrak{Z}(\mathcal{S})$ with two boundaries. In the condensed matter literature, the similar idea of SymTFT has been proposed as symmetry/topological order correspondence~\cite{Ji:2019jhk,Kong:2020cie,Pace:2023kyi}.  

The power of SymTFT is that the information of symmetry $\mathcal{S}$ and the dynamics are separately stored in the two boundaries. 
The left boundary is the topological boundary $\mathfrak{B}^{\textrm{sym}}_{\mathcal{S}}$ supporting the symmetry $\mathcal{S}$ and all symmetry manipulations take place on this boundary. The symmetry manipulations are implemented by fusing a co-dimension one symmetry defect of the SymTFT to the topological boundary. The right boundary is the dynamical (physical) boundary $\mathfrak{B}^{\textrm{phys}}_{\mathfrak{T}_{\mathcal{S}}}$ that depends on the details of $\mathfrak{T}_{\mathcal{S}}$. As a concrete example, we give a review of the $(2+1)$ BF theory as a SymTFT in Appendix~\ref{sec:ordinaryBF}. 

In this paper, we will propose a SymTFT for subsystem symmetry. We will focus on subsystem $\mathbb Z_2$ symmetry in $(2+1)$d, which is a 2-foliated theory with one-dimensional layers foliated in all spacial directions $x,y$. The natural candidate for the SymTFT is a theory with the same foliation structure but with an extra topological direction, which turns out to be the 2-foliated BF theory in $(3+1)$d~\cite{Gorantla:2020jpy,Spieler:2023wkz}\footnote{Strictly speaking, it is not a topological field theory in the ordinary sense since the theory is only topological in the directions without foliation. }. This principle to construct subsystem SymTFT can apply to theories in higher dimensions, like the X-cube model, which we leave for future investigation.

Here is the organization of this paper. In Sec.~\ref{sec:sub}, we review the $(2+1)$d subsystem $\mathbb{Z}_2$  symmetry on the lattice and subsystem KW/JW duality transformation. In Sec.~\ref{sec:2foli}, we propose the $(3+1)$d SymTFT for subsystem $\mathbb{Z}_2$ symmetry in $(2+1)$d and study the topological boundary conditions. In Sec.~\ref{sec:sl2z}, we consider the $SL(2,\mathbb{Z}_2)$ symmetry of the subsystem SymTFT and the duality web of the boundary theories. In Sec.~\ref{sec:defects}, we construct the condensation defects and twist defects of $S$-transformation in the subsystem $SL(2,\mathbb Z_2)$.
Finally, we conclude and point out interesting future directions in Sec.~\ref{sec:con}.

\section{Subsystem symmetry and duality in \texorpdfstring{$(2+1)$}\ d }\label{sec:sub}

In this section, we will review the subsystem $\mathbb{Z}_2$ symmetry in $(2+1)$d regularized on a 2d square lattice and the duality transformations including the subsystem Kramers-Wannier (KW) transformation~\cite{Cao:2023doz} and the subsystem Jordan-Wigner (JW) transformation~\cite{Cao:2022lig}.  

\subsection{Subsystem \texorpdfstring{$\mathbb Z_2\ $}\ symmetry on lattice}
\begin{figure}[htbp]
\begin{tikzpicture}
        \draw[thick] (-1,1) -- (0, 2);
        \draw[ultra thick, dgreen] (0.5,1) -- (1.5, 2);
        \draw[thick] (2,1) -- (3, 2);
        \draw[thick] (-1,1) -- (2, 1);
        \draw[ultra thick, red] (-0.5,1.5) -- (2.5, 1.5);
        \draw[thick] (0,2) -- (3, 2);
        \node[left] at (-0.7,1.5) {$j$};
        \node[right] at (2.7,1.5) {$U^x_j$};
        \node[below] at (0.5,1) {$i$};
        \node[above] at (1.5,2) {$U^y_i$};
        \node[] at (1,0) {symmetry operator $U^x_j,U^y_i$};
        \draw[thick] (5,1) -- (6, 2);
        \draw[thick] (8,1) -- (9, 2);
        \draw[thick] (5,1) -- (8, 1);
        \draw[] (5.5,1.5) -- (8.5, 1.5);
        \draw[ultra thick, blue] (7,0.5) -- (7, 1);
        \draw[ultra thick, dotted, blue] (7,1) -- (7, 1.5);
        \draw[ultra thick, blue] (7,1.5) -- (7, 2.5);
        \draw[thick] (6,2) -- (9, 2);
        \node[left] at (5.3,1.5) {$j$};
        \node[] at (7,0) {defect operator $U^x_j$};
        \draw[thick] (10,1) -- (11, 2);
        \draw[thick] (13,1) -- (14, 2);
        \draw[thick] (10,1) -- (13, 1);
        \draw[thick] (11,2) -- (14, 2);
        \draw[ultra thick, red] (10.5,1.5) -- (12, 1.5);
        \node[left] at (10.3,1.5) {$j$};
        \node[] at (12,0) {twist operator $U^{xz}_{0,j}$};
    \end{tikzpicture}
    \caption{Examples of subsystem $\mathbb Z_2$ symmetry operators, defect operators and twist operators.}
    \label{fig:subZ2ex}
\end{figure}
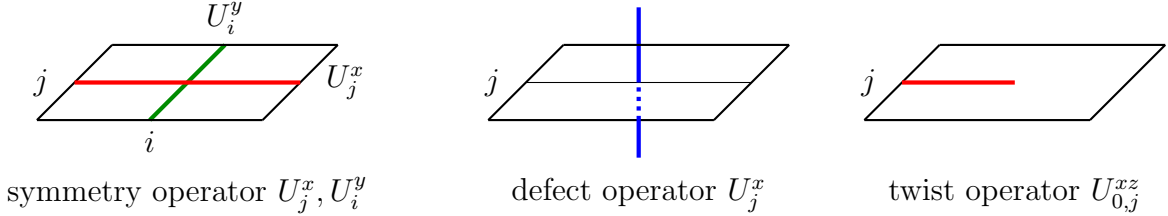

Consider a closed $L_x\times L_y$ square lattice. On each site there is a spin-$1/2$ state $|s\rangle_{i,j}$ where $s = \pm 1$, $i=1,\cdots,L_x$ and $j=1,\cdots,L_y$. Denote the Pauli matrices at each site as $X_{i,j},Y_{i,j},Z_{i,j}$ and they act on the site in a canonical way
    \begin{equation}
        X_{i,j}|s\rangle_{i,j} = |-s\rangle_{i,j},\quad Z_{i,j}|s\rangle_{i,j} = s |s\rangle_{i,j}.
    \end{equation}
The generators of subsystem $\mathbb{Z}_2$  global symmetry are line operators acting on each row and column
\begin{equation}\label{Subsystem-Review-U-operators}
    U_j^x = \prod_{i=1}^{L_x} X_{i,j},\quad U_i^y = \prod_{j=1}^{L_y} X_{i,j}.
\end{equation}
They satisfy $(U_j^x)^2 = (U_i^y)^2 =1$ and flip the spin of all sites of $j$th-row or $i$th-column as illustrated in Fig. \ref{fig:subZ2ex}. We will denote the eigenvalues of $U_j^x$, $U_i^y$ as $(-1)^{u_j^x}, (-1)^{u_i^y}$ where $u_j^x,u_i^y =0,1$ are $\mathbb{Z}_2$-valued integers. These $L_x+L_y$ operators are not independent and they are restricted by the constraint
\begin{equation}\label{Subsystem-Review-Conatraints-of-U}
    \prod_{j=1}^{L_y}U^x_j\prod_{i=1}^{L_x}U^y_i= \prod_{j=1}^{L_y}(-1)^{u_j^x}\prod_{i=1}^{L_x}(-1)^{u_i^y} =1,
\end{equation}
and there are $L_x+L_y-1$ independent symmetry generators.

One can also insert the subsystem $\mathbb{Z}_2$ defects along the time direction (represented by $z$) as shown in the middle diagram in Fig.~\ref{fig:subZ2ex}. If the lattice is infinite, they are implemented by the $\mathbb{Z}_2$ twist operators (e.g. $U^{xz}_{0j}$ in Fig.~\ref{fig:subZ2ex}) on half line
    \begin{equation}\label{Subsystem-Review-U-defects}
        U_{0,j}^{xz} = \prod_{i'<0} X_{i',j},\quad U_{i,0}^{yz} = \prod_{j'<0} X_{i,j'}.
    \end{equation}
The operator $U_{0,j}^{xz}$ is mobile along the $x$-direction and is not mobile along the $y$-direction. Similarly, $U_{i,0}^{yz}$ is mobile along the $y$-direction and is not mobile along the $x$-direction. For periodic lattice, inserting defects on the lattice will twist the boundary condition for each row and column by
\begin{equation}
    |s_{i+L_x,j}\rangle = |(-1)^{t_j^x} s_{i,j} \rangle,\quad |s_{i,j+L_y}\rangle = |(-1)^{t_i^y} s_{i,j}\rangle,\quad |s_{i+L_x,i+L_y}\rangle = |(-1)^{t^{xy}+t_j^x+t_i^y} s_{i,j}\rangle,
\end{equation}
where $t^x_j,t^y_i =0,1$ are twist variables and $t^{xy} = 0,1$ is the boundary condition of the twist variables
    \begin{equation}
        t_{i+L_x}^y = t_i^y+t^{xy},\quad t_{j+L_y}^x = t_{j}^x+t^{xy}.
    \end{equation}
Although there are $L_x+L_y+1$ twist parameters but the Hamiltonian with subsystem $\mathbb{Z}_2$ symmetry depends only on the combinations $\mathbf{t}^x_{j+\frac12},\mathbf{t}^y_{i+\frac12}$~\cite{Cao:2022lig,Cao:2023doz},
\begin{equation}\label{Subsystem-Review-Twist-Variables}
    \mathbf{t}^x_{j+\frac12} \equiv t^x_j + t^x_{j+1},\quad \mathbf{t}^y_{i+\frac12} \equiv t^y_i + t^y_{i+1}, \quad \sum_{j=1}^{L_y} \mathbf{t}^y_{i+\frac12} = \sum_{i=1}^{L_x} \mathbf{t}^x_{j+\frac12} = t^{xy},
\end{equation}
and only $L_x+L_y-1$ twist variables are independent.

Given a $(2+1)$d theory $\mathfrak{T}_{\textrm{sub}}$ with the subsystem $\mathbb{Z}_2$ symmetry, the eigenvalues of subsystem symmetry and twist boundary conditions will divide the Hilbert space into sectors with $\mathbb Z_2$-valued symmetry-twist labels $(\{u^x_j\},\{u^y_i\},\{\mathbf{t}^x_{j+\frac12}\},\{\mathbf{t}^y_{i+\frac12}\})$.
Here $\{ \cdots \}$ denotes the collection of variables for all $j =1,\cdots,L_y$ and $i = 1,\cdots,L_x$. 
The symmetry-twist labels have overall constraints
\begin{equation}\label{Subsystem-Review-Symmetry-Twist-Constraints}
    \prod_{j=1}^{L_y}(-1)^{u^x_j}\prod_{i=1}^{L_x}(-1)^{u^y_i}=1,\quad  \prod_{j=1}^{L_y}(-1)^{\mathbf{t}^x_{j+\frac12}}\prod_{i=1}^{L_x}(-1)^{\mathbf{t}^y_{i+\frac12}}=1.
\end{equation}
With the above constraints, the Hilbert space is divided into $2^{2(L_x+L_y-1)}$ different sectors and the partition function for each sector is
\begin{equation}\label{Subsystem-Review-Symmetry-Twist-Partition-Function}
    Z_{\mathfrak{T}_{\textrm{sub}}}[\{u^x_j\},\{u^y_i\},\{\mathbf{t}^x_{j+\frac12}\},\{\mathbf{t}^y_{i+\frac12}\}]=\text{Tr}_{\mathcal{H}_{\mathbf{t}}}\left(\prod_{i=1}^{L_x}\frac{1+(-1)^{u^y_i}U^y_i}{2}\right)\left(\prod_{j=1}^{L_y}\frac{1+(-1)^{u^x_j}U^x_j}{2}\right)e^{-\beta H},
\end{equation}
 where  $\mathcal{H}_{\mathbf{t}}$ is the Hilbert space of the twist sector with label $(\{\mathbf{t}^x_{j+\frac12}\},\{\mathbf{t}^y_{i+\frac12}\})$.

For simplicity, we will write any quartet $(\{u^x_j\},\{u^y_i\},\{\mathbf{t}^x_{j+\frac12}\},\{\mathbf{t}^y_{i+\frac12}\})$ or doublet $(\{\mathbf{t}^x_{j+\frac12}\},\{\mathbf{t}^y_{i+\frac12}\})$ as $(u^x_j,u^y_i,\mathbf{t}^x_{j+\frac12},\mathbf{t}^y_{i+\frac12})$ and $(\mathbf{t}^x_{j+\frac12},\mathbf{t}^y_{i+\frac12})$ in the following discussion.

\subsubsection*{Coupling to background field}
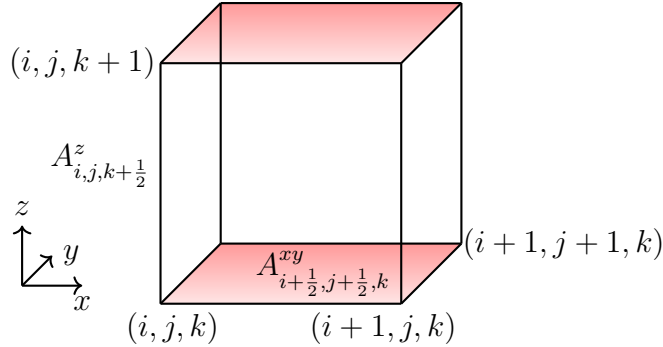
\begin{figure}[htbp]
	\centering
	\[{\begin{tikzpicture}[scale=0.8,baseline=-20]
\draw[thick, ->] (-3,-1) -- (-2, -1); 
\draw[thick, ->] (-3,-1) -- (-3, -1+1); 
\draw[thick, ->] (-3,-1) -- (-3+0.5, -1+0.5); 
\node[below] at (-2, -1) {$x$};
\node[above] at (-3, -1+1) {$z$};
\node[right] at (-3+0.5, -1+0.5) {$y$};
  \shade[top color=red!40, bottom color=red!10]  (-0.7,2.7) -- (3.3,2.7) -- (4.3,3.7)-- (0.3,3.7);
  \draw[thick] (-0.7,2.7) -- (3.3,2.7);
 \draw[thick] (3.3,2.7)-- (4.3,3.7);
  \draw[thick] (4.3,3.7) --(0.3,3.7);
  \draw[thick] (0.3,3.7) -- (-0.7,2.7);
\shade[top color=red!40, bottom color=red!10]  (-0.7,-1.3) -- (3.3,-1.3) -- (4.3,-0.3)-- (0.3,-0.3);
  \draw[thick] (-0.7,-1.3) -- (3.3,-1.3);
 \draw[thick] (3.3,-1.3)-- (4.3,-0.3);
  \draw[thick] (4.3,-0.3) --(0.3,-0.3);
  \draw[thick] (0.3,-0.3) -- (-0.7,-1.3);
\draw[thick] (-0.7,-1.3) -- (-0.7,2.7);
\draw[thick] (3.3,-1.3) -- (3.3,2.7);
\draw[thick] (4.3,-0.3) -- (4.3,3.7);
\draw[thick] (0.3,-0.3) -- (0.3,3.7);
 \node[] at (-2,2.7) {$(i,j,k+1)$};
 \node[] at (-1.7,1) {$A^{z}_{i,j,k+\frac12}$};
 \node[] at (-0.5,-1.7) {$(i,j,k)$};
\node[] at (3,-1.7) {$(i+1,j,k)$};
\node[] at (6,-0.3) {$(i+1,j+1,k)$};
 \node[] at (2,-0.8) {$A^{xy}_{i+\frac12,j+\frac12,k}$};
\end{tikzpicture}}
	\]
	\caption{Background gauge field for subsystem $\mathbb Z_2$ symmetry on lattice.}
	\label{fig:bgg}
\end{figure}

We can introduce background subsystem $\mathbb{Z}_2$ symmetry gauge field $(A^z,A^{xy})$ on the lattice. Consider a cubic spacetime lattice $M_3$ with $L_x\times L_y \times L_z$ sites and the topological $z$-direction is the time direction. The space component of the gauge field $A^{xy}$ lives on the $xy$-plaquette and the time component $A^z$ lives on the $z$-link, as shown in Fig.~\ref{fig:bgg}. 

The $\mathbb{Z}_2$-valued holonomies are regularized by summing the gauge fields along different cycles on the lattice. The holonomy of $A^z$ along the time direction is
\begin{equation}\label{Subsystem-Review-Holonomies-time}
    w_{z;i,j}=\sum_{k=1}^{L_z}A^z_{i,j,k+\frac{1}{2}}=w_{z,x;j}+w_{z,y;i},
\end{equation}
which is highly reducible and we can decompose it into $w_{z,x;j}, w_{z,y;i}$ detecting the insertion of symmetry operator $(U^x_j)^{w_{z,x;j}}$ and $(U^y_i)^{w_{z,y;i}}$ respectively.  The constraint \eqref{Subsystem-Review-Conatraints-of-U} on the symmetry operators imposes a gauge redundancy
\begin{equation}\label{Subsystem-Review-Holonomies-time-gauge-redundancy}
    (w_{z,x;j},w_{z,y;i}) \sim (w_{z,x;j}+1,w_{z,y;i}+1).
\end{equation}
On the other hand, the holonomy of $A^{xy}$ along $x$ and $y$ directions are
\begin{equation}\label{Subsystem-Review-Holonomies-space}
            w_{x;j+\frac12}=\sum_{i=1}^{L_x}A^{xy}_{i+\frac{1}{2},j+\frac{1}{2},k}=\mathbf{t}^x_{j+\frac{1}{2}}\, \quad
        w_{y;i+\frac{1}{2}}=\sum_{j=1}^{L_y}A^{xy}_{i+\frac{1}{2},j+\frac{1}{2},k}=\mathbf{t}^y_{i+\frac{1}{2}}\, .
\end{equation}
They detect the insertion of symmetry defects along the $z$-direction and are the same as the twist variables $\mathbf{t}^x_{j+\frac{1}{2}}, \mathbf{t}^y_{i+\frac{1}{2}}$ introduced in \eqref{Subsystem-Review-Twist-Variables}. They obey the same constraint
    \begin{equation}\label{Subsystem-Review-Holonomies-space-Constraints}
        \prod_{j=1}^{L_y}(-1)^{w_{x;j+\frac12}}\prod_{i=1}^{L_x}(-1)^{w_{y;i+\frac{1}{2}}}=1.
    \end{equation}

For a generic subsystem $\mathbb{Z}_2$ symmetry background $(w_{z,x;j}, w_{z,y;i},w_{x;j+\frac12},w_{y;i+\frac12})$, the partition function is
\begin{equation}\label{Subsystem-Review-Partition-function}
    Z_{\mathfrak{T}_{\textrm{sub}}}[w_{z,x;j}, w_{z,y;i},w_{x;j+\frac12},w_{y;i+\frac12}]=\text{Tr}_{\mathcal{H}_{\mathbf{t}}}\left(\prod_{j=1}^{L_y}(U^x_j)^{w_{z,x;j}}\right)\left(\prod_{i=1}^{L_x}(U^y_i)^{w_{z,y;i}}\right)e^{-\beta H}.
\end{equation}
It is related to the partition function in the sector with symmetry-twist label \eqref{Subsystem-Review-Symmetry-Twist-Partition-Function} by a discrete Fourier transformation
\begin{align}\label{Subsystem-Review-Partition-function-Fourier}
    Z_{\mathfrak{T}_{\textrm{sub}}}[w_{z,x;j}, w_{z,y;i},w_{x;j+\frac12},w_{y;i+\frac12}] = \sum_{u^y_i,u^x_j=0,1}(-1)^{\sum_{i}u^y_iw_{z,y;i}+\sum_{j}u^x_jw_{z,x;j}}Z_{\mathfrak{T}_{\textrm{sub}}}[u^x_j,u^y_i,\mathbf{t}^x_{j+\frac12},\mathbf{t}^y_{i+\frac12}],
\end{align}
where $w_{x;j+\frac{1}{2}} = \mathbf{t}^x_{j+\frac{1}{2}},w_{y;i+\frac{1}{2}} = \mathbf{t}^y_{i+\frac{1}{2}}$ and the summation over $(u^y_i,u^x_j)$ should obey the constraint \eqref{Subsystem-Review-Symmetry-Twist-Constraints}.

\subsection{Subsystem KW transformation}
\begin{figure}[htbp]
    \centering
    \begin{tikzpicture}
        \draw[thick, dashed, red] (0.5,0) -- (0.5, 3.5);
        \draw[thick, dashed, red] (1.5,0) -- (1.5, 3.5);
        \draw[thick, dashed, red] (2.5,0) -- (2.5, 3.5);
        \draw[thick] (1,0) -- (1, 3.5);
        \draw[thick] (2,0) -- (2, 3.5);
        \draw[thick] (3,0) -- (3, 3.5);
        \draw[thick, dashed, red] (0,0.5) -- (3.5, 0.5);
        \draw[thick, dashed, red] (0,1.5) -- (3.5, 1.5);
        \draw[thick, dashed, red] (0,2.5) -- (3.5, 2.5);
        \draw[thick] (0,1) -- (3.5, 1);
        \draw[thick] (0,2) -- (3.5, 2);
        \draw[thick] (0,3) -- (3.5, 3);
        \node[below] at (1,0) {$i-1$};
        \node[below] at (2,0) {$i$};
        \node[below] at (3,0) {$i+1$};
        \node[left] at (0,1) {$j-1$};
        \node[left] at (0,2) {$j$};
        \node[left] at (0,3) {$j+1$};
        \node[above] at (1,3.5) {$u^y_{i-1}$};
        \node[above] at (2.5,3.5) {$\color{red}{\hat{u}^y_{i+\frac{1}{2}}}$};
        \draw[thick, red, decorate,decoration={brace,amplitude=5pt,raise=4ex}]
  (0.5,3.5) -- (1.5,3.5) node[midway,yshift=3em]{$\hat{\mathbf{t}}^y_{i-1}$};
        \draw[thick, decorate,decoration={brace,amplitude=5pt,raise=4ex}]
  (2,3.5) -- (3,3.5) node[midway,yshift=3em]{$\mathbf{t}^y_{i+\frac{1}{2}}$};
        \node[right, red] at (3.5, 2.5) {$\hat{u}^x_{j+\frac{1}{2}}$};
        \node[right] at (3.5, 1) {${u}^x_{j-1}$};
        \draw[thick, decorate,decoration={brace,amplitude=5pt,mirror,raise=5ex}]
  (3.5,2) -- (3.5,3) node[midway,xshift=3.8em]{$\mathbf{t}^x_{j+\frac{1}{2}}$};
        \draw[thick, red, decorate,decoration={brace,amplitude=5pt,mirror,raise=5ex}]
  (3.5,0.5) -- (3.5,1.5) node[midway,xshift=3.8em]{$\hat{\mathbf{t}}^x_{j-1}$};
    \end{tikzpicture}
    \caption{Mapping of symmetry-twist sectors. The original lattice is in black while the dual lattice is in red. For example, the symetry variable $u^x_{j-1}$ is mapped to the dual twist variable $\hat{\mathbf{t}}^x_{j-1}=\hat{t}^x_{j-\frac32}+\hat{t}^x_{j-\frac12}$.}
    \label{fig:mapping}
\end{figure}
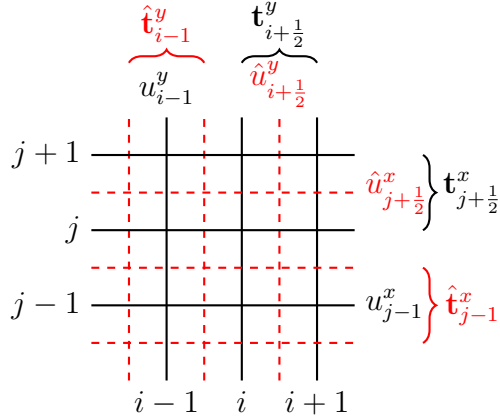

We can gauge the subsystem $\mathbb Z_2$ symmetry by doing a subsystem KW transformation $\mathcal{N}^{\text{sub}}$ \cite{Cao:2023doz} which maps the original lattice with spin $\{|s\rangle_{i,j} \}$ to the dual lattice with spin $\{|\hat{s}\rangle_{i+\frac{1}{2},j+\frac{1}{2}} \}$ living on the plaquette of the original lattice. In terms of Pauli operators, the explicit transformation of $\mathcal N^{\text{sub}}$ is
\begin{align}\label{eq:subsysdifflattice}
   \begin{split}
   &\mathcal{N}^{\text{sub}}Z_{i,j}Z_{i,j+1}Z_{i+1,j}Z_{i+1,j+1} =\hat{X}_{i+\frac12,j+\frac12}\mathcal{N}^{\text{sub}},\\
   &\mathcal{N}^{\text{sub}}X_{i,j}=\hat{Z}_{i-\frac12,j-\frac12}\hat{Z}_{i+\frac12,j-\frac12}\hat{Z}_{i-\frac12,j+\frac12}\hat{Z}_{i+\frac12,j+\frac12}\mathcal{N}^{\text{sub}},
   \end{split}
\end{align}
where $\hat{X}_{i+\frac12,j+\frac12},\hat{Z}_{i+\frac12,j+\frac12}$ are Pauli operators acting on the dual lattice. After gauging, the dual theory $\hat{\mathfrak{T}}_{\textrm{sub}}$ lives on the dual lattice and has a dual subsystem $\mathbb Z_2$ symmetry. The Hilbert space of the dual theory  $\hat{\mathfrak{T}}_{\textrm{sub}}$ is similarly divided into sectors labelled by the dual symmetry-twist variables $(\hat{u}^x_{j+\frac{1}{2}}, \hat{u}^y_{i+\frac{1}{2}},\hat{\mathbf{t}}^x_j,\hat{\mathbf{t}}^y_i)$ with the constraints
\begin{equation}
    \prod_{j=1}^{L_y}(-1)^{\hat{u}^x_{j+\frac{1}{2}}}\prod_{i=1}^{L_x}(-1)^{\hat{u}^y_{i+\frac{1}{2}}}=1,\quad  \prod_{j=1}^{L_y}(-1)^{\hat{\mathbf{t}}^x_j}\prod_{i=1}^{L_x}(-1)^{\hat{\mathbf{t}}^y_i}=1.
\end{equation}
They are related to the symmetry-twist variables $(u^x_j,u^y_i,\mathbf{t}^x_{j+\frac12},\mathbf{t}^y_{i+\frac12})$ in the original theory $\mathfrak{T}_{\textrm{sub}}$ as
\begin{equation}\label{Subsystem-Review-Symmetry-Twist-Exchange}
   \hat{u}^x_{j+\frac12}=\mathbf{t}^x_{j+\frac12},\quad \hat{u}^y_{i+\frac12}=\mathbf{t}^y_{i+\frac12},\quad \hat{\mathbf{t}}^x_{j}=u^x_{j}, \quad \hat{\mathbf{t}}^y_i=u^y_i.
\end{equation}
where symmetry/twist sectors are exchanged as shown in Fig. \ref{fig:mapping}.

The holonomy variables of the dual gauge fields $(\hat{A}^z,\hat{A}^{xy})$ are $(\hat{w}_{z,x;j+\frac12}, \hat{w}_{z,y;i+\frac12},\hat{w}_{x;j},\hat{w}_{y;i})$, with the gauge redundancy and constraints
\begin{equation}
    (\hat{w}_{z,x;j+\frac{1}{2}},\hat{w}_{z,y;i+\frac{1}{2}}) \sim (\hat{w}_{z,x;j+\frac{1}{2}}+1,\hat{w}_{z,y;i+\frac{1}{2}}+1),\quad \prod_{j=1}^{L_y}(-1)^{\hat{w}_{x;j}}\prod_{i=1}^{L_x}(-1)^{\hat{w}_{y;i}}=1.
\end{equation}
As before, one has $\hat{w}_{x;j} = \hat{\mathbf{t}}^x_j, \hat{w}_{y,i} = \hat{\mathbf{t}}^y_i$ and $(\hat{w}_{z,x;j+\frac{1}{2}},\hat{w}_{z,y;i+\frac{1}{2}})$ are the Fourier partners of $(\hat{u}^x_{j+\frac12},\hat{u}^x_{j+\frac12})$ as in \eqref{Subsystem-Review-Partition-function-Fourier}. Implied by \eqref{Subsystem-Review-Symmetry-Twist-Exchange}, the partition function of the dual theory $\hat{\mathfrak{T}}_{\textrm{sub}}$ is related to the partition of  the original theory $\mathfrak{T}_{\textrm{sub}}$ in \eqref{Subsystem-Review-Partition-function} as
\begin{align}\label{Subsystem-Review-Partition-function-KWdual}
    \begin{split}
    &Z_{\hat{\mathfrak{T}}_{\textrm{sub}}}[\hat{w}_{z,x;j+\frac12}, \hat{w}_{z,y;i+\frac12},\hat{w}_{x;j},\hat{w}_{y;i}]\\
    &= \frac{1}{2^{L_x+L_y-1}}\sum_{w_{z,x;j}, w_{z,y;i},w_{x;j+\frac12},w_{y;i+\frac12}=0,1}Z_{\mathfrak{T}_{\textrm{sub}}}[w_{z,x;j}, w_{z,y;i},w_{x;j+\frac12},w_{y;i+\frac12}]\\
    &\times(-1)^{\sum_{i}(\hat{w}_{z,y;i+\frac12}w_{y;i+\frac12}+\hat{w}_{y;i}w_{z,y;i})+\sum_{j}(\hat{w}_{z,x;j+\frac12}w_{x;j+\frac12}+\hat{w}_{x;j}w_{z,x;j})}\, .
    \end{split}
\end{align}
The summation of $(w_{z,x;j}, w_{z,y;i},w_{x;j+\frac12},w_{y;i+\frac12})$ should obey the restrictions in \eqref{Subsystem-Review-Holonomies-time-gauge-redundancy} and \eqref{Subsystem-Review-Holonomies-space-Constraints}.

Suppose the theory $\mathfrak{T}_{\textrm{sub}}$ is invariant under the subsystem KW transformation, which means $\hat{\mathfrak{T}}_{\textrm{sub}} = \mathfrak{T}_{\textrm{sub}}$. The subsystem KW transformation becomes a symmetry and we can insert the KW operator/defect $\mathcal{N}^{\textrm{sub}}$ along a 2-dimensional surface $M_2$ by gauging half of the spacetime. If $M_2$ is the $x$-$y$ plane, $\mathcal{N}^{\text{sub}}$ is an operator acting on the Hilbert space. The fusion between the symmetry operator $\mathcal{N}^{\text{sub}}$ and its orientation reversal $\mathcal{N}^{\text{sub}}{}^{\dagger}$ is
\begin{equation}\label{Subsystem-Review-Fusion-Rule-Operators}
    \mathcal{N}^{\text{sub}}{}^{\dagger} \times \mathcal{N}^{\text{sub}} = \frac{1}{2} \prod_{i=1}^{L_x} \left( 1 + (-1)^{\hat{\mathbf{t}}^y_i} U^y_i\right) \prod_{j=1}^{L_y} \left( 1 + (-1)^{\hat{\mathbf{t}}^x_j} U^x_j\right).
\end{equation}
On the other hand, if $M_2$ is the $z$-$x$ (or $z$-$y$) plane then $\mathcal{N}^{\text{sub}}$ is a defect twisting the boundary condition. The fusion rule of the subsystem KW defect on the $z$-$x$ plane is
\begin{equation}\label{Subsystem-Review-Fusion-Rule-Defects}
    \mathcal{N}^{\text{sub}}{}^{\dagger} \times \mathcal{N}^{\text{sub}} = \sum_{t^y_i = 0,1} \prod_i^{L_x} (U^{yz}_{0,i})^{t^y_i}.
\end{equation}
The fusion rules are first derived in~\cite{Cao:2023doz}. We give an alternative derivation in Appendix~\ref{sec:deri} following \cite{Kaidi:2022cpf}.

\subsubsection*{Subsystem KW transformation on one lattice}
The subsystem KW transformation \eqref{eq:subsysdifflattice} maps from the lattice to the dual lattice \cite{Cao:2023doz}. We can also define another subsystem KW transformation on one lattice
\begin{align}
   \begin{split}
   &\bar{\mathcal{N}}^{\text{sub}}Z_{i,j}Z_{i,j+1}Z_{i+1,j}Z_{i+1,j+1} =X_{i+1,j+1}\bar{\mathcal{N}}^{\text{sub}},\\
   &\bar{\mathcal{N}}^{\text{sub}}X_{i,j}=Z_{i,j}Z_{i+1,j}Z_{i,j+1}Z_{i+1,j+1}\bar{\mathcal{N}}^{\text{sub}},
   \end{split}
\end{align}
and the fusion rule of $\bar{\mathcal{N}}^{\text{sub}}\times \bar{\mathcal{N}}^{\text{sub}}$ will mix with the one-site translation in the diagonal direction $\mathcal T$ 
\begin{align}
    \begin{split}
\bar{\mathcal{N}}^{\text{sub}} \times \bar{\mathcal{N}}^{\text{sub}} &= \frac{1}{2} \prod_{i=1}^{L_x} \left( 1 + (-1)^{\hat{\mathbf{t}}^y_i} U^y_i\right) \prod_{j=1}^{L_y} \left( 1 + (-1)^{\hat{\mathbf{t}}^x_j} U^x_j\right)\mathcal{T},\\
\bar{\mathcal{N}}^{\text{sub}}{}^{\dagger} \times \bar{\mathcal{N}}^{\text{sub}} &= \frac{1}{2} \prod_{i=1}^{L_x} \left( 1 + (-1)^{\hat{\mathbf{t}}^y_i} U^y_i\right) \prod_{j=1}^{L_y} \left( 1 + (-1)^{\hat{\mathbf{t}}^x_j} U^x_j\right).
\end{split}
\end{align}
This is a natural generalization of the ordinary KW transformation \cite{Seiberg:2023cdc} whose fusion rule on lattice is different from the fusion rule in the continuum theory by a one-site translation.

\subsection{Subsystem JW transformation}
Besides the subsystem KW transformation that maps a bosonic $\mathfrak{T}_{\textrm{sub}}$ theory to another bosonic theory $\hat{\mathfrak{T}}_{\textrm{sub}}$, we also have the subsystem JW transformation that maps the bosonic theory $\mathfrak{T}_{\textrm{sub}}$ to a fermionic theory $\mathfrak{T}_{F,\textrm{sub}}$~\cite{Cao:2022lig}. 

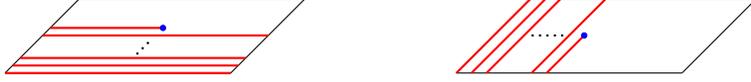
\begin{figure}[htbp]
\centering
\begin{tikzpicture}
        \draw[] (-1,1) -- (0, 2);
        \draw[] (2,1) -- (3, 2);
        \draw[] (-1,1) -- (2, 1);
        \draw[] (0,2) -- (3, 2);
        \draw[thick, red] (-1,1) -- (2, 1);
        \draw[thick, red] (-0.9,1.1) -- (2.1, 1.1);
        \draw[thick, red] (-0.8,1.2) -- (2.2, 1.2);
        \draw[thick, red] (-0.5,1.5) -- (2.5, 1.5);
        \draw[thick, red] (-0.4,1.6) -- (1.1, 1.6);
        \filldraw [blue] (1.1,1.6) circle (1pt);
        \draw[thick, dotted, black] (0.75,1.25) -- (0.95, 1.45);
        \draw[] (5,1) -- (6, 2);
        \draw[] (8,1) -- (9, 2);
        \draw[] (5,1) -- (8, 1);
        \draw[] (6,2) -- (9, 2);
        \draw[thick, red] (5,1) -- (6, 2);
        \draw[thick, red] (5.2,1) -- (6.2, 2);
        \draw[thick, red] (5.4,1) -- (6.4, 2);
        \draw[thick, dotted, black] (6,1.5) -- (6.5, 1.5);
        \draw[thick, red] (6,1) -- (7, 2);
        \draw[thick, red] (6.2,1) -- (6.7, 1.5);
        \filldraw [blue] (6.7,1.5) circle (1pt);
    \end{tikzpicture}
    \caption{Subsystem JW transformation winding around $x$ and $y$ directions.}
    \label{fig:subJW}
\end{figure}

The subsystem JW transformation maps Pauli operators $X_{i,j},Y_{i,j},Z_{i,j}$ to Majorana fermion operators $\gamma_{i,j},\gamma'_{i,j}$ and vice versa. To preserve the standard anticommutation relation among Majorana fermions, one must attach a 1d JW tail (product of Pauli $X$ operators) whose winding directions will lead to different choices of subsystem JW transformation. In Fig.~\ref{fig:subJW}, we give examples where the tail winds around the $x$ direction and $y$ direction and we will denote the two fermionic theories after each transformation separately as $\mathfrak{T}_{F,x,\textrm{sub}}$ and $\mathfrak{T}_{F,y,\textrm{sub}}$.

For the first choice, the explicit transformation is 
\begin{align}\label{JWxIsing}
\begin{split}
\gamma_{i,j}&=\left(\prod_{i'=1}^{L_x}\prod_{j'=1}^{j-1}X_{i',j'}\right)\left(\prod_{i'=1}^{i-1}X_{i',j}\right)Z_{i,j},\\
\gamma_{i,j}'&=-\left(\prod_{i'=1}^{L_x}\prod_{j'=1}^{j-1}X_{i',j'}\right)\left(\prod_{i'=1}^{i-1}X_{i',j}\right)Y_{i,j}.
\end{split}
\end{align}
The fermionic theory has subsystem $\mathbb Z_2$ fermion parity symmetry $(-1)^F$. Considering the symmetry operators and twists of $(-1)^F$, the Hilbert space is divided into $2^{2(L_x+L_y-1)}$ sectors with labels $(u^x_{f,j},u^y_{f,i},\mathbf{t}^x_{f,j+\frac12},\mathbf{t}^y_{i+\frac12})$. Using the transformation \eqref{JWxIsing}, one can work out the mapping between symmetry-twist sectors in the bosonic and fermionic theory
\begin{equation}\label{eq:fersymsector}
    u^x_{f,j}=u^x_{f},\quad  u^y_{f,i}=u^y_i,\quad \mathbf{t}^x_{f,j+\frac12}=\mathbf{t}^x_{j+\frac12}+u^x_{j}+u^x_{j+1},\quad \mathbf{t}^y_{f,i+\frac12}=\mathbf{t}^y_{i+\frac12}.
\end{equation}
We can introduce the background fields for subsystem $\mathbb Z_2$ fermion parity symmetry $(-1)^F$ and define the corresponding holonomy variables as $(s_{z,x;j}, s_{z,y;i},s_{x;j+\frac12},s_{y;i+\frac12})$. Similar to the bosonic case, the space direction holonomy has the following identification
\begin{equation}
    s_{x;j+\frac12} = \mathbf{t}^x_{f,j+\frac12},\quad s_{y;i+\frac12}=\mathbf{t}^y_{f,i+\frac12},
\end{equation}
From the sector correspodence~\eqref{eq:fersymsector}, we can derive the relation between the partition functions of the bosonic theory $\mathfrak{T}_{\textrm{sub}}$ and the fermionic theory $\mathfrak{T}_{F,x,\textrm{sub}}$
\begin{align}\label{Subsystem-Review-JW-Transformation-x-direction}
    &Z_{\mathfrak{T}_{F,x,\textrm{sub}}}[s_{z,x;j}, s_{z,y;i},s_{x;j+\frac12},s_{y;i+\frac12}]\nonumber\\ =& \sum_{u^x_f,u^y_f=0,1} (-1)^{\sum_i u^y_{f,i} s_{z,y;i} +\sum_j u^x_{f,j} s_{z,x;j}} Z_{\mathfrak{T}_{F,x,\textrm{sub}}}[u^x_f,u^y_f,s_{x;j+\frac12},s_{y;i+\frac12}]\nonumber\\
    =&\sum_{u^x,u^y=0,1} (-1)^{\sum_i u^y_{i} s_{z,y;i} +\sum_j u^x_{j} s_{z,x;j}} Z_{\mathfrak{T}_{\textrm{sub}}}[u^x,u^y,s_{x;j+\frac12}+ u^x_{j}+ u^x_{j+1},s_{y;i+\frac12}]\nonumber\\
    =&\frac{1}{2^{L_x+L_y-1}}\sum_{u^x,u^y,w_{z,x},w_{z,y}=0,1}(-1)^{\sum_i u^y_{i} (s_{z,y;i}+w_{z,y;i}) +\sum_j u^x_{j} (s_{z,x;j}+w_{z,x;j})}\nonumber\\
    & \times  Z_{\mathfrak{T}_{\textrm{sub}}}[w_{z,x;j},w_{z,y;i},s_{x;j+\frac12}+ u^x_{j}+ u^x_{j+1},s_{y;i+\frac12}].
\end{align}

If the subsystem JW transformation winds along the $y$ direction, we have a different transformation and a different symmetry-twist sector mapping
\begin{equation}
    u^x_{f,j}=u^x_{f},\quad  u^y_{f,i}=u^y_i,\quad \mathbf{t}^x_{f,j+\frac12}=\mathbf{t}^x_{j+\frac12},\quad \mathbf{t}^y_{f,i+\frac12}=\mathbf{t}^y_{i+\frac12}+u^y_{i}+u^y_{i+1}.
\end{equation}
Moreover, one can first perform a JW transformation winds along the $x$ direction and then do an inverse JW transformation winds along the $y$ direction, which ends to another bosonic theory $\mathfrak{T}_{xy,\textrm{sub}}$. One can easily check that now the symmetry-twist sector labels $( u'^x_{j}, u'^y_{i},\mathbf{t}'^x_{j+\frac12},\mathbf{t}'^y_{i+\frac12})$ in this new bosonic theory are
\begin{equation}
    u'^x_{j}=u^x_j,\quad u'^y_{i}=u^y_i,\quad \mathbf{t}'^x_{j+\frac12}=\mathbf{t}^x_{j+\frac12}+u^x_{j}+u^x_{j+1},\quad \mathbf{t}'^y_{i+\frac12}=\mathbf{t}^y_{i+\frac12}+u^y_{i}+u^y_{i+1}.
\end{equation}

Combining different subsystem JW transformation, we get a duality web relating two bosonic theoies and two fermionic theories. A simple realization of the duality web starts from the plaquette Ising model
\begin{equation}\label{eq:plaqising}
   H_{\text{PlaqIsing}}=-\sum_{i,j}Z_{i,j}Z_{i+1,j}Z_{i,j+1}Z_{i+1,j+1}-h\sum_{i,j}X_{i,j}.
\end{equation}
Applying the subsystem JW transformation winding along $x$ and $y$ direction seperately, we get two different plaquette fermion models
\begin{align}
\begin{split}
H_{\text{Pfer},x}=&\sum_{i,j} \gamma'_{i,j} \gamma_{i+1,j} \gamma'_{i,j+1} \gamma_{i+1,j+1} +ih\sum_{i,j}\gamma_{i,j}\gamma'_{i,j}, \\
H_{\text{Pfer},y}=&\sum_{i,j} \gamma'_{i,j}\gamma_{i,j+1}  \gamma'_{i+1,j} \gamma_{i+1,j+1} +ih\sum_{i,j}\gamma_{i,j}\gamma'_{i,j}.
\end{split}
\end{align}
Further applying the inverse subsystem JW transformation along $y$ direction to $H_{\text{Pfer},x}$, or along $x$ direction to $H_{\text{Pfer},y}$, we will get another bosonic theory
\begin{equation}\label{eq:bos}
    H'_{bos}=-\sum_{i,j}Z_{i,j}Y_{i+1,j}Y_{i,j+1}Z_{i+1,j+1}-h\sum_{i,j}X_{i,j}.
\end{equation}
The duality web will be enlarged by further considering the subsystem KW transformation, which is elaborated in Sec.~\ref{sec:sl2z}.

\section{2-foliated theory as the subsystem SymTFT}\label{sec:2foli}
In this section, we will give the analogy of SymTFT for subsystem $\mathbb Z_N$ symmetry in $(2+1)$d. The candidate theory is the $(3+1)$d 2-foliated BF theory with level $N$~\eqref{eq:2foliated} where the foliation is along $x,y$ directions. The theory is topological along the remaining directions $z,\tau$. From the exotic-foliated duality~\cite{Ohmori:2022rzz,Spieler:2023wkz}, we will focus on the dual formulation, the exotic tensor gauge theory~\eqref{eq:exotic} where the subsystem symmetry is more obvious. We will quantize the theory by picking the topological direction $\tau$ as the time direction. After quantization, we will see this theory supports a topological boundary $\mathfrak{B}^{\textrm{sym}}_{\textrm{sub}}$ with a $(2+1)$d  subsystem $\mathbb{Z}_N$ symmetry. We will explore various bosonic and fermionic topological boundaries of the bulk theory. As an application, the subsystem KW and JW transformations have a subsystem SymTFT interpretation as switching between different topological boundaries.

\subsection{2-foliated BF theory revisited}

The candidate for subsystem SymTFT of our interest is the $(3+1)$d 2-foliated BF theory with level $N$
\begin{equation}\label{eq:2foliated}
    S_{2\textrm{-foliated}} = \frac{N}{2\pi}\int b \wedge dc + \sum_{k=1,2} d B^k \wedge C^k \wedge d x^k + \sum_{k=1,2} b \wedge C^k \wedge d x^k .
 \end{equation}
The first term is a usual 4d BF theory where $b$ is a 2-form gauge field and $c$ is a 1-form gauge field, the second term gives a foliation of 3d BF theories along $x^1,x^2$ direction where $B^1,B^2,C^1,C^2$ are 1-form gauge fields, and the third term is the interaction term that couples the foliated fields and the bulk fields. In the following we will label the coordinates $(x^0,x^1,x^2,x^3)$ as $(\tau,x,y,z)$.

The 2-foliated BF theory~\eqref{eq:2foliated} is equivalent to the exotic tensor gauge theory~\cite{Gorantla:2020jpy,Spieler:2023wkz}
\begin{equation}\label{eq:exotic}
    S_{\textrm{exotic}} = \frac{N}{2\pi} \int \left[A^{\tau} (\partial_z \hat{A}^{xy} - \partial_x \partial_y \hat{A}^z) - A^z (\partial_{\tau} \hat{A}^{xy} - \partial_x \partial_y \hat{A}^{\tau})- A^{xy} (\partial_{\tau} \hat{A}^z - \partial_z \hat{A}^{\tau})  \right].
\end{equation}
The foliated-exotic duality is sketched in Appendix~\ref{sec:duality} by integrating out some auxiliary fields and redefining the others. In the action~\eqref{eq:exotic}, $A = \{A^{\tau},A^z,A^{xy} \}$ and $\hat{A} = \{\hat{A}^{\tau},\hat{A}^z,\hat{A}^{xy} \}$ are electric and magnetic gauge fields with the following gauge transformations
\begin{align}
    \begin{split}
         &A^{\tau} \sim A^{\tau} + \partial_{\tau} \lambda,\quad A^{z} \sim A^{z} + \partial_{z} \lambda,\quad A^{xy}\sim A^{xy} + \partial_{x}\partial_y \lambda,\\
         &\hat{A}^{\tau} \sim \hat{A}^{\tau} + \partial_{\tau} \hat{\lambda},\quad \hat{A}^z \sim \hat{A}^z + \partial_z \hat{\lambda},\quad \hat{A}^{xy} \sim \hat{A}^{xy} + \partial_x \partial_y \hat{\lambda},
    \end{split}
\end{align}
where $\lambda,\hat{\lambda}$ are gauge parameters. 
The equations of motion for gauge fields $A$ and $\hat{A}$ are
 \begin{align}\label{Foliated-Theory-EOM}
 \begin{split}
     &\partial_z A^{\tau} - \partial_{\tau} A^z=0,\quad \partial_{\tau} A^{xy} - \partial_x \partial_y A^{\tau}=0,\quad \partial_{z} A^{xy} - \partial_x \partial_y A^{z}=0,\\
      &\partial_z \hat{A}^{\tau} - \partial_{\tau} \hat{A}^z=0,\quad \partial_{\tau} \hat{A}^{xy} - \partial_x \partial_y \hat{A}^{\tau}=0,\quad \partial_{z} \hat{A}^{xy} - \partial_x \partial_y \hat{A}^{z}=0.
 \end{split}
 \end{align}
In the exotic theory~\eqref{eq:exotic}, there exists a naive $SL(2,\mathbb{Z}_N)$ symmetry
\begin{align}\label{eq:sym}
    \begin{split}
        &S:\quad A\rightarrow \hat{A},\quad \hat{A}\rightarrow -A,\\
        &  T:\quad A\rightarrow A,\quad \hat{A} \rightarrow \hat{A} + A.
    \end{split}
\end{align}
with $S^2 = C$ the charge conjugation symmetry
\begin{equation}
    C: \quad A \rightarrow -A,\quad \hat{A}\rightarrow -\hat{A},
\end{equation}
which is hard to see in the original 2-foliated formulation. We will elaborate more on this $SL(2,\mathbb{Z}_N)$ symmetry regularized on the lattice in the next section.

The gauge invariant operators have restricted mobility due to the foliation. There exist the electric/magnetic line operators that are topological in the $z$-$\tau$ plane but cannot move freely along the $x,y$ directions
\begin{align}\label{Foliated-Theory-Line-Operators}
    \begin{split}
    W(C_{z,\tau}(x,y))  =& \exp\left(i \oint_{C_{z,\tau}(x,y)} A^{\tau} d\tau + A^z dz\right),\\ \hat{W}(C_{z,\tau}(x,y)) =& \exp\left(i \oint_{C_{z,\tau}(x,y)} \hat{A}^{\tau} d\tau + \hat{A}^z dz\right),        
    \end{split}
\end{align}
where $C_{z,\tau}(x,y)$ is a curve in the $z$-$\tau$ plane and is localized at $(x,y)$ in the ambient space. 
The exotic theory also has gauge invariant strip operators spanned along $x$ or $y$ directions
\begin{align}\label{Foliated-Theory-Strip-Operators}
    \begin{split}
    W(x_1,x_2,C_{y,z,\tau}(x)) =& \exp\left(i\int_{x_1}^{x_2}dx\oint_{C_{y,z,\tau}(x)} A^{xy} dy + \partial_x A^{z} dz + \partial_x A^{\tau} d\tau \right),  \\  W(y_1,y_2,C_{x,z,\tau}(y)) =& \exp\left(i\int_{y_1}^{y_2}dy\oint_{C_{x,z,\tau}(y)} A^{xy} dx + \partial_y A^{z} dz + \partial_y A^{\tau} d\tau \right),  
    \end{split}
\end{align}
for electric gauge field $A$. There are also hat versions for magnetic gauge field $\hat{A}$. Here $C_{x,z,\tau}(y)$ is a curve in the $x$-$z$-$\tau$ plane with fixed $y$, and $C_{y,z,\tau}(x)$ is a curve in the $y$-$z$-$\tau$ plane with fixed $x$. The curve $C_{x,z,\tau}(y)$ can be deformed in $x$-$z$-$\tau$ plane but not along $y$ direction and the similar restricted mobility for $C_{y,z,\tau}(x)$. The above properties of  restricted mobility follow from the equations of motion \eqref{Foliated-Theory-EOM} 
of the gauge fields $A$ and $\hat{A}$.

\subsection*{Quantization}
We can quantize the exotic theory~\eqref{eq:exotic} by picking $\tau$ as the time direction with the Coulomb gauge $A_{\tau} = \hat{A}_{\tau}=0$. The action~\eqref{eq:exotic} becomes
\begin{equation}\label{Foliated-Theory-Action-Coulomb-Gauge}
    S_{\textrm{exotic}} = \frac{N}{2\pi} \int \left[- A^{xy} (\partial_{\tau} \hat{A}^{z}) - A^z (\partial_{\tau} \hat{A}^{xy}) \right].
\end{equation}
with the canonical commutation relations between conjugate fields $A$ and $\hat{A}$
\begin{align}
    \left[ A^{xy} (x,y,z) , \hat{A}^z (x',y',z')\right] &= \frac{2\pi i}{N} \delta_3(x-x',y-y',z-z'),\\ \nonumber \left[ A^z(x,y,z), \hat{A}^{xy} (x',y',z') \right] &= \frac{2\pi i}{N} \delta_3(x-x',y-y',z-z').
\end{align}
The Gauss laws
\begin{equation}\label{Foliated-Theory-Gauss-Law}
    \partial_x \partial_y \hat{A}^{z} - \partial_z \hat{A}^{xy}=0,\quad \partial_x \partial_y A^z - \partial_z A^{xy} = 0,
\end{equation}
imply the flat condition.

We will consider the 2-foliated theory (or exotic tensor theory) on a spatial manifold $M_3 = T^2 \times S^1$, where $(x,y)$ parameterize the torus $T^2$ and $z$ is the coordinate of $S^1$. The gauge invariant operators \eqref{Foliated-Theory-Line-Operators},\eqref{Foliated-Theory-Strip-Operators} restricting to $M_3$ gives the electric line/strip operators
\begin{align}\label{Foliated-Theory-Operators-electric}
    \begin{split}
        &W(x,y)=\exp\left(i\oint dz A^z\right),\\
        &W(x_1,x_2)=\exp\left(i\int_{x_1}^{x_2}dx\oint dy A^{xy}\right),\\
        &W(y_1,y_2)=\exp\left(i\int_{y_1}^{y_2}dy\oint dx A^{xy}\right),
    \end{split}
\end{align}
and the magnetic line/strip operators
\begin{align}\label{Foliated-Theory-Operators-magnetic}
    \begin{split}
        &\hat{W}(x,y)=\exp\left(i\oint dz \hat{A}^{z}\right),\\
        &\hat{W}(x_1,x_2)=\exp\left(i\int_{x_1}^{x_2}dx\oint dy \hat{A}^{xy}\right),\\
        &\hat{W}(y_1,y_2)=\exp\left(i\int_{y_1}^{y_2}dy\oint dx\hat{A}^{xy}\right).
    \end{split}
\end{align}
They are $\mathbb Z_N$ valued operators 
\begin{equation}
    W^N = \hat{W}^N = 1,
\end{equation} 
with the following commutation relations 
\begin{align}\label{Foliated-Theory-Operators-Albetra-1}
    \begin{split}
    W(x_1,x_2) \hat{W} (x,y) &=\exp (2\pi i/N) \hat{W} (x,y) W(x_1,x_2),\quad \textrm{if}\ x_1<x<x_2,\\
    W(y_1,y_2) \hat{W} (x,y) &= \exp (2\pi i/N) \hat{W} (x,y) W(y_1,y_2),\quad \textrm{if}\ y_1<y<y_2,     
    \end{split}
\end{align}
and,
\begin{align}\label{Foliated-Theory-Operators-Albetra-2}
    \begin{split}
    \hat{W}(x_1,x_2) W (x,y) &= \exp (-2\pi i/N) W (x,y) \hat{W}(x_1,x_2),\quad \textrm{if}\ x_1<x<x_2,\\
    \hat{W}(y_1,y_2) W (x,y) &= \exp (-2\pi i/N) W (x,y) \hat{W}(y_1,y_2),\quad \textrm{if}\ y_1<y<y_2,        
    \end{split}
\end{align}
where the extra phase $\exp (\pm 2\pi i/N)$ indicates a mixed t' Hooft anomaly between the two sets of subsystem $\mathbb Z_N$ symmetry generated by the electric and magnetic line/strip operators.

\subsection{Topological boundaries with subsystem symmetry}

In this subsection, we will study the topological
boundaries of the exotic theory~\eqref{eq:exotic}, which are also the topological boundaries of the 2-foliated theory because of the foliated-exotic duality. The boundary theory has subsystem $\mathbb Z_N$ symmetry. For simplicity, we will present the case for $N=2$ which is straightforward to be extended to general $N$. We will study the bosonic topological boundaries corresponding to the Dirichlet boundary condition for $A$ and $\hat{A}$ and the fermionic boundary from the subsystem JW transformation on the bosonic boundary. In addition, we will give a bulk-boundary point of view of subsystem KW and JW transformation.

As reviewed in Sec.~\ref{sec:sub}, it is natural to regularize theories with subsystem symmetry on a lattice. On a finite lattice, the Gauss laws impose nontrivial constraints between gauge invariant operators.  For example, using the Gauss laws \eqref{Foliated-Theory-Gauss-Law}, the holonomy of electric gauge field $A_z$ can be split as
\begin{equation}\label{eq:split}
    \oint dz A^z = \mathcal{A}^y(x) + \mathcal{A}^x (y), 
\end{equation}
where $\mathcal{A}^y(x)$ and $\mathcal{A}^x(y)$ are operators only depend on $x$ and $y$. The split of holonomy \eqref{eq:split} implies the decomposition of the line operator
\begin{equation}
    W(x,y) = W_{z,y}(x) W_{z,x}(y),
\end{equation}
where $W_{z,y}(x),W_{z,x}(y)$ are two line operators along $z$-directions that are separately mobile along $y$ and $x$ directions. However, this decomposition is not unique because of the gauge redundancy
\begin{equation}\label{eq:gaugere}
    \mathcal{A}^y(x) \rightarrow \mathcal{A}^y(x) + \pi,\quad \mathcal{A}^x(y) \rightarrow \mathcal{A}^x(y) + \pi,
\end{equation}
which leaves $\oint dz A^z$ invariant modulo $2\pi$. Both $W_{z,y}(x)$ and $W_{z,x}(x)$ flip the sign under the transformation but the combination $W(x,y)$ is invariant. On the other hand, the strip operators $W(x_1,x_2)$ and $W(y_1,y_2)$ are mobile along $z$ directions with the constraint
\begin{equation}\label{eq:gaugecon}
    W(x,x+L_x) = W(y,y+L_y) = \exp\left( i \oint dx dy A^{xy} \right).
\end{equation}
There are similar gauge redundancy and constraint for magnetic operators $\hat{W}$.

\subsubsection*{Discretization on a lattice}
Discretizing the boundary manifold $M_3$ as a $L_x\times L_y\times L_z$ periodic lattice with label $\{x_i,y_j,z_k\}$, we have in total $2(L_x+L_y)$ electric operators: line operators $W_{z,y}(x_i),W_{z,x}(y_j)$ and strip operators $W(x_i,x_{i+1}),W(y_j,y_{j+1})$ with $i=1,\cdots,L_x, j=1,\cdots,L_y$. On the lattice, the gauge redundancy~\eqref{eq:gaugere} and the constraint~\eqref{eq:gaugecon} become
\begin{equation}\label{Foliated-Theory-Operators-electric-Gauge-Redundancy-electric}
    (W_{z,y}(x_i),W_{z,x}(y_j)) \sim (-W_{z,y}(x_i),-W_{z,x}(y_j)),
\end{equation}
and,
\begin{equation}\label{Foliated-Theory-Operators-electric-Constraints-electric}
    \prod_{i=1}^{L_x} W(x_i,x_{i+1}) \prod_{j=1}^{L_y} W(y_j,y_{j+1}) = 1,
\end{equation}
leaving only $2(L_x+L_y-1)$ operators independent. Similarly, there are $2(L_x+L_y-1)$ independent magnetic $\hat{W}$ operators: line operators $\hat{W}_{z,y}(x_{i+\frac{1}{2}}),\hat{W}_{z,x}(y_{j+\frac{1}{2}})$ and strip operators $\hat{W}(x_{i-\frac{1}{2}},x_{i+\frac{1}{2}}),\hat{W}(y_{j-\frac{1}{2}},y_{j+\frac{1}{2}})$ on the dual lattice with similar gauge redundancy and constraint.

The discretized version of the algebras between $W$ and $\hat{W}$ \eqref{Foliated-Theory-Operators-Albetra-1},\eqref{Foliated-Theory-Operators-Albetra-2} is
\begin{align}\label{Foliated-Theory-Operators-Albetra-Discrete-1}
    \begin{split}
    W(x_{i},x_{i+1}) \hat{W}_{z,y}(x_{i+\frac{1}{2}}) &= - \hat{W}_{z,y}(x_{i+\frac{1}{2}}) W(x_{i},x_{i+1}),\\
    W(y_{i},y_{i+1}) \hat{W}_{z,x}(y_{j+\frac{1}{2}}) &= - \hat{W}_{z,x}(y_{j+\frac{1}{2}}) W(y_{i},y_{i+1}),
    \end{split}
\end{align}
and
\begin{align}\label{Foliated-Theory-Operators-Albetra-Discrete-2}
    \begin{split}
    \hat{W}(x_{i-\frac{1}{2}},x_{i+\frac{1}{2}}) W_{z,y}(x_i) &= - W_{z,y}(x_i) \hat{W}(x_{i-\frac{1}{2}},x_{i+\frac{1}{2}}),\\
    \hat{W}(y_{j-\frac{1}{2}},y_{j+\frac{1}{2}}) W_{z,x}(y_j) &= - W_{z,x}(y_j) \hat{W}(y_{j-\frac{1}{2}},y_{j+\frac{1}{2}}).    
    \end{split}
\end{align}

\subsubsection*{Dirichlet boundary condition for gauge field $A$}
The gauge redundancy \eqref{Foliated-Theory-Operators-electric-Gauge-Redundancy-electric} and constraint \eqref{Foliated-Theory-Operators-electric-Constraints-electric} for electric operators $W$ are consistent to those satisfied by the holonomies $(w_{z,x;j},w_{z,y;i},w_{x;j+\frac{1}{2}},w_{y,i+\frac{1}{2}})$ introduced in \eqref{Subsystem-Review-Holonomies-time},\eqref{Subsystem-Review-Holonomies-time-gauge-redundancy},\eqref{Subsystem-Review-Holonomies-space} and \eqref{Subsystem-Review-Holonomies-space-Constraints} with the following correspondence
\begin{equation}
    W(x_i,y_j) \leftrightarrow (-1)^{w_{z;i,j}},\quad W_{z,x}(y_j) \leftrightarrow (-1)^{w_{z,x;j}},\quad W_{z,y}(x_i) \leftrightarrow (-1)^{w_{z,y;i}}
\end{equation}
and
\begin{equation}
W(y_j,y_{j+1}) \leftrightarrow (-1)^{w_{x;j+\frac12}},\quad W(x_i,x_{i+1}) \leftrightarrow (-1)^{w_{y;i+\frac12}}.
\end{equation}
Therefore, we can introduce a canonical basis of the Hilbert space of the 2-foliated BF theory on the boundary $M_3$
\begin{equation}\label{Foliated-Theory-Boundary-State-A}
    \ket{\mathbf{w}}:=\ket{w_{z,x;j},w_{z,y;i},w_{x;j+\frac12},w_{y;i+\frac12}},
\end{equation}
and the electric operators $W$ are diagonalized as
    \begin{equation}\label{Foliated-Theory-Boundary-State-A-Eigenvaleus}
        \left\{ \begin{array}{l}
            W_{z,x}(y_{j}) \ket{\mathbf{w}} = (-1)^{w_{z,x;j}} \ket{\mathbf{w}}\\
            W_{z,y}(x_{i}) \ket{\mathbf{w}} = (-1)^{w_{z,y;i}} \ket{\mathbf{w}}\\
            W(y_{j},y_{j+1}) \ket{\mathbf{w}} = (-1)^{w_{x;j+\frac12}} \ket{\mathbf{w}}\\
            W(x_i,x_{i+1}) \ket{\mathbf{w}} = (-1)^{w_{y;i+\frac12}} \ket{\mathbf{w}}
        \end{array}\right. .
    \end{equation}
This canonical basis~\eqref{Foliated-Theory-Boundary-State-A} defines the Dirichlet boundary condition for gauge field $A$ where the values of $A$ are fixed at the boundary. 

On the other hand, the magnetic operators $\hat{W}$ conjugate to electric operators $W$ will shift the eigenvalues when acting on the state $\ket{\mathbf{w}}$
\begin{equation}\label{Foliated-Theory-Boundary-State-A-Shift}
        \left\{ \begin{array}{l}
            \hat{W}(y_{j'-\frac{1}{2}},y_{j'+\frac{1}{2}})\ket{\mathbf{w}} = \ket{w_{z,x;j} + \delta_{j,j'},w_{z,y;i},w_{x;j+\frac12},w_{y;i+\frac12}}\\
            \hat{W}(x_{i'-\frac{1}{2}},x_{i'+\frac{1}{2}})\ket{\mathbf{w}} = \ket{w_{z,x;j},w_{z,y;i} + \delta_{i,i'},w_{x;j+\frac12},w_{y;i+\frac12}}\\            
            \hat{W}_{z,x}(y_{j'+\frac{1}{2}})\ket{\mathbf{w}} = \ket{w_{z,x;j},w_{z,y;i},w_{x;j+\frac12} + \delta_{j,j'},w_{y;i+\frac12}}\\
            \hat{W}_{z,y}(x_{i'+\frac{1}{2}})\ket{\mathbf{w}} = \ket{w_{z,x;j},w_{z,y;i},w_{x;j+\frac12},w_{y;i+\frac12} + \delta_{i,i'}}            
        \end{array} \right.
\end{equation}
which follows from the algebras~\eqref{Foliated-Theory-Operators-Albetra-Discrete-1} and~\eqref{Foliated-Theory-Operators-Albetra-Discrete-2}. 
Because the magnetic operators $\hat{W}$ along the spatial/temporal cycle shift the temporal/spatial holonomies $\mathbf{w}$ of electric gauge field $A$,  they are identified one-to-one to the subsystem $\mathbb Z_2$ symmetry and twist operators in \eqref{Subsystem-Review-U-operators},\eqref{Subsystem-Review-U-defects}
\begin{align}\label{Foliated-Theory-Operators-Correspondence}
    \begin{split}
       & \hat{W}(y_{j-\frac{1}{2}},y_{j+\frac{1}{2}}) \leftrightarrow U^x_j,\quad \hat{W}(x_{i-\frac{1}{2}},x_{i+\frac{1}{2}}) \leftrightarrow U^y_i,\\
    &\hat{W}_{z,x}(y_{j+\frac{1}{2}}) \leftrightarrow \prod_{j'\leq j} U^{xz}_{0,j'} ,\quad \hat{W}_{z,y}(x_{i+\frac{1}{2}}) \leftrightarrow \prod_{i'\leq i} U^{yz}_{0,i'}.
    \end{split}
\end{align}

Therefore, the boundary represented by the $\ket{\mathbf{w}}$ basis is a topological boundary supporting the subsystem $\mathbb{Z}_2$ symmetry generated by the magnetic operators $\hat{W}$. 
The general boundary state $ \ket{\mathbf{w}}$ with nontrivial $W$-holonomies is created by acting magnetic operators $\hat{W}$ on the vacuum state $\ket{0}$ where all $W$-holonimies are trivial
\begin{align}\label{Foliated-Theory-Boundary-State-A-Creation}
    \begin{split}
        \ket{\mathbf{w}}=&\prod_{i}\left(\hat{W}_{z,y}(x_{i+\frac12})\right)^{w_{y;i+\frac12}}\left(\hat{W}(x_{i-\frac12},x_{i+\frac12})\right)^{w_{z,y;i}}\\
        &\times\prod_{j}\left(\hat{W}_{z,x}(y_{j+\frac12})\right)^{w_{x;j+\frac12}}\left(\hat{W}(y_{j-\frac12},y_{j+\frac12})\right)^{w_{z,x;j}}\ket{0}.
    \end{split}
\end{align}
As a consistency check, the invariance of $\ket{\mathbf{w}}$ under the gauge redundancy of magnetic operators $\hat{W}$ implies the constraints~\eqref{Subsystem-Review-Holonomies-space-Constraints} and the constraint among magnetic operators $\hat{W}$ requires the invariance of the state $\ket{\mathbf{w}}$ under the gauge transformation~\eqref{Subsystem-Review-Holonomies-time-gauge-redundancy}.

\subsubsection*{Dirichlet boundary condition for gauge field $\hat{A}$}
Alternatively, one can consider the dual basis
\begin{equation}\label{Foliated-Theory-Boundary-State-A-hat}
    \ket{\hat{\mathbf{w}}}:=\ket{\hat{w}_{z,x;j+\frac12}, \hat{w}_{z,y;i+\frac12},\hat{w}_{x;j},\hat{w}_{y;i}},
\end{equation}
where $\hat{W}$ operators are diagonalized
    \begin{equation}\label{Foliated-Theory-Boundary-State-A-hat-Eigenvaleus}
        \left\{ \begin{array}{l}
            \hat{W}_{z,x}(y_{j+\frac{1}{2}}) \ket{\hat{\mathbf{w}}} = (-1)^{\hat{w}_{z,x;j+\frac{1}{2}}} \ket{\hat{\mathbf{w}}}\\
            \hat{W}_{z,y}(x_{i+\frac{1}{2}}) \ket{\hat{\mathbf{w}}} = (-1)^{\hat{w}_{z,y;i+\frac{1}{2}}} \ket{\hat{\mathbf{w}}}\\
            \hat{W}(y_{j-\frac{1}{2}},y_{j+\frac{1}{2}}) \ket{\hat{\mathbf{w}}} = (-1)^{\hat{w}_{x;j}} \ket{\hat{\mathbf{w}}}\\
            \hat{W}(x_{i-\frac{1}{2}},x_{i+\frac{1}{2}}) \ket{\hat{\mathbf{w}}} = (-1)^{\hat{w}_{y;i}} \ket{\hat{\mathbf{w}}}
        \end{array}\right. .
    \end{equation}
The dual basis~\eqref{Foliated-Theory-Boundary-State-A-hat} defines the Dirichlet boundary condition for the gauge field $\hat{A}$. Acting on the state $\ket{\hat{\mathbf{w}}}$, the electric operators $W$ will shift the dual holonomies
    \begin{equation}\label{Foliated-Theory-Boundary-State-A-hat-shift}
        \left\{ \begin{array}{l}
            W(y_{j'},y_{j'+1})\ket{\hat{\mathbf{w}}} = \ket{\hat{w}_{z,x;j+\frac{1}{2}} + \delta_{j,j'},\hat{w}_{z,y;i+\frac{1}{2}},\hat{w}_{x;j},\hat{w}_{y;i}}\\
            W(x_{i'},x_{i'+1})\ket{\hat{\mathbf{w}}} = \ket{\hat{w}_{z,x;j+\frac{1}{2}},\hat{w}_{z,y;i+\frac{1}{2}} + \delta_{i,i'},\hat{w}_{x;j},\hat{w}_{y;i}}\\            
            W_{z,x}(y_{j'})\ket{\hat{\mathbf{w}}} = \ket{\hat{w}_{z,x;j+\frac12}, \hat{w}_{z,y;i+\frac12},\hat{w}_{x;j} + \delta_{j,j'},\hat{w}_{y;i}}\\
            W_{z,y}(x_{i'})\ket{\hat{\mathbf{w}}} = \ket{\hat{w}_{z,x;j+\frac12}, \hat{w}_{z,y;i+\frac12},\hat{w}_{x;j},\hat{w}_{y;i}+ \delta_{i,i'}}            
        \end{array} \right. .
    \end{equation}
Therefore, the electric operators $W$ can be identified as the symmetry and twist operators. The boundary state $\ket{\hat{\mathbf{w}}}$ corresponds to a topological boundary supporting the subsystem $\mathbb{Z}_2$ symmetry generated by electric operators $W$.

The dual state $|\hat{\mathbf{w}}\rangle$ is related to original state $|\mathbf{w}\rangle$ via a discrete Fourier transformation,
\begin{equation}\label{Foliated-Theory-Boundary-State-KW-Relation}
    \ket{\hat{\mathbf{w}}}= \frac{1}{2^{(L_x+L_y-1)}}\sum_{\mathbf{w}\in M_{v}}(-1)^{\sum_{i}(\hat{w}_{z,y;i+\frac12}w_{y;i+\frac12}+\hat{w}_{y;i}w_{z,y;i})+\sum_{j}(\hat{w}_{z,x;j+\frac12}w_{x;j+\frac12}+\hat{w}_{x;j}w_{z,x;j})}\ket{\mathbf{w}},
\end{equation}
where we introduce $M_{v}$ as the set of $\mathbb{Z}_2$-valued vector $\mathbf{w}$ satisfying the gauge redundancy and constraint,
\begin{equation}\label{Foliated-Theory-Holonomies-Restriction}
    M_{v} = \left\{\mathbf{w} \Big{|} \prod_{j=1}^{L_y}(-1)^{w_{x;j+\frac12}}\prod_{i=1}^{L_x}(-1)^{w_{y;i+\frac{1}{2}}}=1 ;  (w_{z,x;j},w_{z,y;i}) \sim (w_{z,x;j}+1,w_{z,y;i}+1) \right\}.
\end{equation}
The restrictions in \eqref{Foliated-Theory-Holonomies-Restriction} for $\mathbf{w}$ automatically impose restrictions for $\hat{\mathbf{w}}$.

\subsubsection*{Subsystem KW transformation}
Based on the SymTFT picture, we consider the 2-foliated BF theory on the 4-dimensional manifold $M_3 \times [0,1]$ where $\tau$ is the coordinate of the time interval. The initial state at $\tau=0$ is the dynamical boundary state $|\chi\rangle$ and the final state at $\tau=1$ is the topological boundary state. Given any $(2+1)$-dimensional theory $\mathfrak{T}_{\textrm{sub}}$ with a  subsystem $\mathbb{Z}_2$ symmetry, we can write down the dynamical boundary state as,
    \begin{equation}\label{Foliated-Theory-Dynamical-Boundary-State}
        |\chi\rangle = \sum_{\mathbf{w}\in M_{v}} Z_{\mathfrak{T}_{\textrm{sub}}}[\mathbf{w}] |\mathbf{w}\rangle,
    \end{equation}
where the coefficient is the partition function of $\mathfrak{T}_{\textrm{sub}}$ on $M_3$ coupled with the subsystem $\mathbb{Z}_2$ symmetry background $\mathbf{w}$.

Choosing $|\mathbf{w}\rangle$ as the topological boundary state at $\tau=1$, one has,
    \begin{equation}
        Z_{\mathfrak{T}_{\textrm{sub}}} = \langle \mathbf{w}| \chi \rangle, 
    \end{equation}
which projects back to the partition function of $\mathfrak{T}_{\textrm{sub}}$. Alternatively, choosing the dual boundary state $|\hat{\mathbf{w}}\rangle$ at $\tau=1$ reproduces the partition function of the dual theory 
\begin{align}
        \begin{split}
        Z_{\hat{\mathfrak{T}}_{\textrm{sub}}}(\hat{\mathbf{w}}) =& \langle \hat{\mathbf{w}} |\chi\rangle\\ 
        =& \frac{1}{2^{(L_x+L_y-1)}} \sum_{\mathbf{w}\in M_{v}}(-1)^{\sum_{i}(\hat{w}_{z,y;i+\frac12}w_{y;i+\frac12}+\hat{w}_{y;i}w_{z,y;i})+\sum_{j}(\hat{w}_{z,x;j+\frac12}w_{x;j+\frac12}+\hat{w}_{x;j}w_{z,x;j})} Z_{\mathfrak{T}_{\textrm{sub}}}(\mathbf{w})             
        \end{split}
\end{align}
The change of boundary conditions in the 2-foliated BF theory recovers the subsystem KW transformation \eqref{Subsystem-Review-Partition-function-KWdual} between the boundary theories.

\subsubsection*{Fermionic boundary conditions}
Based on the discussion of the subsystem JW transformation in the previous section, we can further consider the fermionic topological state $|\mathbf{s}\rangle = |s_{z,x;j}, s_{z,y;i},s_{x;j+\frac12},s_{y;i+\frac12}\rangle$ and write the partition function of $(2+1)$d fermionic theory with subsystem symmetry 
as the path integral $\langle \mathbf{s} |\chi\rangle$. 

For example, the fermionic topological boundary state corresponding to the fermionic theory $\mathfrak{T}_{F,x,\textrm{sub}}$ after the subsystem JW transformation~\eqref{Subsystem-Review-JW-Transformation-x-direction} is
\begin{equation}\label{Foliated-Theory-JW-Transformation-State}
    |\mathbf{s}\rangle = \frac{1}{2^{L_x+L_y-1}} \sum_{(u,w_z) \in M_{u,w_z}}(-1)^{\sum_i u^y_{i} (s_{z,y;i}+w_{z,y;i}) +\sum_j u^x_{j} (s_{z,x;j}+w_{z,x;j})} | w_{z,x;j},w_{z,y;i},w_{x;j+\frac12},w_{y;i+\frac12} \rangle,
\end{equation}
with $w_{x;j+\frac12} =s_{x;j+\frac12} + u^x_{j}+ u^x_{j+1},  w_{y;i+\frac12} = s_{y;i+\frac12} $ and $M_{u,w_z}$ the set,
\begin{equation}
    M_{u,w_z} = \left\{ (u^y_{i},u^x_{j},w_{z,x;j},w_{z,y;i}) \Big{|} \prod_{j=1}^{L_y}(-1)^{u^x_j}\prod_{i=1}^{L_x}(-1)^{u^y_i}=1, (w_{z,x;j},w_{z,y;i}) \sim (w_{z,x;j}+1,w_{z,y;i}+1) \right\}.
\end{equation}
The fermionic state $|\mathbf{s}\rangle$ diagonalizes the electric operators $W$ along the $y$ direction, and the composite operators along $x$ direction made up by the electric operators $W$ sandwiched by a pair of magnetic operators $\hat{W}$ nearby
\begin{equation}\label{Foliated-Theory-Fermionic-State-Operators}
    \left\{ \begin{array}{l}
     \hat{W}_{z,x}(y_{j-\frac{1}{2}})W_{z,x}(y_j)\hat{W}_{z,x}(y_{j+\frac{1}{2}})|\mathbf{s}\rangle = (-1)^{s_{z,x;j}} |\mathbf{s}\rangle\\
     W_{z,y}(x_i)|\mathbf{s}\rangle= (-1)^{s_{z,y;i}}|\mathbf{s}\rangle\\
     \hat{W}(y_{j-\frac{1}{2}},y_{j+\frac{1}{2}})W(y_j,y_{j+1})\hat{W}(y_{j+\frac{1}{2}},y_{j+\frac{3}{2}})|\mathbf{s}\rangle = (-1)^{s_{x;j+\frac{1}{2}}}|\mathbf{s}\rangle\\
     W(x_i,x_{i+1})|\mathbf{s}\rangle = (-1)^{s_{y;i+\frac{1}{2}}}|\mathbf{s}\rangle
    \end{array}\right.
\end{equation}
The fermionic subsystem $\mathbb{Z}_2$ parity symmetry is generated by magnetic operators $\hat{W}$
    \begin{equation}
        \left\{ \begin{array}{l}
            \hat{W}(y_{j'-\frac{1}{2}},y_{j'+\frac{1}{2}})\ket{\mathbf{s}} = \ket{s_{z,x;j} + \delta_{j,j'},s_{z,y;i},s_{x;j+\frac12},s_{y;i+\frac12}}\\
            \hat{W}(x_{i'-\frac{1}{2}},x_{i'+\frac{1}{2}})\ket{\mathbf{s}} = \ket{s_{z,x;j},s_{z,y;i} + \delta_{i,i'},s_{x;j+\frac12},s_{y;i+\frac12}}\\            
            \hat{W}_{z,x}(y_{j'+\frac{1}{2}})\ket{\mathbf{s}} = \ket{s_{z,x;j},s_{z,y;i},s_{x;j+\frac12} + \delta_{j,j'},s_{y;i+\frac12}}\\
            \hat{W}_{z,y}(x_{i'+\frac{1}{2}})\ket{\mathbf{s}} = \ket{s_{z,x;j},s_{z,y;i},s_{x;j+\frac12},s_{y;i+\frac12} + \delta_{i,i'}}            
        \end{array} \right. .
    \end{equation}

There exists another fermionic topological state $|\mathbf{s}'\rangle = |s'_{z,x;j}, s'_{z,y;i},s'_{x;j+\frac12},s'_{y;i+\frac12}\rangle$ which produces the fermionic theory $\mathfrak{T}_{F,y,\textrm{sub}}$ after the subsystem JW transformation along $y$ direction. The fermionic topological state $|\mathbf{s}'\rangle$ diagonalizes the line operators,
\begin{equation}
    W_{z,x}(y_j),\quad \hat{W}_{z,y}(x_{i-\frac{1}{2}})W_{z,y}(x_i)\hat{W}_{z,y}(x_{i+\frac{1}{2}}), \nonumber
\end{equation}
and strip operators,
\begin{equation}
    W(y_j,y_{j+1}),\quad \hat{W}(x_{i-\frac{1}{2}},x_{i+\frac{1}{2}})W(x_i,x_{i+1})\hat{W}(x_{i+\frac{1}{2}},x_{i+\frac{3}{2}}), \nonumber
\end{equation}
where $W_{z,y}(x_i),W(x_i,x_{i+1})$ are sandwiched by a pair of $\hat{W}$ operators instead. The fermionic subsystem $\mathbb{Z}_2$ parity symmetry is still generated by magnetic operators $\hat{W}$.

\subsubsection*{Subsystem JW transformation}
Consider the subsystem SymTFT with the dynamical boundary state~\eqref{Foliated-Theory-Dynamical-Boundary-State} at $\tau=0$ given by the $(2+1)$-dimensional bosonic theory $\mathfrak{T}_{\textrm{sub}}$. Implementing the fermionic topological boundaries $|\mathbf{s}\rangle,|\mathbf{s}'\rangle$ at $\tau=0$ and shrinking the slab gives two fermionic theories $\mathfrak{T}_{F,x,\textrm{sub}}$ and $\mathfrak{T}_{F,y,\textrm{sub}}$ whose partition functions are,
\begin{equation}
    Z_{\mathfrak{T}_{F,x,\textrm{sub}}}(\mathbf{s}) = \langle \mathbf{s}| \chi\rangle,\quad Z_{\mathfrak{T}_{F,y,\textrm{sub}}}(\mathbf{s}') = \langle \mathbf{s}'| \chi\rangle.
\end{equation}
They are related to the bosonic theory $\mathfrak{T}_{\textrm{sub}}$ by performing the subsystem JW transformations along $x$ and $y$ directions respectively.

\section{Subsystem \texorpdfstring{$SL(2,\mathbb{Z}_2)\ $}\ transformation and the duality web}\label{sec:sl2z}

In the previous section, we propose the 2-foliated BF theory in $(3+1)$d 
as the subsystem SymTFT for subsystem $\mathbb Z_N$ symmetry in $(2+1)$d and explore various bosonic and fermionic topological boundaries. In this section, we will see that different topological boundaries are transformed from one to the other via the topological operators associated with the global symmetries of the bulk theory. 

In the exotic theory~\eqref{eq:exotic}, we identify a naive 0-form $SL(2,\mathbb Z_2)$ symmetry
\begin{align}\label{eq:sym2}
    \begin{split}
        &S:\quad A\rightarrow \hat{A},\quad \hat{A}\rightarrow -A,\\
        &  T:\quad A\rightarrow A,\quad \hat{A} \rightarrow \hat{A} + A.
    \end{split}
\end{align}
There should exist corresponding co-dimension one symmetry defects implementing this symmetry. Here, we will mainly focus on the co-dimension one symmetry defects extended along the manifold $M_3'$  parallel to the boundary manifold $M_3$ such that they act on the Hilbert space as operators. Fusing the topological operators with the boundary implements the $SL(2,\mathbb Z_N)$ transformation of the boundary theory. We will see the $S$-transformation generates the subsystem KW transformation, while the $T$-transformation stacks a phase 
\begin{equation}\label{eq:ssptp}
    \exp(-\frac{iN}{2\pi}\int dxdydz A^zA^{xy})
\end{equation}
to the boundary theory. 
The phase~\eqref{eq:ssptp} is the subsystem symmetry protected topological (SSPT) phase~\cite{Burnell:2021reh}~\footnote{
In \cite{Burnell:2021reh}, the Lagrangian of this SSPT is
\begin{equation}\label{eq:sspt}
    \mathcal L_{\text{SSPT}}= \frac{iN}{2\pi} \Phi^{xy} (\partial_z A^{xy} - \partial_{x}\partial_y A^{z})-\frac{iN}{2\pi} A^z A^{xy} ,
\end{equation}
where the auxiliary field $\Phi^{xy}$ guarrentees the flat condition of the gauge field $A$.} in $(2+1)$d.

However, the naive $SL(2,\mathbb Z_2)$ transformation~\eqref{eq:sym2} has ambiguities on the lattice.
For example, when we do $S$-transformation on line operators $\sum_k A^z_{i,j,k+\frac{1}{2}}$, the holonomy of electric gauge field $A$ along the $z$-direction, one expects to map the electric gauge field $A^z$ operators to the nearby magnetic gauge field $\hat{A}^z$ on the dual lattice.  This leads to four inequivalent choices $\sum_z A^z_{i\pm\frac{1}{2},j\pm\frac{1}{2},k}$ because the line operators cannot move freely at the $x$-$y$ plane.
We also need to make a smart choice to avoid the following inconsistencies.
\subsubsection*{Inconsistency with the quantum algebra} 
Consider the following choice of regularized $S$-transformation between the line operators,
\begin{equation}\label{SL2Z-Duality-S-Transformation-Fake-1}
    W_{z,y}(x_i) \leftrightarrow \hat{W}_{z,y}(x_{i+\frac{1}{2}}),\quad W_{z,x}(y_j) \leftrightarrow \hat{W}_{z,x}(y_{j+\frac{1}{2}}),
\end{equation}
and strip operators,
\begin{equation}\label{SL2Z-Duality-S-Transformation-Fake-2}
    W(x_i,x_{i+1}) \leftrightarrow \hat{W}(x_{i+\frac{1}{2}},x_{i+\frac{3}{2}}),\quad W(y_j,y_{j+1}) \leftrightarrow \hat{W}(y_{j+\frac{1}{2}},y_{j+\frac{3}{2}}).
\end{equation}
It maps between the site $(i,j)$ and dual site $(i+\frac{1}{2},j+\frac{1}{2})$. However, the quantum algebras~\eqref{Foliated-Theory-Operators-Albetra-Discrete-1} and~\eqref{Foliated-Theory-Operators-Albetra-Discrete-2} are not preserved under the transformation. For example, consider the following commutation relation,
    \begin{equation}
        W(x_{i},x_{i+1}) \hat{W}_{z,y}(x_{i+\frac{1}{2}}) = - \hat{W}_{z,y}(x_{i+\frac{1}{2}}) W(x_{i},x_{i+1}).
    \end{equation}
If we apply the $S$-transformation given above, we have,
    \begin{equation}
        \hat{W}(x_{i+\frac{1}{2}},x_{i+\frac{3}{2}}) W_{z,y}(x_i) = - W_{z,y}(x_i) \hat{W}(x_{i+\frac{1}{2}},x_{i+\frac{3}{2}})
    \end{equation}
which is clearly wrong because the nontrivial phase only appears after the exchange of electric operators and magnetic operators with intersection. 
\subsubsection*{Inconsistency with the topological property} 
For another choice, we can keep \eqref{SL2Z-Duality-S-Transformation-Fake-1} and modify \eqref{SL2Z-Duality-S-Transformation-Fake-2} to,
    \begin{equation}
        W(x_i,x_{i+1}) \leftrightarrow \hat{W}(x_{i-\frac{1}{2}},x_{i+\frac{1}{2}}),\quad W(y_j,y_{j+1}) \leftrightarrow \hat{W}(y_{j-\frac{1}{2}},y_{j+\frac{1}{2}}),
    \end{equation}
 and we will denote this choice as $\widetilde{S}$. It is straightforward to check $\widetilde{S}$ preserve the quantum algebras \eqref{Foliated-Theory-Operators-Albetra-Discrete-1} and~\eqref{Foliated-Theory-Operators-Albetra-Discrete-2} and it satisfies $\widetilde{S}^2=1$. However, $\widetilde{S}$ assumes different site transformations for line operators and strip operators: it maps $(i,j)$ to $(i+\frac{1}{2},j+\frac{1}{2})$ for line operators and to $(i-\frac{1}{2},j-\frac{1}{2})$ for strip operators. This is inconsistent with the fact that we can bend the strip operators to a pair of line operators. 
 
$\widetilde{S}$ will generate the subsystem KW transformation on the boundary by mapping the topological boundary state $|\mathbf{w}\rangle$ to the dual state $|\hat{\mathbf{w}}\rangle$. For example, applying $\widetilde{S}$ on $|\mathbf{w}\rangle$ leads to 
\begin{align}\label{SL2Z-Duality-S-Transformation-on-States-Fake}
    \begin{split}
       \widetilde{S} \ket{\mathbf{w}}=&\ \widetilde{S}\left(\prod_{i}\left(\hat{W}_{z,y}(x_{i+\frac12})\right)^{w_{y;i+\frac12}}\left(\hat{W}(x_{i-\frac12},x_{i+\frac12}) )\right)^{w_{z,y;i}}\right.\\
        &\ \left. \times\prod_{j}\left(\hat{W}_{z,x}(y_{j+\frac12})\right)^{w_{x;j+\frac12}}\left(\hat{W}(y_{j-\frac12},y_{j+\frac12}) \right)^{w_{z,x;j}}\ket{0}\right)\\
        =&\ \prod_{i}\left(W_{z,y}(x_i)\right)^{w_{y;i+\frac12}}\left( W(x_i,x_{i+1})\right)^{w_{z,y;i}}\\
        &\ \times\prod_{j}\left(W_{z,x}(y_j)\right)^{w_{x;j+\frac12}}\left( W(y_j,y_{j+1})\right)^{w_{z,x;j}}\ket{\hat{0}}\\
        =&\ \ket{\hat{\mathbf{w}}},
    \end{split}
\end{align}
where the dual holonomies $\hat{\mathbf{w}}$ is the same to the original ones $\mathbf{w}$ in value,
    \begin{equation}
        \hat{w}_{z,x;j+\frac{1}{2}} = w_{z,x;j},\quad \hat{w}_{z,y;i+\frac{1}{2}} = w_{z,y;i},\quad \hat{w}_{x;j} = w_{x,j+\frac{1}{2}},\quad \hat{w}_{y;i} = w_{y,i+\frac{1}{2}}.
    \end{equation}
Here $|\hat{0}\rangle = \widetilde{S}|0\rangle$ is the vacuum of the dual state and it is the eigenstate of the operators $\hat{W}$ with trivial eigenvalues.

\subsubsection*{Subsystem $SL(2,\mathbb Z_2$) transformation on the lattice}
\begin{figure}[htbp]
    \centering
    \includegraphics[scale=0.5]{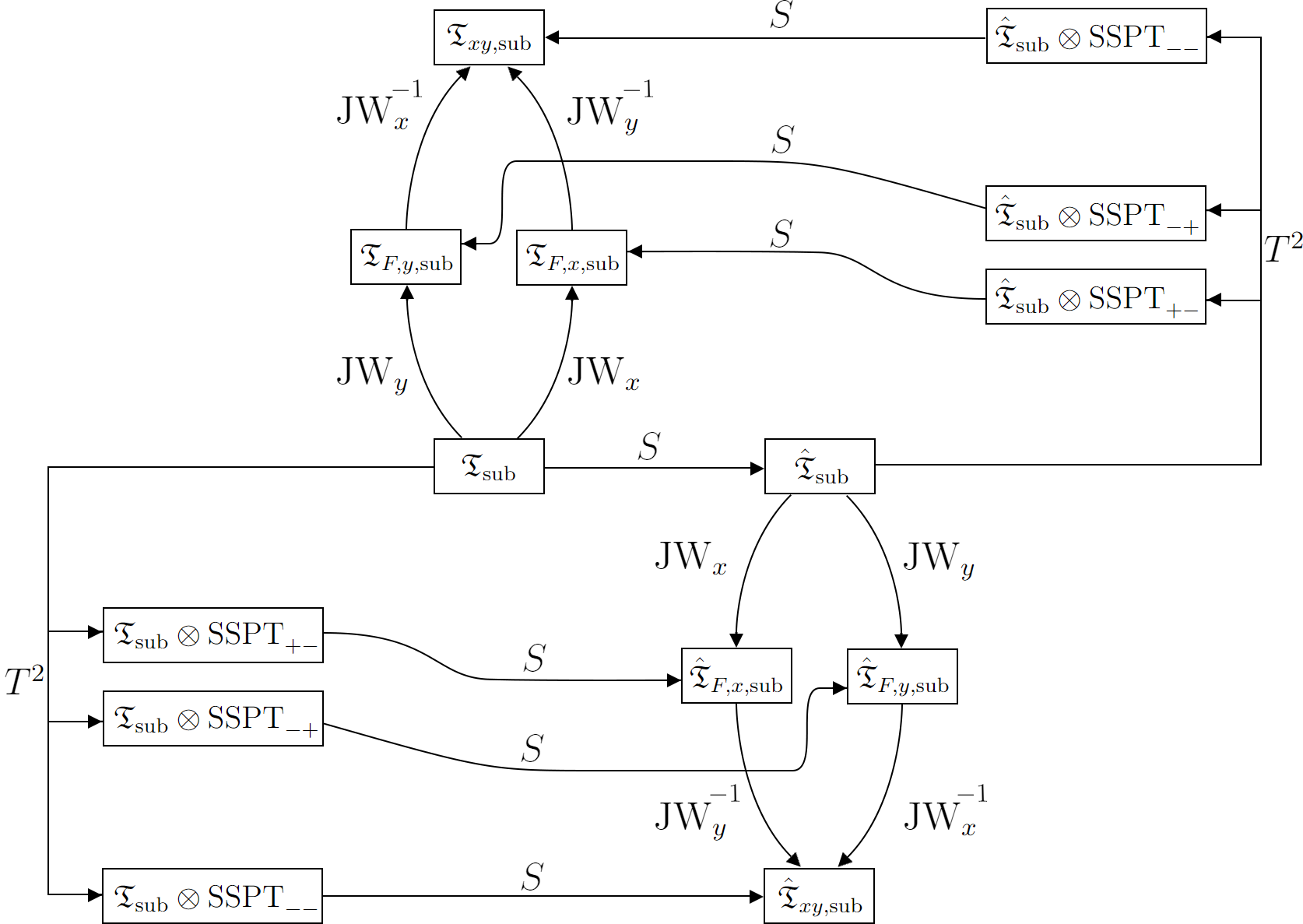}
    \caption{The duality web between four bosonic theories $\mathfrak{T}_{\textrm{sub}},\mathfrak{T}_{xy,\textrm{sub}},\hat{\mathfrak{T}}_{\textrm{sub}},\hat{\mathfrak{T}}_{xy,\textrm{sub}}$ and four fermionic theories $\mathfrak{T}_{F,x,\textrm{sub}},\mathfrak{T}_{F,y,\textrm{sub}},\hat{\mathfrak{T}}_{F,x,\textrm{sub}},\hat{\mathfrak{T}}_{F,y,\textrm{sub}}$ with susbsytem $\mathbb Z_2$ symmetry. The duality transformation is generated by subsystem $SL(2,\mathbb Z_2)$ transformation on the lattice: (1) The subsystem $S$-transformation implements the subsystem KW transformation. (2) 
    There exist nontrivial compositions of $T$-transformations $T^2_{+-},T^2_{-+},T^2_{--}$ generate the phase $\text{SSPT}_{+-},\text{SSPT}_{-+},\text{SSPT}_{--}$. (3) Subsystem JW transformation is a composition of subsystem $SL(2,\mathbb Z_2)$ transformations. For example, the bosonic theory $\mathfrak{T}_{\textrm{sub}}$ and the fermionic theory $\mathfrak{T}_{F,x,\textrm{sub}}(\mathfrak{T}_{F,y,\textrm{sub}})$ are related by subsystem JW transformation, which is equivalent to performing $S^{-1}$, $T^2_{+-}(T^2_{-+})$ and $S$ transformation sequentially.}
    \label{fig:duality-web}
\end{figure}

In this section, we will formulate the proper $S$- and $T$-transformations on the lattice and study their action on the operators and topological states with a focus on $N=2$. 
The proper $SL(2,\mathbb{Z}_2)$ symmetry transformation after discretization should have the following properties
\begin{enumerate}
    \item It should be a \emph{symmetry} of the discretized version of the exotic action~\eqref{eq:exotic} and preserve quantum algebras~\eqref{Foliated-Theory-Operators-Albetra-Discrete-1} and~\eqref{Foliated-Theory-Operators-Albetra-Discrete-2}.
    \item It should be consistent with the topological property of the operators, for example, the bending of strip operators in $z$ direction \eqref{Foliated-Theory-Strip-Operators}.\footnote{Thanks to Wilbur Shirley for raising this issue to us.}
\end{enumerate}  
We will denote the $SL(2,\mathbb Z_2)$ transformation on the lattice as \emph{subsystem} $SL(2,\mathbb Z_2)$ transformation. Besides recovering the subsystem KW and JW transformations, we will find more duality transformations by implementing the subsystem $SL(2,\mathbb Z_2)$ transformation on the boundary. The whole duality transformations are summarized in the duality web (Fig.~\ref{fig:duality-web}).

\subsection{Subsystem \texorpdfstring{$S$}\ -transformation}
The proper $S$-transformation on the lattice is implemented by changing the double-headed arrows in \eqref{SL2Z-Duality-S-Transformation-Fake-1} and \eqref{SL2Z-Duality-S-Transformation-Fake-2} to one-headed arrows and let the $S$-transformation maps the dual site $(i+\frac{1}{2},j+\frac{1}{2})$ to $(i+1,j+1)$ for line operators,
    \begin{equation}
        \hat{W}_{z,y}(x_{i+\frac{1}{2}}) \rightarrow W_{z,y}(x_{i+1}),\quad \hat{W}_{z,x}(y_{j+\frac{1}{2}}) \rightarrow W_{z,x}(y_{j+1}),
    \end{equation}
and for strip operators,
    \begin{equation}
        \hat{W}(x_{i-\frac{1}{2}},x_{i+\frac{1}{2}}) \rightarrow W(x_{i},x_{i+1}),\quad \hat{W}(y_{j-\frac{1}{2}},y_{j+\frac{1}{2}}) \rightarrow W(y_j,y_{j+1}),
    \end{equation}
such that $S^2 = \mathcal{T}$ is the translation $\mathcal{T} : (i,j) \rightarrow (i+1,j+1)$ on lattice. This choice preserves the quantum algebra and is also consistent with the bending of operators at the expense of giving up $S^2=1$. Similarly, it will also generate the subsystem KW transformation on the boundary by mapping the topological boundary state $|\mathbf{w}\rangle$ to the dual state $|\hat{\mathbf{w}}\rangle = S |\mathbf{w}\rangle$ as
    \begin{equation}\label{SL2Z-Duality-S-Transformation-on-States}
        \hat{w}_{z,x;j+\frac{1}{2}} = w_{z,x;j},\quad \hat{w}_{z,y;i+\frac{1}{2}} = w_{z,y;i},\quad \hat{w}_{x;j} = w_{x,j-\frac{1}{2}},\quad \hat{w}_{y;i} = w_{y,i-\frac{1}{2}},
    \end{equation}
and also
    \begin{align}
        S^2 |\mathbf{w}\rangle = \mathcal{T} |\mathbf{w}\rangle,
    \end{align}
where $\mathcal{T}$ will shift the holonomies as
    \begin{equation}
        w_{z,x;j}\rightarrow w_{z,x;j-1},\quad w_{z,y;i}\rightarrow w_{z,y;i-1},\quad w_{x;j+\frac{1}{2}} \rightarrow w_{x;j-\frac{1}{2}},\quad w_{y;i+\frac{1}{2}}\rightarrow w_{y;i-\frac{1}{2}}.
    \end{equation}

As a summary, we have two possible definitions of $S$-transformation on lattice denoted as $\widetilde{S}$ and $S$ and both of them preserve the quantum algebra. The first one satisfies the naive relation $\widetilde{S}^2=1$ but is not consistent with the bending of operators. Therefore it is not a suitable choice on the lattice. We will denote $\widetilde{S}$ as \emph{field theory $S$-transformation} since it implements the naive $S$-transformation of the exotic tensor theory in \eqref{eq:sym2}. The second one is consistent with the bending on the lattice but $S^2$ is a translation $\mathcal{T}$ instead. We will also denote $S$ as \emph{lattice S-transformation}. Both of them will map $|\mathbf{w}\rangle$ to dual state $|\hat{\mathbf{w}}\rangle$ with different assignments of $\hat{\mathbf{w}}$.

\subsection{Subsystem \texorpdfstring{$T$}\ -transformation}
For the subsystem $T$-transformation in~\eqref{eq:sym2}, one needs to dress every magnetic operator $\hat{W}$ with a nearby electric operator $W$. Again we need to avoid the following inconsistencies.

\subsubsection*{Inconsistency with the quantum algebra}
Naively one could have for example,
\begin{equation}\label{SL2Z-Duality-T-Transformation-Fake}
    \hat{W}_{z,y}(x_{i+\frac{1}{2}}) \rightarrow  \hat{W}_{z,y}(x_{i+\frac{1}{2}}) W_{z,y}(x_{i+1}) ,\quad \hat{W}(x_{i-\frac{1}{2}},x_{i+\frac{1}{2}}) \rightarrow \hat{W}(x_{i-\frac{1}{2}},x_{i+\frac{1}{2}}) W (x_i,x_{i+1}),
\end{equation}
where the $W$-operators are on the right of $\hat{W}$-operators. However, just as the $S$-transformation cases the quantum algebras \eqref{Foliated-Theory-Operators-Albetra-Discrete-1} and~\eqref{Foliated-Theory-Operators-Albetra-Discrete-2} are not preserved and $T$-transformation is not a good symmetry on the lattice.

\subsubsection*{Inconsistency with the topological property} 
One can try to modify the transformation \eqref{SL2Z-Duality-T-Transformation-Fake} in a way consistent with the algebra
\begin{align}\label{eq:subT}
    \begin{split}
        &+:\quad \hat{W}_{z,y}(x_{i+\frac{1}{2}}) \rightarrow W_{z,y}(x_i) \hat{W}_{z,y}(x_{i+\frac{1}{2}}) ,\quad \hat{W}(x_{i-\frac{1}{2}},x_{i+\frac{1}{2}}) \rightarrow \hat{W}(x_{i-\frac{1}{2}},x_{i+\frac{1}{2}}) W (x_i,x_{i+1}), \\
        & -:\quad \hat{W}_{z,y}(x_{i+\frac{1}{2}}) \rightarrow  \hat{W}_{z,y}(x_{i+\frac{1}{2}}) W_{y}(x_{i+1}), \quad \hat{W}(x_{i-\frac{1}{2}},x_{i+\frac{1}{2}}) \rightarrow  W (x_{i-1},x_{i}) \hat{W}(x_{i-\frac{1}{2}},x_{i+\frac{1}{2}}),
    \end{split}
\end{align}
and there are two similar choices for operators depending on $y$: $\hat{W}(y_{j-\frac{1}{2}},y_{j+\frac{1}{2}})$ and $\hat{W}_{x}(y_{j+\frac{1}{2}})$. In total, we have four choices and they are denoted as $T_{++},T_{+-},T_{-+},T_{--}$. However, none of the four choices $T_{\pm \pm}$ are compatible with the bending of operators.  Nevertheless, we will elaborate on their actions because they are useful later when we consider the proper $T^2$ transformation on the lattice.

The transformations \eqref{eq:subT} will stack an extra phase when acting on the topological boundary. For example, with the expression~\eqref{Foliated-Theory-Boundary-State-A-Creation} of the topological boundary state $|\mathbf{w}\rangle$, applying $T_{++}$ leads to a new topological boundary state 
\begin{align}
    \begin{split}
        T_{++} \ket{\mathbf{w}}=&\ T_{++} \prod_{i}\left(\hat{W}_{z,y}(x_{i+\frac12})\right)^{w_{y;i+\frac12}}\left(\hat{W}(x_{i-\frac12},x_{i+\frac12}) \right)^{w_{z,y;i}}\\
        &\ \times\prod_{j}\left(\hat{W}_{z,x}(y_{j+\frac12})\right)^{w_{x;j+\frac12}}\left(\hat{W}(y_{j-\frac12},y_{j+\frac12}) \right)^{w_{z,x;j}}\ket{0}\\      
        =&\ \prod_{i}\left(W_{z,y}(x_i)\hat{W}_{z,y}(x_{i+\frac12})\right)^{w_{y;i+\frac12}}\left(\hat{W}(x_{i-\frac12},x_{i+\frac12}) W(x_i,x_{i+1})\right)^{w_{z,y;i}}\\
        &\ \times\prod_{j}\left(W_{z,x}(y_j)\hat{W}_{z,x}(y_{j+\frac12})\right)^{w_{x;j+\frac12}}\left(\hat{W}(y_{j-\frac12},y_{j+\frac12}) W(y_j,y_{j+1})\right)^{w_{z,x;j}}T_{++}\ket{0}\\
        =&\ (-1)^{\sum_j w_{z,x;j} w_{x;{j+\frac{1}{2}}} + \sum_i w_{z,y;i} w_{y;i+\frac{1}{2}} } |\mathbf{w}\rangle       
    \end{split}
\end{align}
where $T_{++} |0\rangle \sim |0\rangle$ because they satisfy the same operators equation \eqref{Foliated-Theory-Boundary-State-A-Eigenvaleus}, and we will assume $|0\rangle$ is invariant under the action of $T_{++}$. 
In general, acting 
$T_{\pm \pm}$ on the topological boundary $|\mathbf{w}\rangle$ will stack the phase
\begin{equation}\label{SL2Z-Duality-SPT}
    (-1)^{\sum_j w_{z,x;j} w_{x;{j\pm\frac{1}{2}}} + \sum_i w_{z,y;i} w_{y;i\pm\frac{1}{2}} }.
\end{equation}
As we mentioned before, $T_{\pm \pm}$ are not good transformations on lattice and we should not take those phases seriously.

\subsubsection*{Proper $T^2$ transformation on the lattice} 

When we compose different $T_{\pm \pm}$ on the lattice, there exist  $T^2$ -transformations which are consistent with both the algebras and the bending.
In the naive $SL(2,\mathbb Z_2)$ transformation~\eqref{eq:sym2} of the field theory, acting $T$-transformation twice is the identity transformation. However, on the lattice, composing different $T$-transformations will lead to four distinct operations
\begin{align}\label{eq:t2}
   \begin{split}
   T^2_{++} &\equiv T_{++} T_{++} = T_{--} T_{--} = T_{+-} T_{+-} = T_{-+} T_{-+},\\
   T^2_{--} &\equiv T_{++} T_{--} = T_{--} T_{++} = T_{+-} T_{-+} = T_{-+} T_{+-},\\
   T^2_{+-} &\equiv T_{++} T_{+-} = T_{--} T_{-+} = T_{+-} T_{++} = T_{-+} T_{--},\\
   T^2_{-+} &\equiv T_{++} T_{-+} = T_{--} T_{+-} = T_{+-} T_{--} = T_{-+} T_{++},
   \end{split}
\end{align}
where the indices follow the sign rule
\begin{equation}
        T^2_{pp',qq'} = T_{p,q} T_{p',q'},\quad p,q=\pm.
\end{equation}
We can also write down       the transformation of operators for $T^2_{\pm \pm}$. $T^2_{++}$ is the identity transformation and $T^2_{--}$ is realized by,
\begin{equation}\label{SL2Z-Duality-T--}
    T^2_{--}: \quad \left\{ \begin{array}{l}
        \hat{W}_{z,y}(x_{i+\frac{1}{2}}) \rightarrow W_{z,y}(x_i) \hat{W}_{z,y}(x_{i+\frac{1}{2}}) W_{z,y}(x_{i+1})\\
        \hat{W}_{z,x}(y_{j+\frac{1}{2}}) \rightarrow W_{z,x}(y_j) \hat{W}_{z,x}(y_{j+\frac{1}{2}}) W_{z,x}(y_{j+1})\\
        \hat{W}(x_{i-\frac{1}{2}},x_{i+\frac{1}{2}}) \rightarrow W (x_{i-1},x_{i}) \hat{W}(x_{i-\frac{1}{2}},x_{i+\frac{1}{2}}) W (x_i,x_{i+1})\\
        \hat{W}(y_{j-\frac{1}{2}},y_{j+\frac{1}{2}}) \rightarrow W (y_{j-1},y_{j}) \hat{W}(y_{j-\frac{1}{2}},y_{j+\frac{1}{2}}) W (y_j,y_{j+1})\\
    \end{array} \right. 
\end{equation}
where all $\hat{W}$ operators are sandwiched by a pair of $W$ operators in a symmetric way. The other two choices $T^2_{+-}$ and $T^2_{-+}$ are given by,
\begin{equation}\label{SL2Z-Duality-T+-}
    T^2_{+-}: \quad \left\{ \begin{array}{l}
        \hat{W}_{z,y}(x_{i+\frac{1}{2}}) \rightarrow \hat{W}_{z,y}(x_{i+\frac{1}{2}}) \\
        \hat{W}_{z,x}(y_{j+\frac{1}{2}}) \rightarrow W_{z,x}(y_j) \hat{W}_{z,x}(y_{j+\frac{1}{2}}) W_{z,x}(y_{j+1})\\
        \hat{W}(x_{i-\frac{1}{2}},x_{i+\frac{1}{2}}) \rightarrow \hat{W}(x_{i-\frac{1}{2}},x_{i+\frac{1}{2}}) \\
        \hat{W}(y_{j-\frac{1}{2}},y_{j+\frac{1}{2}}) \rightarrow W (y_{j-1},y_{j}) \hat{W}(y_{j-\frac{1}{2}},y_{j+\frac{1}{2}}) W (y_j,y_{j+1})\\
    \end{array} \right. 
\end{equation}
and,
\begin{equation}\label{SL2Z-Duality-T-+}
    T^2_{-+}: \quad \left\{ \begin{array}{l}
        \hat{W}_{z,y}(x_{i+\frac{1}{2}}) \rightarrow W_{z,y}(x_i) \hat{W}_{z,y}(x_{i+\frac{1}{2}}) W_{z,y}(x_{i+1})\\
        \hat{W}_{z,x}(y_{j+\frac{1}{2}}) \rightarrow  \hat{W}_{z,x}(y_{j+\frac{1}{2}}) \\
        \hat{W}(x_{i-\frac{1}{2}},x_{i+\frac{1}{2}}) \rightarrow W (x_{i-1},x_{i}) \hat{W}(x_{i-\frac{1}{2}},x_{i+\frac{1}{2}}) W (x_i,x_{i+1})\\
        \hat{W}(y_{j-\frac{1}{2}},y_{j+\frac{1}{2}}) \rightarrow  \hat{W}(y_{j-\frac{1}{2}},y_{j+\frac{1}{2}}) \\
    \end{array} \right. 
\end{equation}
where only part of $\hat{W}$ operators are sandwiched by $W$ operators. Obviously, we have $T^2_{-+} T^2_{+-} = T^2_{--}$ and they all satisfy $\left(T^2_{\pm \pm}\right)^2=1$. The corresponding subsystem symmetric protected topological (SSPT) phases are
\begin{align}
    \begin{split}
        \text{SSPT}_{++}(\mathbf{w})= &\  1,\\
        \textrm{SSPT}_{+-}(\mathbf{w}) =&\ (-1)^{\sum_j w_{z,x;j} (w_{x;{j-\frac{1}{2}}}+w_{x;{j+\frac{1}{2}}})},\\
        \textrm{SSPT}_{-+}(\mathbf{w}) =&\ (-1)^{\sum_i w_{z,y;i} (w_{y;i-\frac{1}{2}} + w_{y;i+\frac{1}{2}})},\\
        \textrm{SSPT}_{--} (\mathbf{w}) =&\  (-1)^{\sum_j w_{z,x;j} (w_{x;{j-\frac{1}{2}}}+w_{x;{j+\frac{1}{2}}}) + \sum_i w_{z,y;i} (w_{y;i-\frac{1}{2}} + w_{y;i+\frac{1}{2}}) }.
    \end{split}
\end{align}

\subsection{Duality web from the subsystem \texorpdfstring{$SL(2,\mathbb Z_2)\ $}\ transformation}

The duality web (Fig.~\ref{fig:duality-web}) is generated by implementing the subsystem $SL(2,\mathbb Z_2)$ transformations $S$ and $T^2$ consecutively. In particular, the subsystem JW transformation is equivalent to performing $S$, $T^2_{+-}(\textrm{or}\  T^2_{-+})$ and $S^{-1}$ transformation sequentially.
This is easy to see in the transformation of the operators. Begin with the bosonic state $|\mathbf{w}\rangle$ which are eigenstates of $W$ operators. If we do an $S$-transformation the roles of $W$ and $\hat{W}$ are exchanged and we get the dual state $|\hat{\mathbf{w}}\rangle$ which are eigenstates of $\hat{W}$ operators. Then applying $T^2_{+-}$ we find $W_{z,y}(x_i)$ and $W(x_i,x_{i+1})$ should be sandwiched by a pair of $\hat{W}$ operators according to \eqref{SL2Z-Duality-T+-} (notice that the roles of $W$ and $\hat{W}$ have been exchanged due to the $S$ transformation). It will stack the phase $\textrm{SSPT}_{+-}(\hat{\mathbf{w}})$ to $|\hat{\mathbf{w}}\rangle$. If we do another $S^{-1}$ transformation then we will obtain some states which are the eigenstates of line operators,
\begin{equation}\label{eq:STS-line}
   \hat{W}_{z,x}(y_{j-\frac{1}{2}})W_{z,x}(y_j)\hat{W}_{z,x}(y_{j+\frac{1}{2}}),\quad W_{z,y}(x_i),
\end{equation}
and strip operators,
\begin{equation}\label{eq:STS-strip}
    \hat{W}(y_{j-\frac{1}{2}},y_{j+\frac{1}{2}})W(y_j,y_{j+1})\hat{W}(y_{j+\frac{1}{2}},y_{j+\frac{3}{2}}),\quad W(x_i,x_{i+1}).
\end{equation}
According to \eqref{Foliated-Theory-Fermionic-State-Operators} they are the same set of operators that diagonalize the fermion topological state $|\mathbf{s}\rangle$. Therefore we have,
    \begin{equation}
        |\mathbf{s}\rangle = S^{-1} T^2_{+-} S |\mathbf{w}\rangle,
    \end{equation}
with $\mathbf{s} = \mathbf{w}$. By similar argument, if we perform $T^2_{+-}$ and $S$ transformation sequentially the resulting states are eigenvalues of \eqref{eq:STS-line} and \eqref{eq:STS-strip} with $W,\hat{W}$ exchanged. We obtain a new fermionic topological state,
\begin{equation}
     |\hat{\mathbf{s}}\rangle \equiv   ST^2_{+-}\ket{\mathbf{w}}
\end{equation}
where the relation between  $\hat{\mathbf{s}}$ and $w$ are suggested in \eqref{SL2Z-Duality-S-Transformation-on-States}. They are the JW transformations of the dual state $|\hat{\mathbf{w}}\rangle$.

Let's check explicitly the state $|\hat{\mathbf{s}}\rangle$ can be written as a JW transformation of $|\hat{\mathbf{w}}\rangle$ by summing over all sectors of the dual bosonic state with proper phases. Following the definition, we stack the phase $\text{SSPT}_{+-}(\mathbf{w}')$ on the state $|\mathbf{w}'\rangle$ introduced in \eqref{Foliated-Theory-Boundary-State-A} and consider the KW transformation given by,
\begin{align}
\begin{split}
    |\hat{\mathbf{s}}\rangle = &\  \frac{1}{2^{L_x+L_y-1}} \sum_{\mathbf{w}'\in M_{v}}(-1)^{\sum_{i}(\hat{s}_{z,y;i+\frac12}w'_{y;i+\frac12}+\hat{s}_{y;i}w'_{z,y;i})+\sum_{j}(\hat{s}_{z,x;j+\frac12}w'_{x;j+\frac12}+\hat{s}_{x;j}w'_{z,x;j})}\\
    &\times (-1)^{\sum_j w'_{z,x;j} (w'_{x;{j-\frac{1}{2}}}+w'_{x;{j+\frac{1}{2}}})} |\mathbf{w}'\rangle.
\end{split}
\end{align}
We can rewrite $|\mathbf{w}'\rangle$ into $|\hat{\mathbf{w}}'\rangle$ using the KW relation \eqref{Foliated-Theory-Boundary-State-KW-Relation} and get
\begin{align}\label{SL2Z-Duality-Web-Double-T-calculation}
\begin{split}
    |\hat{\mathbf{s}}\rangle = &\  \frac{1}{2^{L_x+L_y-1}} \sum_{\mathbf{w}'\in M_{v}}(-1)^{\sum_{i}(\hat{s}_{z,y;i+\frac12}w'_{y;i+\frac12}+\hat{s}_{y;i}w'_{z,y;i})+\sum_{j}(\hat{s}_{z,x;j+\frac12}w'_{x;j+\frac12}+\hat{s}_{x;j}w'_{z,x;j})}\\
    &\times (-1)^{\sum_j w'_{z,x;j} (w'_{x;{j-\frac{1}{2}}}+w'_{x;{j+\frac{1}{2}}})}\\
    &\times\frac{1}{2^{L_x+L_y-1}} \sum_{\mathbf{\hat{w}}'\in M_{v}}(-1)^{\sum_{i}(\hat{w}'_{z,y;i+\frac12}w'_{y;i+\frac12}+\hat{w}'_{y;i}w'_{z,y;i})+\sum_{j}(\hat{w}'_{z,x;j+\frac12}w'_{x;j+\frac12}+\hat{w}'_{x;j}w'_{z,x;j})} |\hat{\mathbf{w}}'\rangle.
\end{split}
\end{align}
Summing $w'_{z,x;j}$ and $w'_{z,y;i}$  produces two restrictions,  
\begin{equation}
    \hat{w}'_{x;j} = \hat{s}_{x;j} + w'_{x;{j-\frac{1}{2}}}+w'_{x;{j+\frac{1}{2}}},\quad \hat{w}'_{y;i} = \hat{s}_{y;i}.
\end{equation}
After relabelling $\hat{u}^x_{j+\frac{1}{2}} \equiv w'_{x;j+\frac{1}{2}}, \hat{u}^y_{i+\frac{1}{2}} \equiv w'_{y,i+\frac{1}{2}}$,~\eqref{SL2Z-Duality-Web-Double-T-calculation} becomes
\begin{equation}
    |\hat{\mathbf{s}}\rangle = \frac{1}{2^{L_x+L_y-1}} \sum_{\hat{u},\hat{w}'_z\in M_{\hat{u},\hat{w}'_z}} (-1)^{\sum_{i}(\hat{s}_{z,y;i+\frac12} +\hat{w}'_{z,y;i+\frac12}) \hat{u}^y_{i+\frac{1}{2}} + (\hat{s}_{z,x;j+\frac12} + \hat{w}'_{z,x;j+\frac12})\hat{u}^x_{j+\frac{1}{2}}} |\hat{\mathbf{w}}'\rangle
\end{equation}
which shows that the dual fermionic state $|\hat{\mathbf{s}}\rangle$ is the subsystem JW transformation of the dual state $\ket{\hat{\mathbf{w}}'}$ resembling \eqref{Foliated-Theory-JW-Transformation-State}. 

As a summary, begin with a bosonic state $|\mathbf{w}\rangle$ the JW transformation can be written as
\begin{equation}
    S^{-1}T^2_{+-}S=\text{JW}_x,\quad S^{-1} T^2_{-+}S=\text{JW}_y
\end{equation}
acting on the state $|\mathbf{w}\rangle$ and therefore the phases $\textrm{SSPT}_{+-}$ and $\textrm{SSPT}_{-+}$ are both fermionic subsystem SPT phases.\footnote{We thank Kantaro Ohmori and Yunqin Zheng for pointing out this to us.} With these identifications, we can generate other path in the duality web. For example, 
\begin{align}\label{eq:bosonicspt}
    \begin{split}
        (\text{JW}_y)^{-1}\text{JW}_x= S^{-1} (T^2_{-+})^{-1}S S^{-1}T^2_{+-}S
        = S^{-1} T^2_{-+}T^2_{+-}S
        = S^{-1} T^2_{--}S
    \end{split}
\end{align}
shows that the subsystem KW transformation of $\textrm{SSPT}_{--} |\mathbf{w}\rangle$ leads to the bosonic topological boundary state which is obtained by performing inverse subsystem JW transformation along the $y$ direction after a subsystem JW transformation along the $x$-direction. From~\eqref{eq:bosonicspt}, the phase $\textrm{SSPT}^2_{--}$ is a bosonic phase.

Based on the above analysis, we can obtain a duality web as shown in Fig. \ref{fig:duality-web}.

\section{$S$-defects in the subsystem SymTFT}\label{sec:defects}

In this section, we will construct the co-dimensional one symmetry defects generating the $SL(2,\mathbb{Z}_2)$ 0-form symmetry with a focus on $S$-defect. It was shown in~\cite{Fuchs:2012dt,Roumpedakis:2022aik,Kaidi:2022cpf} that in a $(d+1)$-dimensonal TFT, such kind of symmetry defects $\mathcal{D}$ extending along a co-dimension one hypersurface $M_{d}$ are built by condensing certain types of topological defects $\mathcal{L}$ along $M_{d}$.  If the topological defects $\mathcal{
L}$ generate a $q$-form symmetry inside $M_{d}$, the condensation defect $\mathcal{D}$ is equivalently understood as gauging the $q$-form symmetry inside $M_{d}$ which is referred to as $1$-gauging of the $q$-form symmetry. In the appendix~\ref{sec:ordinaryBF}, we give an example of the condensation defect generating the electric-magnetic $\mathbb{Z}_2$ symmetry in $(2+1)$d BF theory with level $N$. 

As discussed in the previous section, the proper $S$-transformation defined on the lattice satisfies $S^2 =\mathcal{T}$ where $ \mathcal{T}$ is the translation $(i,j)\rightarrow(i+1,j+1)$ on the lattice.  
We will construct the condensation defects in 2-foliated BF theory along $M_3$, a 3d manifold parallel to the boundary, by condensing line/strip operators on $M_3$. We will also discuss the twist defects by putting a``Dirichlet" boundary condition for the condensation defects. We will re-derive the subsystem non-invertible fusion rules by the fusion of twist defects.

\subsection{Conventions on operators and algebras}
For later convenience, we introduce $U_I$ and $\hat{U}_I$ as the collection of electric and magnetic line/strip operators respectively
\begin{equation}\label{Condensation-Defect-U}
    U_I = \left\{ \begin{array}{l}
        W(y_I,y_{I+1}) \quad I = 1,\cdots,L_y\\
        W(x_{I-L_y},x_{I-L_y+1}) \quad I = L_y+1,\cdots,L_y+L_x\\
        W_{z,x}(y_{I-L_x-L_y})\quad I = L_y+L_x+1,\cdots,2L_y+L_x\\
        W_{z,y}(x_{I-L_x-2L_y})\quad I = 2L_y + L_x + 1,\cdots, 2L_y + 2L_x
    \end{array} \right.
\end{equation}
\begin{equation}\label{Condensation-Defect-U-hat}
    \hat{U}_I = \left\{ \begin{array}{l}
        \hat{W}(y_{I-\frac{1}{2}},y_{I+\frac{1}{2}}) \quad I = 1,\cdots,L_y\\
        \hat{W}(x_{I-L_y-\frac{1}{2}},x_{I-L_y+\frac{1}{2}}) \quad I = L_y+1,\cdots,L_y+L_x\\
        \hat{W}_{z,x}(y_{I-L_x-L_y+\frac{1}{2}})\quad I = L_y+L_x+1,\cdots,2L_y+L_x\\
        \hat{W}_{z,y}(x_{I-L_x-2L_y+\frac{1}{2}})\quad I = 2L_y + L_x + 1,\cdots, 2L_y + 2L_x
    \end{array} \right.
\end{equation}
where $I=1,\cdots,2L_x+2L_y$. We will use the lattice size integer $n \equiv L_x + L_y$ for simplicity. In this convention, the quantum algebras \eqref{Foliated-Theory-Operators-Albetra-Discrete-1} and \eqref{Foliated-Theory-Operators-Albetra-Discrete-2} between electric and magnetic operators has a compact and symmetric form,
\begin{equation}\label{eq:quanalgcom}
   U_I \hat{U}_J  = - \Omega_{IJ} \hat{U}_J U_I,
\end{equation}
where $\Omega_{IJ}$ is a $2n \times 2n$ symmetric matrix,
\begin{equation}
    \Omega_{IJ} = \left(\begin{array}{cc}
        0 & I_{n\times n} \\
        I_{n\times n} & 0
    \end{array} \right).
\end{equation}

We will then formulate the general operators, the algebras between them and their actions on the boundary states. 
The general operator
\begin{equation}\label{Condensation-Defect-Operators}
    K[\alpha,\hat{\alpha}]:= \prod_{I=1}^{2n} U^{\alpha_I}_I \prod_{J=1}^{2n} \hat{U}^{\hat{\alpha}_J}_J 
\end{equation}
is parametrized by two $2n$-dimensional vectors with $\mathbb{Z}_2$-valued entries
\begin{align}
    \begin{split}
        &\alpha= (a,b):=(a_1,a_2,\cdots,a_n,b_1,b_2,\cdots,b_n),\\
        &\hat{\alpha}=(\hat{a},\hat{b}):= (\hat{a}_1,\hat{a}_2,\cdots,\hat{a}_n,\hat{b}_1,\hat{b}_2,\cdots,\hat{b}_n).
    \end{split}
\end{align}
From the quantum algebra~\eqref{eq:quanalgcom}, the general operators $K[\alpha,\hat{\alpha}],K[\alpha',\hat{\alpha}']$ have the following fusion rule
\begin{align}\label{Condensation-Defect-Algebra}
    K[\alpha,\hat{\alpha}] K[\alpha',\hat{\alpha}'] = (-1)^{-\hat{\alpha}\cdot \alpha'} K[\alpha+\alpha',\hat{\alpha}+\hat{\alpha}'],
\end{align}
together with the commutation algebra
\begin{equation}\label{Condensation-Defect-Commutation-Relation}
    K[\alpha,\hat{\alpha}] K[\alpha',\hat{\alpha}'] = (-1)^{\alpha \cdot \hat{\alpha}' - \hat{\alpha}\cdot \alpha'} K[\alpha',\hat{\alpha}'] K[\alpha,\hat{\alpha}].
\end{equation}
where the symmetric inner product between two $2n$-dimensional vectors is defined by
\begin{equation}\label{eq:sympro}
    \alpha \cdot \hat{\alpha}= \sum_{IJ} \Omega_{IJ} \alpha_I \hat{\alpha}_J.
\end{equation}

Here are some comments about the parameters $\alpha,\hat{\alpha}$. Due to the gauge redundancy~\eqref{Foliated-Theory-Operators-electric-Gauge-Redundancy-electric} and constraint~\eqref{Foliated-Theory-Operators-electric-Constraints-electric} among the electric operators, as well as their magnetic counterparts, these parameters have the identification
\begin{equation}\label{eq:vecred}
    a \sim a+1,\quad \hat{a} \sim \hat{a}+1
\end{equation}
of the $n$-dimensional vectors $a$ and $\hat{a}$ and the constraint
\begin{equation}\label{eq:veccon}
    \sum_{i=1}^n b_i = \sum_{i=1}^n \hat{b}_i = 0 
\end{equation}
among the $n$-dimensional vectors $b$ and $\hat{b}$. 
The inner product~\eqref{eq:sympro} is invariant under the gauge transformation~\eqref{eq:vecred} providing the constraints~\eqref{eq:veccon}.

On the boundary manifold $M_3$, we can identify the vector $\alpha$ with the holonomies
\begin{equation}\label{Condensation-Defect-a-and-w-relation}
    \alpha \equiv ( w_{z,x;j},w_{z,y;i},w_{x;j+\frac12},w_{y;i+\frac12}),
\end{equation}
Similarly, we can define $\hat{\alpha}$ as the dual parameter of $\alpha$, which is 
\begin{equation}
    \hat{\alpha} \equiv (\hat{w}_{z,x;j+\frac12}, \hat{w}_{z,y;i+\frac12},\hat{w}_{x;j},\hat{w}_{y;i}) .
\end{equation}
The lattice $S$-defect will map $|\alpha\rangle = |a,b\rangle$ to the dual state $|\hat{\alpha}\rangle = |\hat{a},\hat{b}\rangle$ with
\begin{equation}
    \hat{a}=a,\quad \hat{b} \equiv ( b_{L_y} , b_{1},\cdots,b_{L_y-1} , b_{L_x+L_y},b_{L_y+1},\cdots,b_{L_x+L_y-1}) \equiv b_T,
\end{equation}
which is equivalent to \eqref{SL2Z-Duality-S-Transformation-on-States}. In the rest of the paper, we will use $b_T$ to denote the shifted vector introduced above. On the other hand, the field theory $S$-defect maps $|\alpha\rangle = |a,b\rangle$ to the dual state $|\hat{\alpha}\rangle = |\hat{a},\hat{b}\rangle$ with $\hat{a}=a$ and $\hat{b}=b$.

From now on, we will use $\ket{\alpha}$ and $\ket{\hat{\alpha}}$ for the boundary state and its subsystem KW dual. 
We can rewrite the actions of electric and magnetic operators on topological boundary states \eqref{Foliated-Theory-Boundary-State-A-Eigenvaleus},\eqref{Foliated-Theory-Boundary-State-A-Shift},\eqref{Foliated-Theory-Boundary-State-A-hat-Eigenvaleus},\eqref{Foliated-Theory-Boundary-State-A-hat-shift} as
\begin{equation}\label{Condensation-Defect-Boundary-State-A}
    K[\gamma,0]|\alpha\rangle = (-1)^{\gamma\cdot \alpha}|\alpha\rangle,\quad K[0,\gamma]|\alpha\rangle = |\alpha+\gamma\rangle.
\end{equation}
\begin{equation}\label{Condensation-Defect-Boundary-State-A-hat}
    K[0,\hat{\gamma}] |\hat{\alpha}\rangle = (-1)^{\hat{\gamma}\cdot \hat{\alpha}}|\hat{\alpha}\rangle,\quad K[\hat{\gamma},0]|\hat{\alpha}\rangle = |\hat{\alpha}+\hat{\gamma}\rangle,
\end{equation}
together with the subsystem KW transformation \eqref{Foliated-Theory-Boundary-State-KW-Relation} between the boundary state $\ket{\alpha}$ and its dual $\ket{\hat{\alpha}}$ as 
\begin{equation}\label{Condensation-Defect-Boundary-State-KW-Relation}
    |\hat{\alpha}\rangle = \frac{1}{2^{L_x+L_y-1}} \sum_{\alpha\in M_v} (-1)^{\hat{\alpha} \cdot \alpha}|\alpha\rangle,
\end{equation}
where $M_{v}$ denote the set of $2n$-dimensional vectors satisfying the restrictions,
\begin{equation}
    M_{v} = \left\{ \alpha = (a,b) | a\sim a+1, \sum_{i=1}^n b_i=0 \right\}.
\end{equation}
With the orthogonality $\langle \beta|\alpha\rangle=\delta_{\alpha\beta}$, one has
\begin{equation}
    \langle \beta|\hat{\alpha}\rangle = \frac{1}{2^{L_x+L_y-1}}(-1)^{\hat{\alpha} \cdot \beta}.
\end{equation}
The inverse transformation is,
\begin{equation}
    |\alpha\rangle = \frac{1}{2^{L_x+L_y-1}} \sum_{\hat{\alpha}\in M_v} (-1)^{\hat{\alpha} \cdot \alpha}|\hat{\alpha}\rangle,
\end{equation}
and for consistency, we should have the orthogonality relation,
    \begin{equation}
        \frac{1}{2^{2(L_x+L_y-1)}} \sum_{\alpha \in M_v} (-1)^{\alpha \cdot \beta} = \delta_{\beta,0}.
    \end{equation}
This is not obviously true because $\alpha,\beta$ are not free and they should satisfy the restrictions given above. Before ending this section, let us check this relation explicitly.  Decompose $\alpha=(a,b),\beta=(c,d)$, one has,
    \begin{equation}
        \frac{1}{2^{2(L_x+L_y-1)}} \sum_{(a,b)\in M_v} (-1)^{a\cdot d } (-1)^{b\cdot c}.
    \end{equation}
In the first factor $(-1)^{a\cdot d}$ we have $a\sim a+1$ and $\sum d = 0$. Therefore we can relax the restriction $a\sim a+1$ and write,
    \begin{equation}
        \frac{1}{2^{L_x+L_y-1}} \sum_{a\sim a+1} (-1)^{a \cdot d } = \frac{1}{2^{L_x+L_y-1}} \times \frac{1}{2} \sum_a (-1)^{a\cdot d} = \delta_{d,0}.
    \end{equation}
In the second factor $(-1)^{b\cdot c}$ we have $\sum b = 0$ and $c \sim c+1$. We can also relax the restriction $\sum b=0$ by adding an Lagrangian multiplier $\lambda \in \mathbb{Z}_2$,
    \begin{align}
        &\frac{1}{2^{L_x+L_y-1}} \sum_{b|\sum b=0} (-1)^{b\cdot c}\nonumber \\
        =& \frac{1}{2^{L_x+L_y-1}} \times \frac{1}{2} \sum_{\lambda} (-1)^{\lambda(\sum b)} \sum_{b} (-1)^{b\cdot c} \nonumber\\
        =& \frac{1}{2^{L_x+L_y}} \sum_{\lambda} \sum_{b} (-1)^{b\cdot (c+\lambda)} = \delta_{c,0} + \delta_{c+1,0},
    \end{align}
where in the last line $\lambda$ is understood as the constant vector $(\lambda,\cdots,\lambda)$. This is also consistent with the fact $c \sim c+1$. Combined everything together we have proven the orthogonality relation.

\begin{figure}[htbp]
\centering
\begin{tikzpicture}
        \shade[top color=blue!40, bottom color=blue!10]  (0,0) -- (0,3) -- (1,2.5) -- (1,-0.5)-- (0,0);
        \draw[] (0,0) -- (0, 3);
        \draw[] (0,3) -- (1, 2.5);
        \draw[] (1,2.5) -- (1,-0.5);
        \draw[] (1,-0.5) -- (0,0);
        \draw[thick,red] (-1+0.75-1,1.5) -- (0+0.75-1,1);
        \node[above] at (-1,1.5) {$U_I$};
        \node[] at (2,1.25) {$=$};
        \node[above] at (-1+1+4.5,1.5) {$\hat{U}_I$};
        \draw[thick,blue] (-1+0.75+4.5,1.5) -- (0+0.75+4.5,1);
        \shade[top color=blue!40, bottom color=blue!10]  (0+4-1,0) -- (0+4-1,3) -- (1+4-1,2.5) -- (1+4-1,-0.5)-- (0+4-1,0);
        \draw[] (0+4-1,0) -- (0+4-1, 3);
        \draw[] (0+4-1,3) -- (1+4-1, 2.5);
        \draw[] (1+4-1,2.5) -- (1+4-1,-0.5);
        \draw[] (1+4-1,-0.5) -- (0+4-1,0);
        \node[above] at (2,3.5) {$S$-defect};
    \end{tikzpicture}
    \caption{Action of $S$-defect on line/strip operators}
    \label{fig:SandT}
\end{figure}
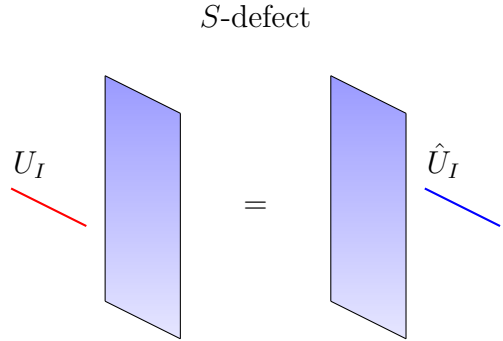

\subsection{\texorpdfstring{$S$}\ -defect}
In the last section, we discuss two kinds of $S$-transformation, the lattice $S$-transformation $S$ and the field theory $S$-transformation $\widetilde{S}$. We will discuss both lattice $S$-defect and field theory $S$-defect in the following.

As shown in Fig.~\ref{fig:SandT}, the lattice $S$-defect (or field theory $S$-defect) maps the electric operator $U_I$ to magnetic operator $\hat{U}_I$ and vice versa.
Fusing to the boundary, it maps the boundary state to its KW dual. As we mentioned before, the lattice $S$-defect will map $|\alpha\rangle \equiv |a,b\rangle$ to the dual state $|\hat{\alpha}\rangle = |\hat{a},\hat{b}\rangle$ with $\hat{a}=a$ and $\hat{b}=b_T$ where $b_T$ is related to $b$ as,
\begin{equation}\label{eq:bT}
    b_T \equiv ( b_{L_y} , b_{1},\cdots,b_{L_y-1} , b_{L_x+L_y},b_{L_y+1},\cdots,b_{L_x+L_y-1}).
\end{equation}
We will construct the condensation defect on $M_3$ in the bulk using the bra-ket trick,
\begin{equation}\label{eq:sdef}
        S = 
        \sum_{(\hat{a},\hat{b})=(a,b_T) \in M_{v}} |\hat{a},\hat{b}\rangle \langle a,b|,
\end{equation}
which manifests its action on the state. We use $(\hat{a},\hat{b})=(a,b_T)$ to emphasize that the values of the holonomies $\hat{b}$ of the dual states are related to the holonomies $b$ of the original state via a shift.
In the following, we will omit the $M_v$ notation and assume every vector $\alpha=(a,b)$ or $\hat{\alpha}=(\hat{a},\hat{b})$ should satisfy the constraints automatically. Notice that $S^{\dagger} S = 1$ by construction.

Before moving on, it is convenient to decompose the $S$-operators as,
    \begin{equation}
        S = \widetilde{\mathcal{T}} \widetilde{S},
    \end{equation}
where $\widetilde{\mathcal{T}}$ and $\widetilde{S}$ are defined as,
    \begin{equation}
        \widetilde{\mathcal{T}} = \sum_{(\hat{a},\hat{b})} | \hat{a} , \hat{b}_T \rangle \langle \hat{a} , \hat{b} | ,\quad \widetilde{S} = 
        \sum_{(a,b)=(\hat{a},\hat{b})} |\hat{a},\hat{b}\rangle \langle a,b|,
    \end{equation}
where $\hat{b}_T$ is defined as the hat version of \eqref{eq:bT}. One can check
    \begin{align}
        \widetilde{T} \widetilde{S} &= \sum_{(\hat{a}',\hat{b}')} \sum_{(a,b)=(\hat{a},\hat{b})} | \hat{a}' , \hat{b}'_T \rangle \langle \hat{a}' , \hat{b}' | \hat{a},\hat{b}\rangle \langle a,b| \nonumber\\
        &=\sum_{(\hat{a}',\hat{b}')} \sum_{(a,b)=(\hat{a},\hat{b})} \delta_{\hat{a}',\hat{a}} \delta_{\hat{b}',\hat{b}}| \hat{a}' , \hat{b}'_T \rangle \langle a,b|\nonumber\\
        &= \sum_{(\hat{a},\hat{b})=(a,b)} | \hat{a} , \hat{b}_T \rangle \langle a,b| = S.
    \end{align}
Here $\widetilde{S}$ is the field theory $S$-defect which implements the field theory $S$-transformation defined in \eqref{SL2Z-Duality-S-Transformation-on-States-Fake} in the previous section and it satisfies
    \begin{equation}
        U_I \widetilde{S} = \widetilde{S} \hat{U}_I,\quad \hat{U}_I \widetilde{S} = \widetilde{S} U_I,
    \end{equation}
using \eqref{Condensation-Defect-Algebra} and \eqref{Condensation-Defect-Commutation-Relation}.  Moreover, one can check $\widetilde{S}^2$ is identity 
\begin{align}
    \widetilde{S}^2 &= 
    \sum_{(a,b)=(\hat{a},\hat{b}),(c,d)=(\hat{c},\hat{d})} |\hat{a},\hat{b}\rangle \langle a,b|\hat{c},\hat{d}\rangle \langle c,d|\nonumber\\
    &= 
    \frac{1}{2^{L_x+L_y-1}}\sum_{(a,b)=(\hat{a},\hat{b}),(c,d)=(\hat{c},\hat{d})} (-1)^{\hat{c}\cdot b + \hat{d}\cdot a} |\hat{a},\hat{b}\rangle \langle c,d|\nonumber\\
    &= 
    \sum_{(c,d)} |c,d\rangle \langle c,d|\equiv\text{I},
\end{align}
where we use the subsystem KW transformation \eqref{Condensation-Defect-Boundary-State-KW-Relation} in the second line. On the other hand, the operator $\widetilde{\mathcal{T}}$ will shift the holonomies 
such that the combination $\widetilde{\mathcal{T}} \widetilde{S}$ implements the transformation \eqref{SL2Z-Duality-S-Transformation-on-States}. To be concrete, let's evaluate and check $(\widetilde{\mathcal{T}} \widetilde{S})^2 = \mathcal{T}$ where $\mathcal{T}$ is the translation. First, we have
\begin{equation}
    \sum_{(\hat{a},\hat{b})} | \hat{a} , \hat{b}_T \rangle \langle \hat{a} , \hat{b} | \widetilde{S} =  \sum_{(a,b)=(\hat{a},\hat{b})} | \hat{a},\hat{b}_T \rangle \langle a , b |.
\end{equation}
Moreover we have the following identities
\begin{equation}
    \sum_{(\hat{a},\hat{b})} | \hat{a} , \hat{b}_T \rangle \langle \hat{a} , \hat{b} | = \sum_{(a,b)} |a_T,b\rangle \langle a,b|,\quad \sum_{(\hat{a},\hat{b})} | \hat{a}_T , \hat{b} \rangle \langle \hat{a} , \hat{b} | = \sum_{(a,b)} |a,b_T\rangle \langle a,b|,
\end{equation}
and also
\begin{equation}
    \sum_{(\hat{a},\hat{b})} | \hat{a}_T , \hat{b}_T \rangle \langle \hat{a} , \hat{b} | = \sum_{(a,b)} | a_T , b_T \rangle \langle a , b |.
\end{equation}
It is easy to check them by acting both sides on a state $|a,b\rangle$. For example, for the first identity, acting the LHS on $|a,b\rangle$ we get
    \begin{align}
        &\sum_{\hat{a},\hat{b}}  | \hat{a} , \hat{b}_T \rangle \langle \hat{a} , \hat{b} | a,b\rangle \nonumber\\
        =&\sum_{\hat{a},\hat{b}}  | \hat{a} , \hat{b}_T \rangle \frac{1}{2^{L_x+L_y-1}} (-1)^{\hat{a} \cdot b + \hat{b}\cdot a}\nonumber\\
        =&\sum_{\hat{a},\hat{b}}  | \hat{a} , \hat{b}_T \rangle \frac{1}{2^{L_x+L_y-1}} (-1)^{\hat{a} \cdot b + \hat{b}_T \cdot a_T} = |a_T,b\rangle,
    \end{align}
where we use the subsystem KW transformation and the fact $\hat{b}_T \cdot a_T = \hat{b} \cdot a$. It exactly matches the result obtained by acting the RHS on $|a,b\rangle$.  Using those identities we can examine
\begin{equation}
   \sum_{(\hat{a},\hat{b})} \widetilde{S}| \hat{a}_T , \hat{b} \rangle \langle \hat{a} , \hat{b} | = \sum_{(a,b)} \widetilde{S}| a , b_T \rangle \langle a , b | = \sum_{(\hat{a},\hat{b})=(a,b)} | \hat{a} , \hat{b}_T \rangle \langle a , b | .
\end{equation}
Therefore we have the following commutation relation
\begin{equation}
    \sum_{(\hat{a},\hat{b})} | \hat{a} , \hat{b}_T \rangle \langle \hat{a} , \hat{b} | \widetilde{S} =  \sum_{(\hat{a},\hat{b})} \widetilde{S}| \hat{a}_T , \hat{b} \rangle \langle \hat{a} , \hat{b} |
\end{equation}
and we have
\begin{align}
    \left( \sum_{\hat{a},\hat{b}} | \hat{a} , \hat{b}_T \rangle \langle \hat{a} , \hat{b} | \widetilde{S} \right)^2 =&  \sum_{\hat{a}',\hat{b}'}  \sum_{\hat{a},\hat{b}}| \hat{a}' , \hat{b}'_T \rangle \langle \hat{a}' , \hat{b}' | \widetilde{S}| \hat{a} , \hat{b}_T \rangle \langle \hat{a} , \hat{b} | \widetilde{S}\nonumber\\
    =&\sum_{\hat{a}',\hat{b}'}  \sum_{\hat{a},\hat{b}}| \hat{a}' , \hat{b}'_T \rangle \langle \hat{a}' , \hat{b}' | \widetilde{S}^2 | \hat{a}_T , \hat{b} \rangle \langle \hat{a} , \hat{b} |\nonumber\\
    =&\sum_{(\hat{a},\hat{b})} | \hat{a}_T , \hat{b}_T \rangle \langle \hat{a} , \hat{b} | = \sum_{(a,b)} | a_T , b_T \rangle \langle a , b | \equiv \mathcal{T}
\end{align}
where $\mathcal{T}$ is an operator which shifts $(a,b)$ to $(a_T,b_T)$. Remember that $a=(w_{z,x;j},w_{z,y;i})$ and $b=(w_{x;j+\frac{1}{2}},w_{y;i+\frac{1}{2}})$ so that $\mathcal{T}$ shifts the holonomies as,
    \begin{equation}
        w_{z,x;j}\rightarrow w_{z,x;j-1},\quad w_{z,y;i}\rightarrow w_{z,y;i-1},\quad w_{x;j+\frac{1}{2}} \rightarrow w_{x;j-\frac{1}{2}},\quad w_{y;i+\frac{1}{2}}\rightarrow w_{y;i-\frac{1}{2}}.
    \end{equation}
which indicates $\mathcal{T}$ is the translation operator.

To write the lattice $S$-defect as a condensation of the line/strip operators, we parameterize it as,
    \begin{equation}
        S = \sum_{\alpha,\hat{\alpha}\in M_{v}} \lambda_{\alpha,\hat{\alpha}} K[\alpha,\hat{\alpha}].
    \end{equation}
We will use the property that the operators $K[\alpha,\hat{\alpha}]$ with different $\alpha=(a,b)$ and $\hat{\alpha}=(\hat{a},\hat{b})$ are orthogonal to each other in the trace
\begin{equation}
        \textrm{Tr} \left(K[\alpha,\hat{\alpha}]^{\dagger} K[\alpha',\hat{\alpha}']\right):=\sum_{\gamma}\bra{\gamma}K[\alpha,\hat{\alpha}]^{\dagger} K[\alpha',\hat{\alpha}']\ket{\gamma} = 2^{2(L_x+L_y-1)} \delta_{\alpha,\alpha'} \delta_{\hat{\alpha},\hat{\alpha}'}
\end{equation}
to project out the coefficients $\lambda_{\alpha,\hat{\alpha}}$,
\begin{align}
        \lambda_{\alpha,\hat{\alpha}} =& \frac{1}{2^{2(L_x+L_y-1)}} \textrm{Tr} \left( K[\alpha,\hat{\alpha}]^{\dagger} \sum_{(c,d)=(\hat{c},\hat{d})}  
        (|\hat{c},\hat{d}_T\rangle) \langle c,d| \right)\nonumber\\
        =&
       \frac{1}{2^{2(L_x+L_y-1)}} \sum_{(c,d)=(\hat{c},\hat{d})} \langle c,d| K[0,(\hat{a},\hat{b})] K[(a,b),0] |\hat{c},\hat{d}_T\rangle\nonumber\\
        =&
       \frac{1}{2^{2(L_x+L_y-1)}}\sum_{(c,d)=(\hat{c},\hat{d})} (-1)^{\hat{a}\cdot (\hat{d}_T+b) + \hat{b}\cdot (\hat{c}+a)} \langle c,d |\hat{c}+a,\hat{d}_T+b \rangle \nonumber\\
        =&
       \frac{1}{2^{3(L_x+L_y-1)}}\sum_{(c,d)} (-1)^{(\hat{a}+c)\cdot (b+d_T) + (\hat{b}+d)\cdot (a+c)} \nonumber\\
        =&
       \frac{1}{2^{3(L_x+L_y-1)}}\sum_{(c,d)} (-1)^{((c,d) + \hat{\alpha}) \cdot ((c,d_T) + \alpha)},
\end{align}
and the condensation defect can be written in the compact form,
\begin{align}
    S &= \frac{1}{2^{3(L_x+L_y-1)}}\sum_{(c,d),\alpha,\hat{\alpha}} (-1)^{((c,d) + \hat{\alpha}) \cdot ((c,d_T) + \alpha)} K[\alpha,\hat{\alpha}] \nonumber \\
    &= \frac{1}{2^{3(L_x+L_y-1)}}\sum_{(c,d),(a,b),(\hat{a},\hat{b})} (-1)^{(c+\hat{a},d+\hat{b}) \cdot (c+a,d_T+b)} K[(a,b),(\hat{a},\hat{b})] \nonumber \\
    &= \frac{1}{2^{3(L_x+L_y-1)}}\sum_{(c,d),(a,b),(\hat{a},\hat{b})} (-1)^{(c+\hat{a},\hat{b}) \cdot (c+a,b)} K[(a,b+d_T),(\hat{a},\hat{b}+d)] \nonumber \\
    &= \frac{1}{2^{2(L_x+L_y-1)}} \sum_{a,\hat{a},b,d} (-1)^{(a+\hat{a})\cdot b } K[(a,b+d_T),(\hat{a},b+d)].
\end{align}
It is also illuminating to write down the condensation defect for $\widetilde{S}$, which is the field theory $S$-defect, and $\widetilde{\mathcal{T}}$. Using the same method, one can obtain
\begin{align}
    \widetilde{S} = 
       \frac{1}{2^{3(L_x+L_y-1)}} \sum_{\gamma , \alpha , \hat{\alpha}} (-1)^{\gamma\cdot(\alpha+\hat{\alpha})} K[\alpha,\hat{\alpha}] =
       \frac{1}{2^{L_x+L_y-1}}
       \sum_{\alpha} K[\alpha,\alpha].
\end{align}
Expanding $K[\alpha,\alpha]$ using $U_I,\hat{U}_J$ gives
\begin{equation}\label{Condensation-Defect-S-tilde}
    \widetilde{S} = \frac{1}{2^{L_x+L_y-1}}
    \sum_{\alpha} \prod_{I=1}^{2n} U^{\alpha_I}_I \prod_{J=1}^{2n} \hat{U}^{\alpha_J}_J =
    \frac{1}{2^{L_x+L_y-1}}
    \sum_{\alpha }(-1)^{\frac{\alpha \cdot \alpha}{2}} \prod_{I=1}^{2n}  (U_I \hat{U}_I)^{\alpha_I},
\end{equation}
which is a condensation of all possible insertions of operators $U_I\hat{U}_I$ built from line and strip operators. The orientation reversal of the field theory $S$-defect is itself $\widetilde{S}^{\dagger} =\widetilde{S}$ and it is also unitary: $\widetilde{S}^{\dagger} \widetilde{S}= 1$. Similarly, one has
\begin{align}
    \widetilde{\mathcal{T}} = \frac{1}{2^{L_x+L_y-1}}  \sum_{\hat{a},b} (-1)^{\hat{a} \cdot b } K[(0,b_T + b),(\hat{a},0)].
\end{align}
One can also check the fusion between $\widetilde{\mathcal{T}}$ and $\widetilde{S}$ straightforwardly,
    \begin{align}
        \widetilde{\mathcal{T}} \times \widetilde{S} &= \frac{1}{2^{2(L_x+L_y-1)}} \sum_{a,\hat{a},d,b}(-1)^{\hat{a}\cdot d} K[(0,d_T + d),(\hat{a},0)] K[(a,b),(a,b)]\nonumber\\
        &= \frac{1}{2^{2(L_x+L_y-1)}} \sum_{a,\hat{a},d,b} (-1)^{\hat{a}\cdot (d+b)} K[(a,b + d_T + d),(a+\hat{a},b)]\nonumber\\
        &= \frac{1}{2^{2(L_x+L_y-1)}} \sum_{a,\hat{a},d,b} (-1)^{(\hat{a}+a)\cdot b} K[(a,b+d_T),(\hat{a},b+d)] = S,
    \end{align}
which reproduces the lattice $S$-defect.

\subsection{Twist \texorpdfstring{$S$}\ -defect}
In this subsection, we will consider twist $S$-defect on a 3d manifold $M_3$ with a boundary $\partial M_3 = M_2$, where the boundary can be $x$-$y$ plane, $x$-$z$ plane or $y$-$x$ plane. In the last section, we 
consider both lattice $S$-defect $S$ and field theory $S$-defect $\widetilde{S}$, and they are related by $S = \widetilde{\mathcal{T}} \times \widetilde{S}$. Here $\widetilde{S}$ satisfies $\widetilde{S}^2=1$ and $\widetilde{\mathcal{T}}$ shifts the holonomies and makes $S^2 = \mathcal{T}$ a translation along $x$-$y$ directions on the lattice.

We will first discuss the twist defects for the field theory $S$-defect $\widetilde{S}$ and then move to the twist defect for the lattice $S$-defect $S$. Also, we will mainly focus on the case where the boundary $M_2$ is the $x$-$y$ plane. Since the lattice $S$-defect involves a translation along $x$-$y$ plane, the discussions of the corresponding twist defects located at a fixed $x$ or $y$ are subtle and we will not consider them. Nevertheless, one can still consider the cases where $M_2$ is $x$-$z$ plane or $y$-$z$ plane for the field theory $S$-defect in a similar way.

From \eqref{Condensation-Defect-S-tilde} we see the field theory $S$-defect is a condensation of the operator $U_I \hat{U}_I$ where $U_I,\hat{U}_I$ are related to line/strip operators according to \eqref{Condensation-Defect-U} and \eqref{Condensation-Defect-U-hat}. We impose the Dirichlet boundary condition for the defects $U_I \hat{U}_I$ condensing along $M_3$. The Dirichlet boundary condition is defined as follows. The operators $\{U_I \hat{U}_I\}$ generate a subsystem $\mathbb{Z}_2$ symmetry along $M_3$ and we denote the corresponding gauge fields as $(A'^z,A'^{xy})$. For $x$-$y$ plane we require the $x$-$y$ component $A'^{xy}$ to vanish at the boundary. The Dirichlet boundary is topological along the normal direction given these boundary conditions and see Appendix~\ref{sec:deri} for a detailed discussion about this. 

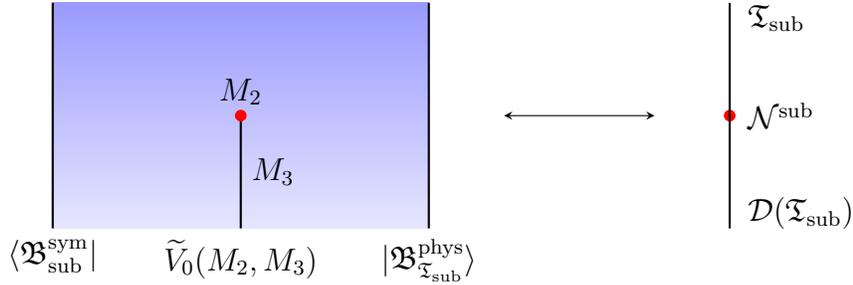
\begin{figure}[htbp]
\centering
\begin{tikzpicture}
        \shade[top color=blue!40, bottom color=blue!10]  (0-1,0) -- (0-1,3) -- (3+1,3) -- (3+1,0)-- (0-1,0);
        \draw[thick] (0-1,0) -- (0-1, 3);
        \draw[thick] (1.5,0) -- (1.5, 1.5);
        \filldraw [red] (1.5,1.5) circle (2pt);
        \node[above] at (1.5,1.5){$M_2$};
        \node[right] at (1.5,0.75){$M_3$};
        \draw[thick] (3+1,0) -- (3+1, 3);
        \node[below] at (1.5,0){$\widetilde{V}_{0}(M_2,M_3)$};
        \node[below] at (0-1,0) {$\bra{\mathfrak{B}^{\textrm{sym}}_{\text{sub}}}$};
        \node[below] at (3+1,0) {$\ket{\mathfrak{B}^{\textrm{phys}}_{\mathfrak{T}_{\text{sub}}}}$};
        \draw [stealth-stealth](4.5+0.5,1.5) -- (6.5+0.5,1.5);
        \filldraw [red] (8,1.5) circle (2pt);
        \draw[thick] (8,0) -- (8, 3);
        \node[right] at (8.1,0.2) {$\mathcal{D}(\mathfrak{T}_{\text{sub}})$};
        \node[right] at (8.1,2.8) {$\mathfrak{T}_{\text{sub}}$};
        \node[right] at (8.1,1.5) {$\mathcal{N}^{\text{sub}}$};
    \end{tikzpicture}
    \caption{Twist $S$-defect $\widetilde{V}_0(M_2,M_3)$ on $M_3$ with a boundary $M_2$. After shinking the slab, the twist defect will create a duality defect as an interface between the original boundary theory $\mathfrak{T}_{\mathcal{S}}$ and the theory $\mathcal{D}(\mathfrak{T}_{\text{sub}})$ after gauging subsystem $\mathbb Z_2$ symmetry.}
    \label{fig:twist}
\end{figure}

We will denote the twist defect as $\widetilde{V}_{0}(M_2,M_3)$ with $M_2$ the $x$-$y$ plane, for example at $z=0$ . As shown in Fig.~\ref{fig:twist}, after shrinking the slab, the twist defect will implement a half-space gauge and create a subsystem KW duality defect $\mathcal{N}^{\text{sub}}$. The $z$-direction is topological and the strip operators can move along the $z$-direction and fuse with the twist defect. Since $M_2$ satisfies the Dirichlet boundary condition $A'^{xy}=0$, the strip operators $K[(a,0),(a,0)]$ can be absorbed by the twist defect,
    \begin{equation}
        K[(a,0),(a,0)] \times \widetilde{V}_{0}[M_2,M_3] = \widetilde{V}_{0}[M_2,M_3],
    \end{equation}
where we split the $2n$-dimensional vector $\alpha,\hat{\alpha}$ into a pair of $n$-dimensional vector
\begin{equation}
    \alpha=(a,b),\quad \hat{\alpha}=(\hat{a},\hat{b}),
\end{equation}
such that $a,\hat{a}$ label the strip operators and $b,\hat{b}$ label the line operators. On the other hand, since $K[(a,0),(0,0)]$ and $K[(0,0),(a,0)]$ do not belong to the condensate, fusing them with $\widetilde{V}_0$ produces new twist defects,
    \begin{equation}
        \widetilde{V}_{a}[M_2,M_3] := K[(a,0),(0,0)] \times \widetilde{V}_{0}[M_2,M_3] = K[(0,0),(a,0)] \times \widetilde{V}_{0}[M_2,M_3].
    \end{equation}
In general, the fusion rule between the strip operators $K[(a,0),(\hat{a},0)]$ with $\widetilde{V}_{a'}[M_2,M_3]$ is,
    \begin{equation}
        K[(a,0),(\hat{a},0)] \times \widetilde{V}_{a'}[M_2,M_3] = \widetilde{V}_{a'+a+\hat{a}}[M_2,M_3].
    \end{equation}

We can also discuss the fusion between twist defects and it is sufficient to discuss the fusion between $\widetilde{V}_0$ and $\widetilde{V}^{\dagger}_0$.
Let's put another twist defect $\widetilde{V}_{0}[M_{2|\epsilon},M_3]$ at $z=\epsilon$ and consider the fusion between $\widetilde{V}_{0}[M_{2|0},M_3] \times \widetilde{V}_{0}[M_{2|\epsilon},M_3]$ with $\epsilon\rightarrow 0$. Here we use $M_{2|0}$ ($M_{2|\epsilon}$) to emphasize $M_2$ is located at $z=0$ ($z=\epsilon$). Since the condensation defects can also be understood as gauging the $(2+1)$d subsystem symmetry on $M_3$, we can derive the fusion rule in a similar way following the discussion in Appendix~\ref{sec:deri} and get,
    \begin{align}\label{eq:fuseoftwist}
        &\lim_{\epsilon \rightarrow 0} \widetilde{V}^{\dagger}_{0}(M_2|_{z=0}, M_3^{z\geq 0}) \times \widetilde{V}_0(M_2|_{z=\epsilon},M_3^{z\geq \epsilon})\nonumber\\
        =& \frac{1}{2}\prod_{i=1}^{L_x} (1+W(y_i,y_{i+1})\hat{W}(y_{i-\frac{1}{2}},y_{i+\frac{1}{2}})) \prod_{j=1}^{L_y} (1+W(x_j,x_{j+1})\hat{W}(x_{j-\frac{1}{2}},x_{j+\frac{1}{2}})).
    \end{align}
where  $\widetilde{V}^{\dagger}_0 = \chi[M_3^{z \geq \epsilon},\mathbb{Z}_2]  \widetilde{V}_{0}$ with $\chi[M_3^{z \geq \epsilon},\mathbb{Z}_2]$ the Euler factor introduced in \eqref{Appendix-Euler-factor}. We have a condensation of strip operators that are mobile along $z$-direction. 
If we put \eqref{eq:fuseoftwist} on the top of the topological boundary $\ket{0}$ at $\tau=1$ where $A=0$ at the boundary, then electric strip operators $W(x_j,x_{j+1}),W(y_i,y_{i+1})$ are absorbed into the boundary and we have,
\begin{equation}\label{eq:fusetwist}
        \widetilde{V}_{0}[M_2,M_3]^{\dagger} \times \widetilde{V}_{0}[M_2,M_3]|_{\tau=1} = \frac12\prod_{i=1}^{L_x} (1+U^y_i) \prod_{j=1}^{L_y} (1+U^x_j),
\end{equation}
where $U^y_i=\hat{W}(y_{i-\frac{1}{2}},y_{i+\frac{1}{2}})$ and $U^x_j=\hat{W}(x_{j-\frac{1}{2}},x_{j+\frac{1}{2}})$ are generators for subsystem $\mathbb{Z}_2$  symmetry at the boundary. The fusion of twist defects \eqref{eq:fusetwist} recovers the fusion rules of the subsystem KW operators in the untwisted sector\cite{Cao:2023doz}\footnote{For twisted sectors, we need to consider more general Dirichlet boundaries of the condensation defects such that corresponding gauge field $A'^{xy}$ is a non-zero fixed value at the boundary. 
}. On the other hand, if we put it on the top of the dual boundary $|\hat{0}\rangle$ where $\hat{A}=0$, then magnetic strip operators $\hat{W}$ are absorbed instead and electric strip operators $W$ serve as the symmetry generators. \eqref{eq:fusetwist} becomes the fusion rule in the untwisted sector of the dual theory.

We then move to the twist defect $V_0$ for the lattice $S$-defect. Put two twist defect $V_{0}[M_{2|0},M_3]$ and $V_{0}[M_{2|\epsilon},M_3]$ at $z=0$ and $z=\epsilon$, the fusion rule is similarly obtained by sending $\epsilon\rightarrow 0$. There are two differences here. First, since $S^2 = \mathcal{T}$ is the translation, the spatial lattice at $z> \epsilon$ is related to that at $z<0$ by the translation $\mathcal{T}$. Second, we need to impose proper Dirichlet boundary condition at the boundary $M_2=\partial M_3$. Recall that the condensation defect for $S$ is,
\begin{align}
    S = \frac{1}{2^{2(L_x+L_y-1)}} \sum_{a,\hat{a},b,d} (-1)^{(a+\hat{a})\cdot b } K[(a,b+d_T),(\hat{a},b+d)].
\end{align}
We will assume the Dirichlet boundary condition is defined such that only the strip operators which are mobile along $z$-direction survive in the limit $\epsilon\rightarrow 0$, then one expects the fusion is,
    \begin{equation}
        \frac{1}{2^{2(L_x+L_y-1)}} \sum_{a,\hat{a}} K[(a,0),(0,0)] K[(0,0),(\hat{a},0)].
    \end{equation}
If we put it on the top of the topological boundary $|0\rangle$ or $|\hat{0}\rangle$ at $\tau=1$, one of the two $K$-operators will be absorbed and we get the same result as before. Combined with the translation $\mathcal{T}$, we have the similar fusion rule,
    \begin{equation}\label{eq:fuse}
        V_{0}[M_2,M_3] \times V_{0}[M_2,M_3]|_{\tau=1} = \frac12\prod_{i=1}^{L_x} (1+U^y_i) \prod_{j=1}^{L_y} (1+U^x_j) \times \mathcal{T}.
\end{equation}
One can also consider the fusion between $V^{\dagger}$ and $V$,
\begin{equation}
        V_{0}[M_2,M_3]^{\dagger} \times V_{0}[M_2,M_3]|_{\tau=1} = \frac12\prod_{i=1}^{L_x} (1+U^y_i) \prod_{j=1}^{L_y} (1+U^x_j).
\end{equation}
where we do not have the translation operator $\mathcal{T}$ on the RHS since $S^{\dagger} S =1$.

\section{Conclusion and discussion}\label{sec:con}

In this paper, we initiate the study of the subsystem symmetry and the associated dualities from a bulk SymTFT point of view. To demonstrate this idea, we study the example of 2-foliated BF theory with level $N$ in $(3+1)$d as the subsystem SymTFT of the subsystem $\mathbb Z_N$ symmetry in $(2+1)$d. We analyze the topological boundaries and construct condensation defects of this specific model with $N=2$.  We interpret the duality transformations of the boundary theory, such as subsystem KW and JW transformation, as the change of topological boundaries which is further implemented by fusing condensation defects of the subsystem $SL(2,\mathbb Z_2)$ symmetry of the bulk subsystem SymTFT on the boundary. On the lattice, the subsystem $SL(2,\mathbb Z_2)$ symmetry has a richer structure than in the field theory. The subsystem $T$ transformation will stack a subsystem SPT phase whose bosonic or fermionic feature depends on the regularization of the lattice. We will leave the detailed study and classification of subsystem SPT phases in the future work.  From the subsystem $SL(2,\mathbb Z_2)$ symmetry, we find new dualities among bosonic and fermionic models with subsystem $\mathbb Z_2$ symmetry. We summarize the duality web in Fig.~\ref{fig:duality-web}.

There are many interesting follow-up directions.     First, it is natural to extend the study of subsystem SymTFT to other models. For the subsystem $\mathbb Z_N$ symmetry in $(2+1)$d, the subsystem SymTFT is expected to have more diverse topological boundaries that can support subsystem parafermionic structures. Furthermore, we can study models with subsystem symmetry in higher dimensions, for example, the X-cube model~\cite{Vijay:2016phm}, where there are fracton excitations. The $\mathbb Z_N$ X-cube model is a 3-foliated theory in $(3+1)$d and the corresponding subsystem SymTFT should be the 3-foliated BF theory with level $N$ in $(4+1)$d
\begin{equation}\label{eq:3foliated}
    S_{3\textrm{-foliated}} = \frac{N}{2\pi}\int b \wedge dc + \sum_{k=1,2,3} d B^k \wedge C^k \wedge d x^k + \sum_{k=1,2,3} b \wedge C^k \wedge d x^k .
 \end{equation}
where the first term is bulk BF term with 3-form gauge field $b$ and 1-form gauge field $c$, the second term is the foliated BF term with 2-form gauge field $B^k$ and 1-form gauge field $C^k$ and the third term is the interaction term. It is interesting to classify the topological boundaries and topological operators of this subsystem SymTFT and explore the duality web of the X-cube model.

Finally, subsystem SymTFT provides a bulk-boundary point of view to study subsystem symmetry. Recently, there are other efforts to study fracton models from bulk-boundary correspondence~\cite{Liu:2022cxc,Fontana:2022zku,Luo:2022mrj,Hsin:2023ooo}. Subsystem SymTFT also provides hints to study fracton statistics\cite{Pai:2019fqg,Song:2023rml}. The quantum algebras~\eqref{Foliated-Theory-Operators-Albetra-1} and~\eqref{Foliated-Theory-Operators-Albetra-2} resemble the braiding statistics in $(2+1)$d $\mathbb Z_N$ gauge theory. Besides, one more topological direction in the bulk will give fracton (or excitations with other restricted mobility) an extra direction to move, which might lead to interesting braiding structures. We will leave these interesting questions for future study.

\section*{Acknowledgements}
We thank Linhao Li, Kantaro Ohmori, Masahito Yamazaki, Shutaro Shimamura and Yunqin Zheng for the discussion when this project is in the early stage.  W.C. is supported by the Global Science Graduate Course (GSGC) program of the University of Tokyo and the JSPS KAKENHI grant numbers JP19H05810, JP20H01896, 20H05860, JP22J21553 and JP22KJ1072. W.C. also acknowledges th USTEP exchange program of the University of Tokyo and the FUTI scholarships for mid-to long-term studies from Friends of UTokyo, Inc. Q.J. is supported by Korea Institute for Advanced Study (KIAS)
Grant PG080802.

\appendix
\section{A review on ordinary BF theory as SymTFT}\label{sec:ordinaryBF}
To illustrate the basic idea of SymTFT, we consider a $(1+1)$d theory $\mathfrak{T}_{\mathbb{Z}_N}$ with $\mathbb{Z}_N$ symmetry. The corresponding SymTFT $\mathfrak{Z}(\mathbb{Z}_N)$ is the $(2+1)$d BF theory with level $N$,
    \begin{equation}
        S_{BF} = \frac{N}{2\pi} \int \hat{A} \wedge d A,
        \end{equation}
where $\hat{A},A$ are 1-form gauge fields. It is a $\mathbb{Z}_N$ gauge theory and is the low energy description of the toric code for $N=2$ in the condensed matter literature~\cite{Kitaev:2005hzj}. 
Fix a gauge $A_0=\hat{A}_0=0$, the canonical quantization gives,
    \begin{equation}
        \left[A_i(x,y) , \hat{A}_j(x',y') \right] =  \frac{2\pi i}{N} \epsilon_{ij} \delta(x-x',y-y').
    \end{equation}
$A$ and $\hat{A}$ are conjugated with each other like position and momentum. 
For simplicity, we place the BF theory on a spatial torus $T^2$, the physical operators are Wilson loops defined as,
    \begin{equation}
        W[\Gamma] = \exp \left(i \oint_{\Gamma} A \right),\quad \hat{W}[\Gamma] = \exp \left(i \oint_{\Gamma} \hat{A} \right),
    \end{equation}
with $\Gamma \in H_1(T^2,\mathbb{Z})$. 
Since the holonomies of $A$ and $\hat{A}$ are periodic, they are quantized as,
\begin{equation}
    \frac{N}{2\pi} \oint_{\Gamma} A =0,1,\cdots,N-1,\quad \frac{N}{2\pi} \oint_{\Gamma} \hat{A} = 0,1,\cdots,N-1,
\end{equation}
and $W^N[\Gamma] = \hat{W}^N[\Gamma] = 1$. The operators satisfy the commutation relation,
\begin{equation}
    W[\Gamma_i] \hat{W}[\Gamma_j] = \omega^{-\int \gamma_i \wedge \gamma_j} \hat{W}[\Gamma_j] W[\Gamma_i],
\end{equation}
where $\gamma \in H^1(T^2,\mathbb{Z})$ is the Poincare dual of $\Gamma$ defined as $\int_{\Gamma} \cdots = \int \gamma \wedge \cdots$.

Let's focus on the partition function $Z_{\mathfrak{T}_{\mathbb{Z}_N}}$ of the theory and see how the SymTFT applies. We can introduce a canonical basis of the Hilbert space of the BF model on $T^2$ where either $W[\Gamma]$ or $\hat{W}[\Gamma]$ are diagonalized. The two different choices give two kinds of topological boundary states $\mathfrak{B}^{\textrm{sym}}_{\mathbb{Z}_N}$ written as a boundary state,
    \begin{itemize}
        \item Dirichlet boundary state $|a\rangle$ for $A$,
            \begin{equation}
                \left\{ \begin{array}{l}
                    W[\Gamma] |a\rangle = \omega^{\int \gamma \wedge a} |a\rangle,\\
                    \hat{W}[\Gamma] |a\rangle = |a - \gamma\rangle,
                \end{array}\right.
            \end{equation}
        \item  Neumann boundary state $|\hat{a}\rangle$ for $A$,
            \begin{equation}
                \left\{ \begin{array}{l}
                    \hat{W}[\Gamma] |\hat{a}\rangle = \omega^{\int \gamma \wedge \hat{a}} |\hat{a}\rangle,\\
                    W[\Gamma] |\hat{a}\rangle = |\hat{a} - \gamma\rangle,
                \end{array}\right.
            \end{equation}
    \end{itemize}
where $a = NA/2\pi, \hat{a} = N\hat{A}/2\pi$ and we have $a,\hat{a}\in H^1(T^2,\mathbb{Z}_N)$. The integration $\int \gamma \wedge a = \int_{\Gamma} a$ gives the holonomy along $\Gamma$. The two bases are related by a discrete Fourier transformation,
    \begin{equation}
        |\hat{a}\rangle = \frac{1}{2} \sum_{a\in H^1(T^2,\mathbb{Z}_N)} \omega^{\int a \wedge \hat{a}}|a\rangle.
    \end{equation}
On the other hand, the physical boundary  $\mathfrak{B}^{\textrm{phys}}_{\mathfrak{T}_{\mathbb{Z}_N}}$ gives a dynamical boundary state $|\chi\rangle$ which depends on the partition function of theory $\mathfrak{T}_{\mathbb{Z}_N}$ with given the $\mathbb{Z}_N$ holonomies of $\mathfrak{T}_{\mathbb{Z}_N}$
    \begin{equation}
        |\chi\rangle = \sum_{a \in H^1(T^2,\mathbb{Z}_N)} Z_{\mathfrak{T}_{\mathbb{Z}_N}}[a] |a\rangle.
    \end{equation}
Choosing different topological boundary states, the path-integral of the BF theory on the slab gives,
    \begin{equation}
        Z[a] = \langle a | e^{iHt} |\chi\rangle = \langle a  |\chi\rangle,\quad Z[\hat{a}] = \langle \hat{a} | e^{iHt} |\chi \rangle = \langle \hat{a} | \chi \rangle.
    \end{equation}
where $Z[a] = Z_{\mathfrak{T}_{\mathbb{Z}_N}}[a] $ agrees with the torus partition function of $\mathfrak{T}_{\mathbb{Z}_N}$ and,
    \begin{equation}
        Z[\hat{a}] = \frac{1}{2} \sum_{a} \omega^{\int \hat{a} \wedge a} Z_{\mathfrak{T}_{\mathbb{Z}_N}}[a] \equiv Z_{\mathfrak{T}_{\mathbb{Z}_N}/\mathbb{Z}_N}[\hat{a}],
    \end{equation}
which is the partition function of the orbifold theory $\mathfrak{T}_{\mathbb{Z}_N}/\mathbb{Z}_N$, the Kramers-Wannier duaity of $\mathfrak{T}_{\mathbb{Z}_N}$. In other words, the $\mathbb{Z}_N$ gauging of $\mathfrak{T}_{\mathbb{Z}_N}$ can be viewed from the SymTFT as switching the topological boundary state from $|a\rangle$ to $|\hat{a}\rangle$.

When $N=2$, there also exists a topological boundary $|s\rangle$, where $s \in H^1(T^2,\mathbb{Z}_2)$ stands for the spin structure, such that JW transformation can be encoded as $Z_{F}[s] = \langle s | \chi \rangle$\footnote{For general $N$, there is a generalized JW transformation that leads to parafermion theories, see\cite{Alicea:2015hja,Yao:2020dqx,Duan:2023ykn}.}. The states $|s\rangle$ are eigenstates of the operators $W_F[\Gamma] \equiv W[\Gamma] \hat{W}[\Gamma]$ and satisfy,
        \begin{equation}
            \left\{ \begin{array}{l}
                W_F[\Gamma] |s\rangle = (-1)^{\textrm{Arf}(s+\gamma) - \textrm{Arf}(s)} |s\rangle\\
                \hat{W}[\Gamma] |s\rangle = |s + \gamma\rangle
              \end{array}\right.
        \end{equation}
where $\textrm{Arf}(s) \equiv s_1 s_2$ is the Arf-invariant where $s_i = \int_{\Gamma_i}s$ is the spin structure along $\Gamma_i$-cycle ($s_i=0$ is chosen to be the NS boundary condition). The topological boundary state $|s\rangle$ can also be expressed as,
    \begin{equation}
        |s\rangle = \frac{1}{2} \sum_{a \in H^1(T^2,\mathbb{Z}_2)} (-1)^{\textrm{Arf}(s+a)}|a\rangle,
    \end{equation}
and the transition amplitude $\langle s | \chi \rangle$ is,
\begin{equation}
    Z_{F}[s] = \langle s | \chi \rangle = \frac{1}{2} \sum_{a\in H^1(T^2,\mathbb{Z}_2)} (-1)^{\textrm{Arf}(s+a)} Z_{\mathfrak{T}_{\mathbb{Z}_N}}[a]
\end{equation}
which gives the partition function of the fermionic theory after JW transformation.

The $(2+1)$d BF theory has a $\mathbb{Z}_2$ symmetry which exchanges the two gauge field,
\begin{equation}
    A\rightarrow \hat{A},\quad \hat{A}\rightarrow A.
\end{equation}
The corresponding symmetry defect $\mathcal{D}_{\mathbb{Z}_2}[M_2]$ along a surface $M_2$ can be constructed as,
    \begin{equation}
    \mathcal{D}_{\mathbb{Z}_2} = \frac{1}{\sqrt{|H^1(M_2,\mathbb{Z}_N)|}} \sum_{\Gamma \in H^1(M_2,\mathbb{Z}_N)} W[\Gamma] \hat{W}[\Gamma]^{-1},
    \end{equation}
which is a condensation of the defect $W\hat{W}^{-1}$ along $M_2$. If $M_2$ is a time slice, one can check,
    \begin{equation}
        D_{\mathbb{Z}_2}[M_2] W[\Gamma] = \hat{W}[\Gamma]D_{\mathbb{Z}_2}[M_2],\quad D_{\mathbb{Z}_2}[M_2] \hat{W}[\Gamma] = W[\Gamma]D_{\mathbb{Z}_2}[M_2],
    \end{equation}
and,
    \begin{equation}
        D_{\mathbb{Z}_2}[M_2] \times D_{\mathbb{Z}_2}[M_2] =1.
    \end{equation}
using the quantum algebra.
\section{Duality between \eqref{eq:2foliated} and \eqref{eq:exotic}}\label{sec:duality}
In this appendix, we sketch the duality between the 2-foliated BF theory ~\eqref{eq:2foliated}
 and the exotic tensor gauge theory~\eqref{eq:exotic}. See also \cite{Spieler:2023wkz}. Begin with the 2-foliated theory~\eqref{eq:2foliated},
 \begin{equation}
    S_{2\textrm{-foliated}} = \frac{N}{2\pi}\int \sum_{k=1,2} (d B^k + b) \wedge C^k \wedge d x^k + b \wedge dc,
 \end{equation}
we split the coordinates $(x^0,x^1,x^2,x^3)$ as $(\tau,x^i)$ with $i=1,2,3$ and denote $(x^1,x^2,x^3)$ as $(x,y,z)$. The action can be written as,
\begin{align}
    S_{2\textrm{-foliated}} =& \frac{N}{2\pi} \int d \tau d^3 x \Big[\epsilon^{ijk}(- B^x_i \partial_{\tau} C^x_j \delta^x_k -   B^y_i \partial_{\tau} C^y_j \delta^y_k + \frac{1}{2}b_{ij}\partial_{\tau} c_k)  \nonumber\\
    & + \epsilon^{ijk} C^x_{\tau} (\partial_i B^x_j + \frac{1}{2} b_{ij}) \delta^x_k + \epsilon^{ijk} C^y_{\tau} (\partial_i B^y_j + \frac{1}{2} b_{ij}) \delta^y_k + \frac{1}{2} \epsilon^{ijk}  c_{\tau} \partial_i b_{jk} \nonumber\\
    &  + \epsilon^{ijk} b_{\tau i} (\partial_j a_k + C^x_j \delta^x_k + C^y_j \delta^y_k) + \epsilon^{ijk}  B^x_{\tau} \partial_i C^x_j \delta^x_k + \epsilon^{ijk}  B^y_{\tau} \partial_i C^y_j \delta^y_k \Big],
\end{align}
up to total derivatives. Integrate $C^x_{\tau},C^y_{\tau},b_{\tau x},b_{\tau y}$, one gets
\begin{equation}
    b_{yz} = \partial_z B^x_y - \partial_y B^x_z,\quad b_{xz} = \partial_z B^y_x - \partial_x B^y_z,
\end{equation}
for $C^x_{\tau},C^y_{\tau}$ and
\begin{equation}
    C^y_z = \partial_y c_z - \partial_z c_y,\quad C^x_z = \partial_x c_z - \partial_z c_x,
\end{equation}
for $b_{\tau x},b_{\tau y}$. We can solve $b_{yz},b_{xz},C^y_z,C^x_z$ and substitute them back to the action. Moreover, integrating $b_{\tau z}$ gives,
\begin{equation}
    C^y_x + \partial_x c_y = C^x_y + \partial_y c_x,
\end{equation}
such that we can define $A^{xy} = C^y_x + \partial_x c_y = C^x_y + \partial_y c_x$.
After renaming other variables,
\begin{equation}
    A^{\tau} \equiv c_{\tau},\quad A^{z} \equiv c_z,
\end{equation}
\begin{equation}
    \hat{A}^{xy} =  \partial_x B^x_y - \partial_y B^y_x + b_{xy},\quad \hat{A}^{\tau} = B^x_{\tau} - B^y_{\tau},\quad \hat{A}^{z} = B^x_z - B^y_z,
\end{equation}
the action is rewritten as,
\begin{equation}
    S_{\textrm{exotic}} = \frac{N}{2\pi} \int \left[A^{\tau} (\partial_z \hat{A}^{xy} - \partial_x \partial_y \hat{A}^z) - A^z (\partial_{\tau} \hat{A}^{xy} - \partial_x \partial_y \hat{A}^{\tau})- A^{xy} (\partial_{\tau} \hat{A}^z - \partial_z \hat{A}^{\tau})  \right].
\end{equation}
which reproduces the exotic tensor gauge theory~\eqref{eq:exotic}.

\section{Derivation of fusion rule of subsystem KW defects}\label{sec:deri}
In this section, we will re-derive the fusion rule between two subsystem KW defects $\mathcal{N}^{\textrm{sub}} \times \mathcal{N}^{\textrm{sub}}$ after the formulation of gauging a subsystem symmetry in a cohomology language\footnote{The original derivation on lattice can be found in \cite{Cao:2023doz}.}. The derivation is a direct generalization from the fusion of  duality defects of guaging 0-form $\mathbb Z_N$ symmetry~\cite{Kaidi:2022cpf}. For simplicity, we will keep $N=2$.

\subsection{Conventions}
Denote the gauge fields $A$ and its dual $\hat{A}$ for  subsystem $\mathbb{Z}_2$ symmetry as the pair,
\begin{equation}
    A = (A^z , A^{xy}),\quad \hat{A} = (\hat{A}^z , \hat{A}^{xy}).
\end{equation}
The gauge transformation is,
\begin{equation}
    A \rightarrow A + \delta \lambda,\quad \hat{A} \rightarrow \hat{A} + \delta \hat{\lambda},
\end{equation}
where the action of $\delta$ on a function is defined as,
\begin{equation}
    \delta f(x,y,z) = (\partial_z f(x,y,z), \partial_x \partial_y f(x,y,z)).
\end{equation}
The flatness condition is written as,
\begin{equation}
    \delta A \equiv \partial_{x} \partial_y A^{z} - \partial_z A^{xy} = 0,
\end{equation}
and one can check $\delta^2 f(x,y,z) = 0$ automatically.

To perform the summation formally, it is useful to introduce the 0-cochain $C^0_{\textrm{sub}}(M_3)$ as the set of functions $f$ on $M_3$, 1-cochain $C^1_{\textrm{sub}}(M_3)$ as the set of the pairs $g = (g^z,g^{xy})$ where $g^z,g^{xy}$ are both functions on $M_3$, and 2-cochain $C^2_{\textrm{sub}}(M_3)$ as the set of functions denoted by $h^{xyz}$ on $M_3$.
The coboundary operator $\delta$ acts on $C^*_{\textrm{sub}}$ as,
\begin{equation}
    \delta f = (\partial_z f, \partial_x \partial_y f),\quad \delta g = \partial_{x}\partial_y g^z - \partial_z g^{xy},\quad \delta h^{xyz} = 0,
\end{equation}
and it satisfies $\delta^2 = 0$. One can define a product $* \cdot *$ which sends $C^m_{\textrm{sub}}(M_3) \times C^n_{\textrm{sub}}(M_3)$ to $C^{m+n}_{\textrm{sub}}(M_3)$ where $C^{m+n}_{\textrm{sub}}(M_3)$ with $m+n >2$ is defined to be trivial. For example, when one of $C^*_{\textrm{sub}}$ is $C^0_{\textrm{sub}}$ whose elements are functions, the product is the usual multiplication; and for $g,g' \in C^1_{\textrm{sub}}(M_3)$ one can assign $g \cdot g' \equiv g^{xy} g'^z + g'^{xy} g^z$.

Let's consider the cohomology\footnote{
Strictly speaking, $H^*$ are not cohomology groups because the closeness condition is not preserved under the product. For example, given $f \in H^0_{\textrm{sub}}(M_3)$ and $g \in H^1_{\textrm{sub}}(M_3)$ one can check,
 \begin{equation}
        \delta(f\cdot g) = \partial_x \partial_y (f g^z) - \partial_z (f g^{xy}) = \partial_x f \partial_y g^z + \partial_y f \partial_x g^z, \nonumber
    \end{equation}
which does not vanish. Nevertheless, we do not need this property in the proof.}
$H^* = Z^*/B^*$ where $Z^*(B^*)$ contains closed (exact) cochains. For example
\begin{equation}
    H^0_{\textrm{sub}}(M_3,\mathbb{Z}_2)=Z^0_{\textrm{sub}}(M_3,\mathbb{Z}_2),
\end{equation}
contains scalar functions that only have $x$ or $y$ dependence. Because of the flatness condition, the subsystem gauge field $A$ and $\hat{A}$ belong to  $H^1_{\textrm{sub}}(M_3,\mathbb{Z}_2)$, closed 1-cochains modulo out the exact 1-cochain (the gauge transformation), and we have $|H^1_{\textrm{sub}}(M_3,\mathbb{Z}_2)|=|M_v|=2^{2(L_x+L_y-1)}$, where the dimension of $H^1_{\textrm{sub}}(M_3,\mathbb{Z}_2)$ is equal to the number gauge invariant holonomies $\mathbf{w}$.

Now we will take a formal, continuous route and only make it discrete at the final step. For example, the subsystem KW transformation
\begin{align}\label{eq:appkw}
    \begin{split}
    &Z_{\hat{\mathfrak{T}}_{\textrm{sub}}}[\hat{w}_{z,x;j+\frac12}, \hat{w}_{z,y;i+\frac12},\hat{w}_{x;j},\hat{w}_{y;i}]\\
    &= \frac{1}{2^{L_x+L_y-1}}\sum_{w_{z,x;j}, w_{z,y;i},w_{x;j+\frac12},w_{y;i+\frac12}=0,1}Z_{\mathfrak{T}_{\textrm{sub}}}[w_{z,x;j}, w_{z,y;i},w_{x;j+\frac12},w_{y;i+\frac12}]\\
    &\times(-1)^{\sum_{i}(\hat{w}_{z,y;i+\frac12}w_{y;i+\frac12}+\hat{w}_{y;i}w_{z,y;i})+\sum_{j}(\hat{w}_{z,x;j+\frac12}w_{x;j+\frac12}+\hat{w}_{x;j}w_{z,x;j})}
    \end{split}
\end{align}
will be written formally as,
\begin{equation}\label{eq:kwformal}
    Z[\hat{A}] = \frac{1}{|H^0_{\textrm{sub}}(M_3,\mathbb{Z}_2)|}\sum_{A\in H^1_{{\textrm{sub}}}(M_3,\mathbb{Z}_2)} Z[A] (-1)^{\int_{M_3} A \cdot \hat{A}},
\end{equation}
where $|H^0_{\textrm{sub}}(M_3,\mathbb{Z}_2)|$ is the dimension of the cohomology group $H^0_{\textrm{sub}}(M_3,\mathbb{Z}_2)$. The labels of partition functions are omitted for more concise expressions. The gauge field $A$ will take values in its gauge equivalent class $H^1_{{\textrm{sub}}}(M_3,\mathbb{Z}_2)$ and the integral is regularized by the sum of holonomies,
\begin{align}
    \begin{split}
        \int_{M_3} A \cdot \hat{A}=& \int_{M_3}(A^{xy} \hat{A}^z + A^{z} \hat{A}^{xy}) \\
        =& \sum_{i}(\hat{w}_{z,y;i+\frac12}w_{y;i+\frac12}+\hat{w}_{y;i}w_{z,y;i})+\sum_{j}(\hat{w}_{z,x;j+\frac12}w_{x;j+\frac12}+\hat{w}_{x;j}w_{z,x;j}),
    \end{split}
\end{align}
where $w$ and $\hat{w}$ are holonomies of $A$ and $\hat{A}$. 
The formal expression~\eqref{eq:kwformal} differs from the regularized one~\eqref{eq:appkw} by the normalization factor  $|H^0_{\textrm{sub}}(M_3,\mathbb{Z}_2)|$ (instead of $\sqrt{|H^1_{\textrm{sub}}(M_3,\mathbb{Z}_2)|}=2^{L_x+L_y-1}$) as suggested in \cite{Kaidi:2022cpf}.

We also need to define relative cohomology $H^1_{\textrm{sub}}(M_3,M_2,\mathbb{Z}_2)$ where $M_2 = \partial M_3$ is the boundary where the gauge fields $A=(A^z,A^{xy})$ should satisfy the ``Dirichlet" boundary condition at the boundary $M_2$. First, we need to define what the ``Dirichlet" boundary condition means for the gauge fields $A^z$ and $A^{xy}$. If $M_2$ is the $x$-$y$ plane we can just set $A^{xy}=0$ at the boundary. The holonomy of $\int A^z dz$ split into $\mathcal{A}^y(x)$ and $\mathcal{A}^x(y)$ and they depend on $x$ and $y$ separately. There also exists a gauge transformation which shifts $\mathcal{A}^y(x) \rightarrow \mathcal{A}^y(x) + \theta$ and $\mathcal{A}^x(y) \rightarrow \mathcal{A}^x(y) - \theta$ by some constant $\theta$ so that $\int A^z dz$ is invariant.  If $M_2$ is the $x$-$z$ plane we will require $\mathcal{A}^y(x)$ to be a constant at the boundary and it is gauge equivalent to zero. Therefore we impose $\partial_x A^z=0$ at the boundary.

In the next subsection, we will consider the KW defects defined by gauging the subsystem $\mathbb{Z}_2$ symmetry in half of the spacetime $M_3$ with the ``Dirichlet" boundary condition imposed at the boundary. To do this, we need to couple the theory to a dynamical subsystem $\mathbb{Z}_2$ gauge theory with flat gauge field $(A^z,A^{xy})$. The subsystem $\mathbb{Z}_2$ gauge theory can be represented as\footnote{This is a generalization that ordinary $q$-form $\mathbb{Z}_N$ gauge theory in $D$-dimenson can be represented by the BF theory with level $N$\cite{Choi:2021kmx,Maldacena:2001ss,Banks:2010zn,Kapustin:2014gua},
    \begin{equation}
        \frac{N}{2\pi} \int d^D x B^{D-q-2} d A^{q+1}
    \end{equation}
where $A^{q+1}$ is the $(q+1)$-form gauge field and $B^{D-q-2}$ is the Lagrange multiplier enforcing $A^{q+1}$ to be $\mathbb{Z}_N$-valued. }
    \begin{equation}
        \frac{1 }{\pi} \int dx dy dz \ \phi \delta A = \frac{1}{\pi} \int dx dy dz \ \phi (\partial_z A^{xy} - \partial_x \partial_y A^z),
    \end{equation}
where $\phi$ is a periodic scalar field that serves as a Lagrangian multiplier enforcing $(A^z, A^{xy})$ to be properly quantized and $\mathbb{Z}_2$-valued. The Dirichlet boundary condition introduced above is topological along the normal direction. To see this, we need to deform the locus of the boundary slightly and see the variation of the action. For example, if the boundary is $x$-$y$ plane at $z=0$ and we deform it to $z=\epsilon$, the difference can be written as the surface integral at $z=0$ and $z=\epsilon$ using Stokes theorem
    \begin{equation}
        \int_{z=0} dx dy \phi A^{xy} - \int_{z=\epsilon} dx dy \phi A^{xy} ,
    \end{equation}
which is zero due to the boundary condition $A^{xy}=0$. On the other hand, if the boundary is $x$-$z$ plane at $y=0$ and we deform it to $y=\epsilon$, the difference is,
    \begin{equation}
        -\int_{y=0} dx dz \phi \partial_x A^z + \int_{y=\epsilon} dx dz \phi  \partial_x A^z, 
    \end{equation}
and the boundary condition $\partial_x A^z=0$ is sufficient to set it zero. 

We will also see how these choices of boundary conditions give the correct fusion rule in the following derivation.

\subsection{Fusion rule of subsystem KW defects}
\begin{figure}[htbp]
\centering
\begin{tikzpicture}
        \shade[top color=blue!40, bottom color=blue!10]  (0,0) -- (0,3) -- (3,3) -- (3,0)-- (0,0);
        \draw[thick] (0,0) -- (0, 3);
        \draw[thick] (3,0) -- (3, 3);
        \draw[thick] (-2,0) -- (5, 0);
        \node[below] at (0,0){$z=0$};
        \node[below] at (3,0){$z=\epsilon$};
        \node[left] at (0,1.5){$\mathcal N$};
         \node[right] at (3,1.5){$\mathcal N$};
    \end{tikzpicture}
    \caption{Fusion of two subsystem KW operators along $z$ direction.}
    \label{fig:fusion}
\end{figure}
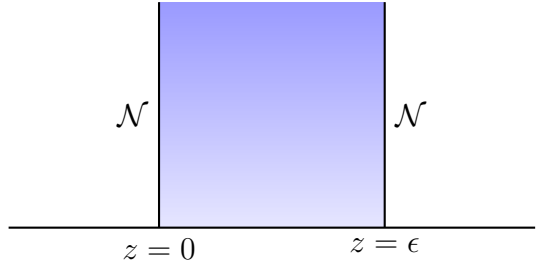

We first consider the case where the defect $\mathcal{N}^{\textrm{sub}}$ is along the $x$-$y$ plane and acts as a symmetry operator. 
Our strategy, as shown in Fig.~\ref{fig:fusion}, is to put two parallel subsystem KW operators with a separation of $\epsilon$ and compute the partition function in the region between two operators. As we take the limit $\epsilon\to 0$, we get the fusion of two operators. It is equivalent to performing 1-gauging on a co-dimension one surface~\cite{Roumpedakis:2022aik}.

Consider the symmetry operator $\mathcal{N}^{\textrm{sub}}$ located at $z=0$ as an example. The theory at $z\geq 0$ is defined to be,
\begin{equation}
    Z[M_3^{\geq 0},\hat{A}] = \frac{1}{|H^0_{\textrm{sub}}(M_3^{\geq 0},M_{2|0},\mathbb{Z}_2)|}\sum_{A\in H^1_{\textrm{sub}}(M_3^{\geq 0},M_{2|0},\mathbb{Z}_2)} Z[M_3^{\geq 0},A] (-1)^{\int_{M_3^{\geq 0}} A \cdot \hat{A}},
\end{equation}
where $H^1_{\textrm{sub}}(M_3^{\geq 0},M_{2|0},\mathbb{Z}_2)$ is the relative cohomology such that $A^{xy}=0$ at the boundary $M_{2|0}$ and we use $M_{2|0}$ to emphasize $M_2$ is located at $z=0$. To compute the fusion $\mathcal{N}^{\textrm{sub}} \times \mathcal{N}^{\textrm{sub}}$, we insert another $\mathcal{N}^{\textrm{sub}}$ at $z = \epsilon$ such that the theory living on $M_3^{\geq 0}$ is given by,
\begin{equation}
    \frac{1}{|H^0_{\textrm{sub}}(M_3^{\geq 0},M_{2|0},\mathbb{Z}_2)||H^0_{\textrm{sub}}(M_3^{\geq \epsilon},M_{2|\epsilon},\mathbb{Z}_2)|}\sum_{\substack{A\in H^1_{{\textrm{sub}}}(M_3^{\geq 0},M_{2|0},\mathbb{Z}_2)\\ \widetilde{A}\in H^1_{{\textrm{sub}}}(M_3^{\geq \epsilon},M_{2|\epsilon},\mathbb{Z}_2)}} Z[M_3^{\geq 0},A] (-1)^{\int_{M_3^{[0,\epsilon)}} A \cdot \hat{A} + \int_{M_3^{\geq \epsilon}} (A - \hat{A}) \cdot \widetilde{A} }.
\end{equation}
Using the relations $|H^1| = |Z^1|/|B^1|$, we can write the sum over cohomologies into a sum over cocycles as  $\sum_{A\in H^1} = \frac{1}{|B^1|} \sum_{A\in Z^1}$. Moreover, notice that $|B^1| = |C^0|/|Z^0|$ and $|B^0| = 1$, we can rewrite the integral as,
\begin{align}
    \frac{1}{|C^0_{\textrm{sub}}(M_3^{\geq 0},M_{2|0},\mathbb{Z}_2)||C^0_{\textrm{sub}}(M_3^{\geq \epsilon},M_{2|\epsilon},\mathbb{Z}_2)|} \sum_{\substack{A\in Z^1_{{\textrm{sub}}}(M_3^{\geq 0},M_{2|0},\mathbb{Z}_2)\\ \widetilde{A}\in Z^1_{{\textrm{sub}}}(M_3^{\geq \epsilon},M_{2|\epsilon},\mathbb{Z}_2)}} Z[M_3^{\geq 0},A] (-1)^{\int_{M_3^{[0,\epsilon)}} A \cdot \hat{A} + \int_{M_3^{\geq \epsilon}} (A - \hat{A}) \cdot \widetilde{A} }.
\end{align}
The cocycle condition can further be relaxed by introducing Lagrange multiplier $\phi\in C^0(M_3^{\geq 0},\mathbb{Z}_2)$ and $\widetilde{\phi} \in C^0(M_3^{\geq \epsilon},\mathbb{Z}_2)$,
\begin{align}
    &\frac{1}{|C^0_{\textrm{sub}}(M_3^{\geq 0},M_{2|0},\mathbb{Z}_2)||C^0_{\textrm{sub}}(M_3^{\geq \epsilon},M_{2|\epsilon},\mathbb{Z}_2)||C^0_{\textrm{sub}}(M_3^{\geq 0},\mathbb{Z}_2)||C^0_{\textrm{sub}}(M_3^{\geq \epsilon},\mathbb{Z}_2)|}\nonumber\\
    &\times \sum_{\substack{A\in C^1_{{\textrm{sub}}}(M_3^{\geq 0},\mathbb{Z}_2),\widetilde{A}\in C^1_{{\textrm{sub}}}(M_3^{\geq \epsilon},\mathbb{Z}_2) \\\phi \in C^0_{{\textrm{sub}}}(M_3^{\geq 0},\mathbb{Z}_2),\widetilde{\phi}\in C^0_{{\textrm{sub}}}(M_3^{\geq \epsilon},\mathbb{Z}_2) }} Z[M_3^{\geq 0},A] (-1)^{\int_{M_3^{[0,\epsilon)}} A \cdot \hat{A} + \int_{M_3^{\geq \epsilon}} (A - \hat{A}) \cdot \widetilde{A} + \int_{M_3^{> 0}} \phi  \delta A - \int_{M_{2|0}} \phi A^{xy} + \int_{M_3^{> \epsilon}} \widetilde{\phi}  \delta \widetilde{A} - \int_{M_{2|\epsilon}} \widetilde{\phi} \widetilde{A}^{xy} }.
\end{align}
Summing over $\phi$ in the bulk $M_3^{\geq 0}$ enforces $A$ to be a cocycle due to the coupling $\int_{M_3^{>0}} \phi  \delta A$, and summing over $\phi$ on the boundary $M_{2|0}$ enforces $A^{xy}=0$, which makes the cocycle relative to $M_{2|0}$. Same to $\hat{A}$.

We can firstly perform the sum over $\widetilde{A}$ and one has,
\begin{equation}
    (-1)^{\int_{M_3^{> \epsilon}} \widetilde{\phi} \delta \widetilde{A} - \int_{M_{2|\epsilon}} \widetilde{\phi} \widetilde{A}^{xy}} = (-1)^{\int_{M_3^{\geq \epsilon}} \delta \widetilde{\phi} \cdot \widetilde{A}},
\end{equation}
where we use integration by part and $\delta \widetilde{\phi} \cdot \widetilde{A} = \partial_z \widetilde{\phi} \widetilde{A}^{xy} + \partial_x \partial_y \widetilde{\phi} \widetilde{A}^z$. Sum over $\widetilde{A}$ will produce a factor $|C^1_{\textrm{sub}}(M_3^{\geq \epsilon} , \mathbb{Z}_2)|$ and enforce $A-\hat{A}-\delta \widetilde{\phi}=0$,
\begin{align}
    & \frac{|C^1_{\textrm{sub}}(M_3^{\geq \epsilon} , \mathbb{Z}_2)|}{|C^0_{\textrm{sub}}(M_3^{\geq 0},M_{2|0},\mathbb{Z}_2)||C^0_{\textrm{sub}}(M_3^{\geq \epsilon},M_{2|\epsilon},\mathbb{Z}_2)||C^0_{\textrm{sub}}(M_3^{\geq 0},\mathbb{Z}_2)||C^0_{\textrm{sub}}(M_3^{\geq \epsilon},\mathbb{Z}_2)|}\nonumber\\
    &\times \sum_{\substack{A\in C^1_{{\textrm{sub}}}(M_3^{\geq 0},\mathbb{Z}_2) \\\phi \in C^0_{{\textrm{sub}}}(M_3^{\geq 0},\mathbb{Z}_2),\widetilde{\phi}\in C^0_{{\textrm{sub}}}(M_3^{\geq \epsilon},\mathbb{Z}_2) }} Z[M_3^{\geq 0},A] (-1)^{\int_{M_3^{[0,\epsilon)}} A \cdot \hat{A} + \int_{M_3^{> 0}} \phi  \delta A - \int_{M_{2|0}} \phi A^{xy} } \delta(A-\hat{A}-\delta \widetilde{\phi})|_{M_3^{\geq \epsilon}}.
\end{align}
We then integrate out $\phi$, which produces a factor $|C^0_{\textrm{sub}}(M_3^{\geq 0} , \mathbb{Z}_2)|$ and enforces $A \in Z^1_{\textrm{sub}}(M_3^{\geq 0},M_{2|0},\mathbb{Z}_2)$,
\begin{align}
    &\frac{|C^1_{\textrm{sub}}(M_3^{\geq \epsilon} , \mathbb{Z}_2)|}{|C^0_{\textrm{sub}}(M_3^{\geq 0},M_{2|0},\mathbb{Z}_2)||C^0_{\textrm{sub}}(M_3^{\geq \epsilon},M_{2|\epsilon},\mathbb{Z}_2)||C^0_{\textrm{sub}}(M_3^{\geq \epsilon},\mathbb{Z}_2)|}\nonumber\\
    &\times \sum_{\substack{A\in Z^1_{{\textrm{sub}}}(M_3^{\geq 0},M_{2|0},\mathbb{Z}_2) \\\widetilde{\phi}\in C^0_{{\textrm{sub}}}(M_3^{\geq \epsilon},\mathbb{Z}_2) }} Z[M_3^{\geq 0},A] (-1)^{\int_{M_3^{[0,\epsilon)}} A \cdot \hat{A} } \delta(A-\hat{A}-\delta \widetilde{\phi})|_{M_3^{\geq \epsilon}}.
\end{align}
The summand is independent of $\widetilde{\phi}$ and we can set $\widetilde{\phi}$ to zero in the delta function and add a normalization factor $|C^0_{\textrm{sub}}(M_3^{\geq \epsilon},\mathbb{Z}_2)|$. The delta function then fixes $A=\hat{A}$ in $M_3^{\geq \epsilon}$ and make $A$ an element of $Z^1_{\textrm{sub}}(M_3^{[0,\epsilon]},M_{2|0} \cup M_{2|\epsilon},\mathbb{Z}_2)$,
\begin{align}\label{Subsystem-Review-N-fusion-Derivation-Normalization}
    \frac{|C^1_{\textrm{sub}}(M_3^{\geq \epsilon} , \mathbb{Z}_2)|}{|C^0_{\textrm{sub}}(M_3^{\geq 0},M_{2|0},\mathbb{Z}_2)||C^0_{\textrm{sub}}(M_3^{\geq \epsilon},M_{2|\epsilon},\mathbb{Z}_2)|} \sum_{A \in Z^1_{\textrm{sub}}(M_3^{[0,\epsilon]},M_{2|0} \cup M_{2|\epsilon},\mathbb{Z}_2)} Z[M_3^{\geq 0},A+\hat{A}_{M_3^{\geq \epsilon}}] (-1)^{\int_{M_3^{[0,\epsilon]}} A \cdot \hat{A} },
\end{align}
where $\hat{A}|_{M_3^{\geq \epsilon}}$ is equal to $\hat{A}$ if we are on $M_3^{\geq \epsilon}$ and vanishes elsewhere.

Let's introduce the Euler factor $\chi[M_3^{\geq \epsilon},\mathbb{Z}_2]$ as,
\begin{equation}\label{Appendix-Euler-factor} 
   \chi[M_3^{\geq \epsilon},\mathbb{Z}_2] \equiv  \frac{|H^2_{\textrm{sub}}(M_3^{\geq \epsilon} , \mathbb{Z}_2)||H^0_{\textrm{sub}}(M_3^{\geq \epsilon} , \mathbb{Z}_2)|}{|H^1_{\textrm{sub}}(M_3^{\geq \epsilon} , \mathbb{Z}_2)|} = \frac{|C^2_{\textrm{sub}}(M_3^{\geq \epsilon} , \mathbb{Z}_2)||C^0_{\textrm{sub}}(M_3^{\geq \epsilon} , \mathbb{Z}_2)|}{|C^1_{\textrm{sub}}(M_3^{\geq \epsilon} , \mathbb{Z}_2)|}.
\end{equation}
where in the second expression we use the fact $H^0 = Z^0$, $Z^2 = C^2$ since $C^2$ is the top one and $|B^{n+1}| = |C^{n}|/|Z^{n}|$. The normalization factor in \eqref{Subsystem-Review-N-fusion-Derivation-Normalization} can be written as,
\begin{equation}
    \frac{|C^2_{\textrm{sub}}(M_3^{\geq \epsilon} , \mathbb{Z}_2)||C^0_{\textrm{sub}}(M_3^{\geq \epsilon} , \mathbb{Z}_2)|}{|C^0_{\textrm{sub}}(M_3^{\geq 0},M_{2|0},\mathbb{Z}_2)||C^0_{\textrm{sub}}(M_3^{\geq \epsilon},M_{2|\epsilon},\mathbb{Z}_2)|} \chi[M_3^{\geq \epsilon},\mathbb{Z}_2]^{-1}.
\end{equation}
The first factor can be further simplified by using $|C^2_{\textrm{sub}}(M_3^{\geq \epsilon} , \mathbb{Z}_2)| = |C^0_{\textrm{sub}}(M_3^{\geq \epsilon},M_{2|\epsilon},\mathbb{Z}_2)| $ since the elements in $C^0_{\textrm{sub}}(M_3^{\geq \epsilon},M_{2|\epsilon},\mathbb{Z}_2)$ and the elements in $C^2_{\textrm{sub}}(M_3^{\geq \epsilon} , \mathbb{Z}_2)$ are Fourier partners under integration over $M_3$. Finally, use the fact that,
\begin{equation}
    |C^n_{\textrm{sub}}(M_3^{\geq 0},M_{2|0},\mathbb{Z}_2)| = |C^n_{\textrm{sub}}(M_3^{\geq \epsilon} , \mathbb{Z}_2)| |C^n_{\textrm{sub}}(M_3^{[0,\epsilon] },M_{2|0} \cup M_{2|\epsilon},\mathbb{Z}_2)|,
\end{equation}
which is a decomposition of cochains on $M_3^{\geq 0}$ into the sum of cochains on $M_3^{[0,\epsilon]}$ with fixed boundary condition at $M_{2|\epsilon}$ and cochains on $M_3^{\geq \epsilon}$ with free boundary conditions. Substituting the simplified normalization into \eqref{Subsystem-Review-N-fusion-Derivation-Normalization} we have,
\begin{equation}
     \frac{\chi[M_3^{\geq \epsilon},\mathbb{Z}_2]^{-1}}{|C^0_{\textrm{sub}}(M_3^{[0,\epsilon] },M_{2|0} \cup M_{2|\epsilon},\mathbb{Z}_2)|} \sum_{A \in Z^1_{\textrm{sub}}(M_3^{[0,\epsilon]},M_{2|0} \cup M_{2|\epsilon},\mathbb{Z}_2)} Z[M_3^{\geq 0},A+\hat{A}_{M_3^{\geq \epsilon}}] (-1)^{\int_{M_3^{[0,\epsilon]}} A \cdot \hat{A} } .
\end{equation}
Write the summation of $Z^1_{\textrm{sub}}(M_3^{[0,\epsilon]},M_{2|0} \cup M_{2|\epsilon},\mathbb{Z}_2)$ back to $H^1_{\textrm{sub}}(M_3^{[0,\epsilon]},M_{2|0} \cup M_{2|\epsilon},\mathbb{Z}_2)$ as $\sum_{A\in Z^1} = |B^1| \sum_{A\in H^1}$, we obtain the finial result,
\begin{equation}
     \frac{\chi[M_3^{\geq \epsilon},\mathbb{Z}_2]^{-1}}{|H^0_{\textrm{sub}}(M_3^{[0,\epsilon] },M_{2|0} \cup M_{2|\epsilon},\mathbb{Z}_2)|} \sum_{A \in H^1_{\textrm{sub}}(M_3^{[0,\epsilon]},M_{2|0} \cup M_{2|\epsilon},\mathbb{Z}_2)} Z[M_3^{\geq 0},A+\hat{A}_{M_3^{\geq \epsilon}}] (-1)^{\int_{M_3^{[0,\epsilon]}} A \cdot \hat{A} },
\end{equation}
where we use the relations $|B^1||Z^0| = |C^0|$ and $|B^0|=1$ again.

Now let's evaluate the integral $\int_{M_3^{[0,\epsilon)}} A \cdot \hat{A}$ with Dirichlet boundary condition $A^{xy}|_{z=0} = A^{xy}|_{z=\epsilon} = 0$.
At this stage, we need to regularize the spacetime and write,
\begin{align}
    \int_{M_2\times [0,\epsilon]} (A^{xy} \hat{A}^z + A^{z} \hat{A}^{xy}) =& \sum_{i=1}^{L_x} \left( w_{y,i+\frac{1}{2}} \left[\sum_{k=1}^{L_\epsilon} \hat{A}^z_{i+\frac{1}{2},j+\frac{1}{2},k} \right]^y\right) + \sum_{j=1}^{L_y} \left( w_{x,j+\frac{1}{2}} \left[ \sum_{k=1}^{L_\epsilon} \hat{A}^z_{i+\frac{1}{2},j+\frac{1}{2},k}\right]^x \right) \nonumber\\
    +&\sum_{i=1}^{L_x}\left(w_{z,y;i} \sum_{j=1}^{L_y} \hat{A}^{xy}_{i,j,k+\frac{1}{2}} \right)+\sum_{j=1}^{L_y}\left(w_{z,x;j} \sum_{i=1}^{L_x}\hat{A}^{xy}_{i,j,k+\frac{1}{2}}\right),
\end{align}
where $L_{\epsilon}$ is the number of sites between $[0,\epsilon)$ and $\omega$ are the holonomies of $A$. $\sum_{j=1}^{L_y} \hat{A}^{xy}_{i,j,k+\frac{1}{2}} ,\sum_{i=1}^{L_x}\hat{A}^{xy}_{i,j,k+\frac{1}{2}}$ are strip operators of the dual field $\hat{A}$ and they generate the subsystem symmetry as mentioned in \eqref{Foliated-Theory-Operators-Correspondence},
\begin{equation}
    U_i^y = \exp \left(i\pi \sum_{j=1}^{L_y} \hat{A}^{xy}_{i,j,k+\frac{1}{2}} \right),\quad U_j^x = \left( i\pi \sum_{i=1}^{L_x}\hat{A}^{xy}_{i,j,k+\frac{1}{2}} \right).
\end{equation}
The line operator $\sum_{k=1}^{\epsilon} \hat{A}^z_{i+\frac{1}{2},j+\frac{1}{2},k}$ of the dual field $\hat{A}^z$ are subsystem symmetry defects. Recall that the line operator can be decomposed into two line operators separately movable along $x$ and $y$ directions and we use the labels $[\cdots]^x$ and $[\cdots]^y$ to represent them. 

We will take the limit $\epsilon \rightarrow 0$ while fixing the holonomies $w$. The first line vanishes in the limit. Another point of view is, since $A^{xy}|_{z=0} = A^{xy}|_{z=\epsilon} = 0$ the holonomies of $A^{xy}$ vanishes and $w_{y,i+\frac{1}{2}} = w_{x,j+\frac{1}{2}}=0$ and first line is trivial. Therefore we only need to consider the second line which implies the fusion rule,
\begin{equation}
    \mathcal{N}^{\text{sub}}{}^{\dagger} \times \mathcal{N}^{\text{sub}} = \sum_{w_{z,y;i},w_{z,x;j}/\sim} (U_i^y)^{w_{z,y;i}} (U_j^x)^{w_{z,x;j}} = \frac{1}{2} \prod_{i=1}^{L_x} \left(1+U^y_i \right)\prod_{j=1}^{L_y} \left(1+U^x_j \right),
\end{equation}
where we have used $|H^0_{\textrm{sub}}(M_3^{[0,\epsilon] },M_{2|0} \cup M_{2|\epsilon},\mathbb{Z}_2)|=1$
\footnote{The elements $f$ in $H^0_{\textrm{sub}}(M_3^{[0,\epsilon] },M_{2|0} \cup M_{2|\epsilon},\mathbb{Z}_2)$ should satisfies $\partial_x \partial_y f = \partial_z f = 0$ and $f=0$ at the boundary. They fix $f$ to be trivial.}
and $\mathcal{N}^{\text{sub}}{}^{\dagger}$ is normalized as $\mathcal{N}^{\text{sub}}{}^{\dagger} = \chi[M_3^{\geq 0},\mathbb{Z}_2]\mathcal{N}^{\text{sub}}$. In the sum, we mod out the gauge redundancy $\sim$ of the holonomies.

Let's then consider the case where the defect $\mathcal{N}^{\textrm{sub}}$ is along the $x$-$z$ plane and acts as a symmetry defect. The derivation of the fusion rule is similar and we have,
\begin{align}
    &\frac{1}{|C^0_{\textrm{sub}}(M_3^{\geq 0},M_{2|0},\mathbb{Z}_2)||C^0_{\textrm{sub}}(M_3^{\geq \epsilon},M_{2|\epsilon},\mathbb{Z}_2)||C^0_{\textrm{sub}}(M_3^{\geq 0},\mathbb{Z}_2)||C^0_{\textrm{sub}}(M_3^{\geq \epsilon},\mathbb{Z}_2)|}\nonumber\\
    \times& \sum_{\substack{A\in C^1_{{\textrm{sub}}}(M_3^{\geq 0},\mathbb{Z}_2),\widetilde{A}\in C^1_{{\textrm{sub}}}(M_3^{\geq \epsilon},\mathbb{Z}_2) \\\phi \in C^0_{{\textrm{sub}}}(M_3^{\geq 0},\mathbb{Z}_2),\widetilde{\phi}\in C^0_{{\textrm{sub}}}(M_3^{\geq \epsilon},\mathbb{Z}_2) }} Z[M_3^{\geq 0},A] (-1)^{\int_{M_3^{[0,\epsilon)}} A \cdot \hat{A} + \int_{M_3^{\geq \epsilon}} (A - \hat{A}) \cdot \widetilde{A} + \int_{M_3^{> 0}} \phi  \delta A + \int_{M_{2|0}} \phi \partial_x A^{z} + \int_{M_3^{> \epsilon}} \widetilde{\phi}  \delta \widetilde{A} + \int_{M_{2|\epsilon}} \widetilde{\phi} \partial_x \widetilde{A}^{z} }.
\end{align}
where the difference is that $M_{2|0}$ and $M_{2|\epsilon}$ are the $x$-$z$ plane located at $y=0$ and $y=\epsilon$, and summing over the Lagrangian multiplier $\phi$ at $M_2$ enforces $\partial_x A^z=0$ as discussed at the beginning of this section. It has the advantage that,
\begin{equation}
    (-1)^{\int_{M_3^{> \epsilon}} \widetilde{\phi} \delta \widetilde{A} + \int_{M_{2|\epsilon}} \widetilde{\phi} \partial_x \widetilde{A}^{z}} = (-1)^{\int_{M_3^{\geq \epsilon}} \delta \widetilde{\phi} \cdot \widetilde{A}},
\end{equation}
which is the same as before. The remaining derivations are exactly the same and we get,
\begin{equation}
     \frac{\chi[M_3^{\geq \epsilon},\mathbb{Z}_2]^{-1}}{|H^0_{\textrm{sub}}(M_3^{[0,\epsilon] },M_{2|0} \cup M_{2|\epsilon},\mathbb{Z}_2)|} \sum_{A \in H^1_{\textrm{sub}}(M_3^{[0,\epsilon]},M_{2|0} \cup M_{2|\epsilon},\mathbb{Z}_2)} Z[M_3^{\geq 0},A+\hat{A}_{M_3^{\geq \epsilon}}] (-1)^{\int_{M_3^{[0,\epsilon]}} A \cdot \hat{A} }.
\end{equation}

We then regularize the integral $\int_{M_3^{[0,\epsilon)}} A \cdot \hat{A}$ in the same way,
\begin{align}
    \int_{M_2\times [0,\epsilon]} (A^{xy} \hat{A}^z + A^{z} \hat{A}^{xy}) =& \sum_{i=1}^{L_x} \left( w_{y,i+\frac{1}{2}} \left[\sum_{k=1}^{L_z} \hat{A}^z_{i+\frac{1}{2},j+\frac{1}{2},k} \right]^y\right) + \sum_{j=1}^{L_{\epsilon}} \left( w_{x,j+\frac{1}{2}} \left[ \sum_{k=1}^{L_z} \hat{A}^z_{i+\frac{1}{2},j+\frac{1}{2},k}\right]^x \right) \nonumber\\
    &+\sum_{i=1}^{L_x}\left(w_{z,y;i} \sum_{j=1}^{L_{\epsilon}} \hat{A}^{xy}_{i,j,k+\frac{1}{2}} \right)+\sum_{j=1}^{L_{\epsilon}}\left(w_{z,x;j} \sum_{i=1}^{L_x}\hat{A}^{xy}_{i,j,k+\frac{1}{2}}\right).
\end{align}
If we take $\epsilon \rightarrow 0$ while fixing the holonomies $w$, we only need to keep the first term. Recall that from \eqref{Foliated-Theory-Operators-Correspondence} we have,
    \begin{equation}
        \exp \left( i \pi \left[\sum_{k=1}^{L_z} \hat{A}^z_{i+\frac{1}{2},j+\frac{1}{2},k} \right]^y \right) \equiv \hat{W}_{z,y}(x_{i+\frac{1}{2}}) \leftrightarrow \prod_{i'\leq i} U^{yz}_{0,i}.
    \end{equation}
The fusion rule is then,
    \begin{equation}
        \mathcal{N}^{\textrm{sub}}{}^{\dagger} \times \mathcal{N}^{\textrm{sub}} = \sum_{w_{y,i+\frac{1}{2}}} \prod_{i=1}^{L_x} \left(\prod_{i'\leq i} U^{yz}_{0,i} \right)^{ w_{y,i+\frac{1}{2}}}.
    \end{equation}
We can recover \eqref{Subsystem-Review-Fusion-Rule-Defects} using $w_{y,i+\frac{1}{2}} = t_i^y + t_{i+1}^y$.

 \bibliographystyle{ytphys}
 \baselineskip=.95\baselineskip
 \bibliography{bib}


\end{document}